%% file: statusreport.tex
\documentclass[12pt,a4paper]{article}
\usepackage{graphicx,color}
\usepackage{amsmath}
\usepackage{amssymb}
\pdfoutput=1
\usepackage{multirow}
\usepackage{url}
\usepackage{subfigure}
\usepackage{authblk}
\usepackage{fullpage}
\usepackage{lscape}
\begin{document}
\vspace*{-0.5in}
\thispagestyle{empty}

\vspace{0.5in}
\begin{center}
\begin{LARGE}
{\bf Status Report (BKG measurement):\\
A Search for Sterile Neutrino at J-PARC MLF\\}
\vspace{5mm}
\end{LARGE}
\begin{large}
November 18, 2014\\
\end{large}
\vspace{5mm}
{\large
M.~Harada, S.~Hasegawa, Y.~Kasugai, S.~Meigo,  K.~Sakai, \\
S.~Sakamoto, K.~Suzuya \\
{\it JAEA, Tokai, Japan}\\
\vspace{2.8mm}
E.~Iwai, T.~Maruyama\footnote{{Spokes person : Takasumi Maruyama (KEK) 
(takasumi.maruyama@kek.jp) }}, 
H.~Monjushiro, K.~Nishikawa, R.~Ohta, M.~Taira\\
{\it KEK, Tsukuba, JAPAN}\\
\vspace{2.8mm}
M.~Niiyama\\
{\it Department of Physics, Kyoto University, JAPAN}\\
\vspace{2.8mm}
S.~Ajimura, T.~Hiraiwa, T.~Nakano, M.~Nomachi, T.~Shima\\
{\it RCNP, Osaka University, JAPAN}\\
\vspace{2.8mm}
T.~J.~C.~Bezerra, E.~Chauveau, T.~Enomoto, H.~Furuta, H.~Sakai, \\
F.~Suekane\\
{\it Research Center for Neutrino Science, Tohoku University, JAPAN}\\
\vspace{2.8mm}
I.~Stancu\\
{\it University of Alabama, Tuscaloosa, AL 35487, USA}\\
\vspace{2.8mm}
M.~Yeh\\
{\it Brookhaven National Laboratory, Upton, NY 11973-5000, USA}\\
\vspace{2.8mm}
H.~Ray\\
{\it University of Florida, Gainesville, FL 32611, USA}\\
\vspace{2.8mm}
G.~T.~Garvey, C.~Mauger, W.~C.~Louis, G.~B.~Mills, R.~Van~de~Water\\
{\it Los Alamos National Laboratory, Los Alamos, NM 87545, USA}\\
\vspace{2.8mm}
J.~Spitz\\
{\it Massachusetts Institute of Technology, Cambridge, MA 02139, USA}
}
%\vspace{1mm}
\end{center}
\renewcommand{\baselinestretch}{2}
\large
\normalsize

\setlength{\baselineskip}{5mm}
\setlength{\intextsep}{5mm}

\renewcommand{\arraystretch}{0.5}

\newpage

\tableofcontents
\newpage

\vspace*{0.5in}
\setcounter{figure}{0}
\setcounter{table}{0}
\indent

\input{1Introduction.tex}
\input{2bkg_michele.tex}

\input{3bkg_prompt_500kg.tex}

\input{4bkg_prompt_tohoku.tex}
\input{5bkg_delayed_500kg.tex}

\input{7bkg_summary_P56.tex}

\input{8sensitivity.tex}

\input{9homework.tex}

\input{10Acknowledgement.tex}

\newpage
\appendix
\input{11detector_calib_500kg.tex}

\input{VetoEff.tex}
\input{12detector_Meigo.tex}

\input{13detector_tohoku24.tex}
\input{14bkg_delayed_tohoku24.tex}

\newpage

%%%%%%%%%%%%%%%%%%%%%%%%%%%%%%%%%%%%%%%%%%%%%%%%%%
% bibliography
%%%%%%%%%%%%%%%%%%%%%%%%%%%%%%%%%%%%%%%%%%%%%%%%%%

\end{document}

%% file: 1Introduction.tex
\section{Introduction}
\indent

At the 17th J-PARC PAC, which was held on September 2013, 
we proposed the sterile neutrino search at J-PARC MLF~\cite{CITE:P56Proposal}.
After reviewing the proposal, PAC recommended to have a background measurement at
the detector's candidate site location in their report~\cite{CITE:PACreport}
to investigate whether the background rates 
can be manageable for the real experiment or not.
Therefore, we have performed the background 
measurements (MLF; 2013BU1301 test experiment) during the summer of
2014, also following the 18th J-PARC PAC recommendations~\cite{CITE:PACreport18th},
and the measurements results are described here.

\section{Background Measurement at the MLF 3rd Floor}
\indent

\subsection{Goal and Strategy}
\indent

The goal of the 2013BU1013 measurements is to check 
the background rate, especially
related to the beam, written
in the P56 proposal~\cite{CITE:P56Proposal}. Table~\ref{tab:efficiency}
is the reprint background summary table from that proposal. Here we assumed 
that the two detectors, with 25 tons liquid scintillator fiducial volume 
(50 tons in total), are set on the MLF 3rd 
floor with the baseline of 17 meters. We also applied the selection criteria 
shown in Table~\ref{TAB:SC} (Reprint of Table~7 in ~\cite{CITE:P56Proposal})
to select the Inverse Beta Decay (IBD) signal events efficiently, which was 
caused by the oscillated signal from the muon decay at rest.
( $\mu^{+} \to e^{+} + \nu_{e} + \bar{\nu_{\mu}}$ (at the MLF mercury target); 
\hspace*{0.1mm} $\bar{\nu_{\mu}} \to$ (oscillation during 17 m)) 
$\to \bar{\nu_{e}}$;   
\hspace*{0.1mm} $\bar{\nu_{e}} + p \to e^{+} + n$ 
(at the liquid scintillator detector). 
The background rates from 
``beam associated fast neutron'' and ``Accidental events'' are 
to be checked. These numbers were estimated by the MLF 1st floor measurements
and the MC simulation in the proposal, therefore the direct measurement was
recommended by the PAC.

\begin{table}[htb]
\begin{center}
\begin{tabular}{|c|c|c|c|}\hline
&Contents&/4years/50tons&Comment\\\hline
&&&$\Delta m^2=3.0eV^{2}$,\\
Signal&$\overline{\nu}_{\mu}\to \overline{\nu}_{e}$ &811&$sin^22\theta=3.0\times10^{-3}$\\
&&&(Best $\Delta m^{2}$ for MLF exp.)\\\hline
&&&$\Delta m^2=1.2eV^{2}$,\\
&&337&$sin^22\theta=3.0\times10^{-3}$\\
&&&(Best fit values of LSND)\\\hline
&$\overline{\nu}_{e}$ from $\mu^{-}$&377& \\
&$^{12}C(\nu_{e},e^{-})^{12}N_{g.s.}$&38& \\
Backgrounds&beam associated fast neutron&0.3&\\
&Cosmic ray induced fast neutron&42& \\
&Total accidental events &37& \\\hline
\end{tabular}
\caption{\setlength{\baselineskip}{4mm}
{\bf A reprint table from the P56 proposal~\cite{CITE:P56Proposal}}.
Numbers of events of the signal and backgrounds with total fiducial mass of 50 tons.
}
\label{tab:efficiency}
\end{center}
\end{table}

\begin{table}[htb]
\begin{center}
\begin{tabular}{|c|c|} \hline
Cut Condition&Cut Efficiency \\\hline
1.0$\le \Delta$t$_{prompt}\le$10$\mu$s&74\% \\\hline
6$\le$E$_{delayed}\le$12MeV&78\% \\\hline
20$\le$E$_{prompt}\le$60MeV&92\% \\\hline
$\Delta$t$_{delayed}\le$100$\mu$s&93\% \\\hline
$\Delta$VTX$\le$60cm&96\% \\\hline\hline
Total&48\% \\\hline
\end{tabular}
\caption{\setlength{\baselineskip}{4mm}
{\bf A reprint};
IBD Selection criteria and efficiencies for the oscillated signals for high $\Delta m^{2}$ case.}
\label{TAB:SC}
\end{center}
\end{table}

The strategy to check the numbers are;
\begin{enumerate}
\item Background rates are measured by 
     a 500 kg plastic scintillator and other small scale (less than 10 kg)
     size detectors described in the appendix. 
\item The rates are scaled to the P56 detector with the fiducial mass of 
     25 tons at first, then multiplied twice for the two detectors. 
     The IBD selection criteria are also applied. 
     The scale is based on the MC simulation since the 25 tons detector 
     has self-shielding effects against gammas and neutrons, and the volume 
     ratio or the surface ratio between the 500 kg (or small size) detectors 
     and the 25 tons detector does not account for the rate alone.
\item In this report, we assume that the 25 ton detector has shielding 
     effects only from the stainless tank and the liquid scintillator 
     inside the tank. There are no additional shields made from
     iron, concrete or mineral oils. 
\end{enumerate}

Figure~\ref{FIG:50ton} shows the current 25 tons detector design described in
the Appendix\footnote{\setlength{\baselineskip}{4mm}
As describe in the proposal, the 3rd floor of the MLF is the maintenance area 
for such as a mercury target, beam monitor equipments, and so forth. Therefore, the detector should be moved to outide of the 
building for the maintenace works, which is held at least once per year.
The detector size is restricted by the entrance of the building.
}. 
Before backgrounds' arriving at the acrylic vessel, there is a 
buffer region made
by the liquid scintillator, whose thickness is 50 cm. This thickness is 
almost the same as the height and width of the 500 kg detector.  

\begin{figure}[h]
 \centering
 \includegraphics[width=1.1 \textwidth]{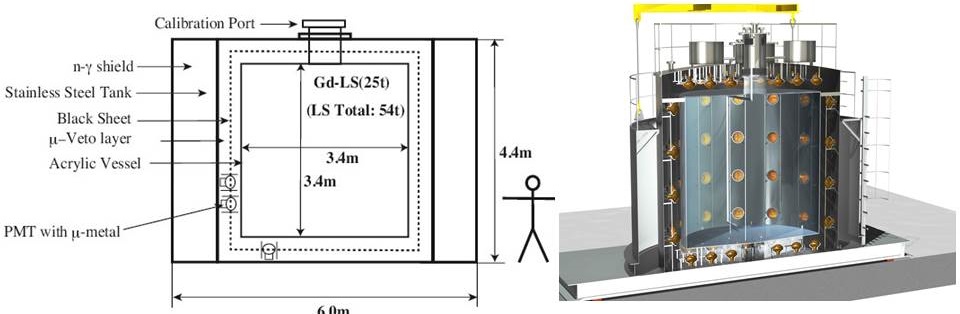}
\caption{\setlength{\baselineskip}{4mm}
Current design of the P56 25 tons detector. Left; schematic view, right;
3D drawings. The ``n-$\gamma$ shield'' in the left cartoon is not 
considered in 
this report, thus only stainless tank and 
scintillator inside the tank are assumed to shield the neutrons and gammas.
No additional shields such as irons, concrete or mineral oils are assumed. 
}
 \label{FIG:50ton}
\end{figure}

\subsection{Measurement Points and Operation Time}
\indent

\begin{figure}[h]
 \centering
 \includegraphics[width=1.1 \textwidth]{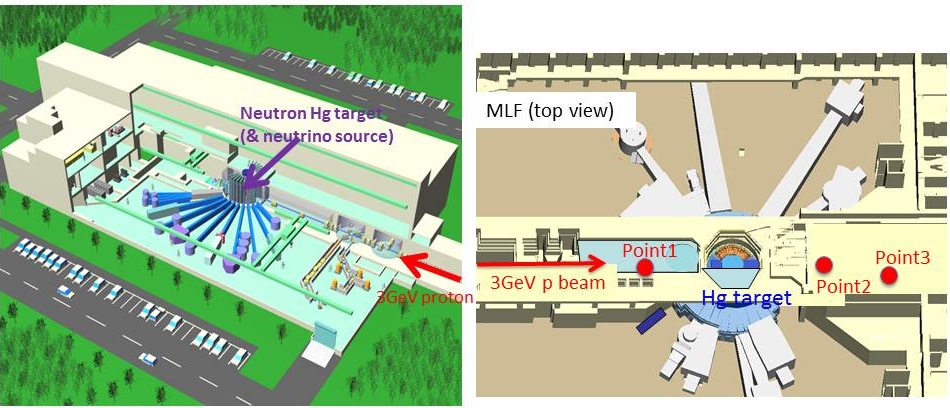}
\caption{\setlength{\baselineskip}{4mm}
A schematic view of the the MLF facility in J-PARC (left), and the
measurement points (red circles) of the 2013BU1301 experiment. (right; 
those are written as
``Point1'', ``Point2'' and ``Point3''))
}
 \label{FIG:MLF}
\end{figure}

Figure~\ref{FIG:MLF} shows the overview of the MLF building (left), and the 
measurement points of the 2013BU1301 experiment (right). The 500~kg 
plastic scintillator detector and their calibrations
are described in Appendix A.  The baselines from the
mercury target are 17m, 20m, 34m for Point1, 2 and 3 in Fig.~\ref{FIG:MLF},
respectively. We performed the background rate measurement for two weeks 
per each point.

Only the result of ``Point2'' is described here since 
the amount of the background rate at ``Point1'' is much larger than the
other points due to the proton beam multiple scattering at the muon 
target of MLF~\cite{JPARCsympo}, and the baseline for ``Point3'' takes 
longer time for the sterile neutrino search in the current configuration
using the 25 tons detectors due to the statistics.

In order to extrapolate the background rates from the test experiment to 
those of the proposed P56 25 tons detector, we assume 24 meters
baseline for the 25 tons detector in this report since 
the realistic two detectors are to be put around Point2
to manage the background rates. However, we cannot put the detectors 
in the area of the beam upstream of the Point2 
since there are many equipments for the MLF facility there. 
{\bf The possible detector location is being discussed in details 
with MLF facility people,  
and will be determined with all constraints.} 

During the test experiment, we found that the default operation 
time of the MLF facility is 5000 hours / year. Therefore we changed to 
the operation time from 4000 hours written in the proposal to 
5000 hours / year for the sensitivity calculation. 

%% file: 2bkg_michele.tex
\subsection{Background from Beam Fast Neutrons}
\label{SEC:ME}
\indent

One of the main purposes of this background measurement is to directly measure the Michel electron background induced by beam fast neutrons, which was indicated by the previous background measurement at BL13~\cite{CITE:P56Proposal}.
According to the Geant4 Monte Carlo simulation, fast neutrons whose kinetic energy are larger than 200 MeV can produce charged pions.
The flux of such fast neutrons at MLF 3rd floor is estimated to be some order of magnitude smaller than that at BL13.
We measured and confirmed that the beam Michel electron background is low enough for the experiment.
We will briefly describe the basic idea and backgrounds of this measurement.

\begin{figure}[hbtp]
	\begin{center}
		\includegraphics[scale=0.38]{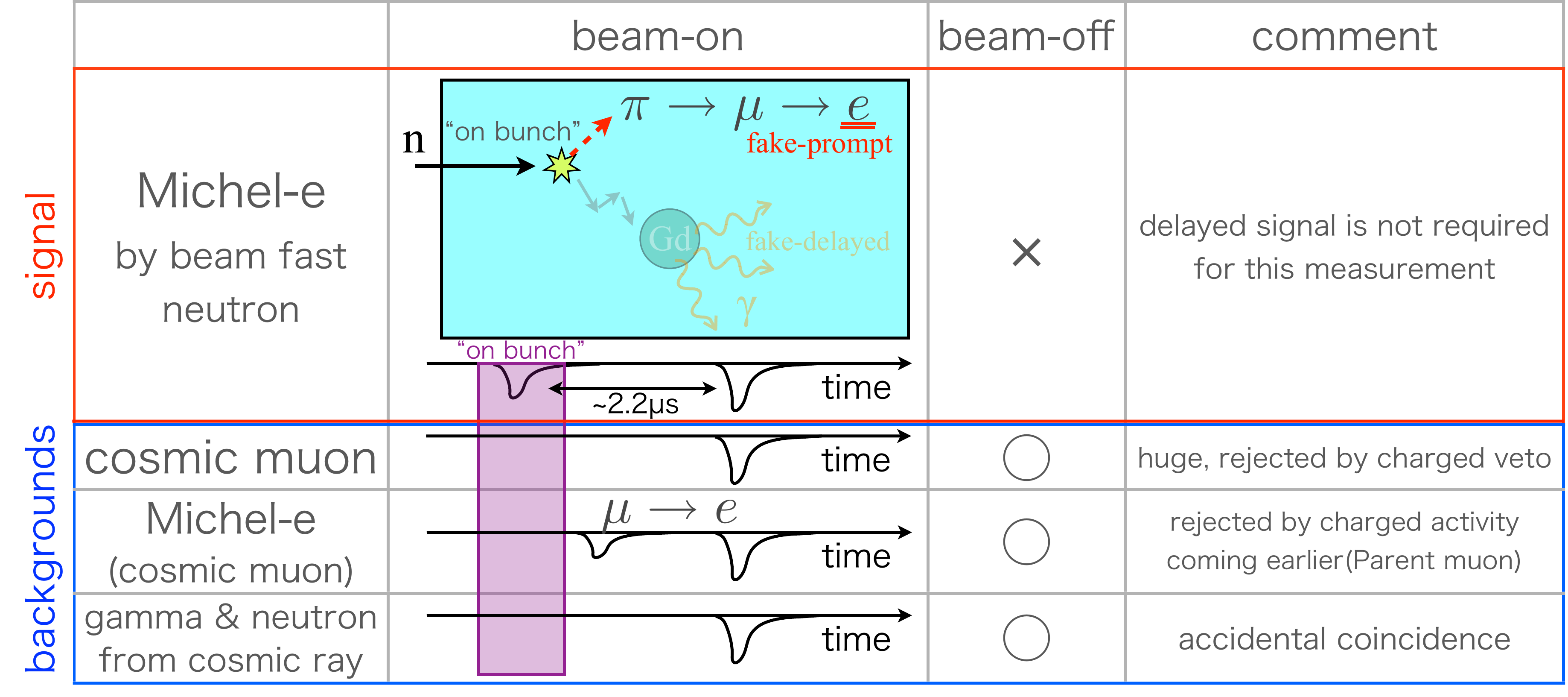}
		\caption{\setlength{\baselineskip}{4mm} The definition of ``signal'' and ``backgrounds'' of this measurement.
		Signal is a Michel electron induced by beam fast neutrons.
		Backgrounds are clipping cosmic muons, Michel electrons from cosmic muons and neutral particles from cosmic rays.}
		\label{500kg_measurement}
	\end{center}
\end{figure}

Figure \ref{500kg_measurement} shows the definitions of the ``signal'' 
and ``background'' of this measurement.
The signal is a Michel electron induced by beam fast neutrons.
Fast neutrons coming on the (proton) bunch timing hit our detector 
and produce pions.
These pions then decay into Michel electrons 
($n + p (or C) \to  X + \pi^+$, then $\pi^+ \to \mu^+ \to e^+$).
The signature of the signal is thus the coincidence between a neutral activity on the bunch timing and a ``prompt signal" about 1 $\mu$s later from the beam timing.
Backgrounds of this measurement are clipping cosmic muons, Michel electrons from cosmic muons and neutral particles (gammas and neutrons) from cosmic rays as was shown in Fig.~\ref{500kg_measurement},
and they come either during beam-on or beam-off.
On the other hand, signal comes only when the beam is on.
The basic idea of this measurement is thus to extract signals from backgrounds by subtracting beam-off activities from beam-on activities.

Based on the concept of the measurement, we took the following data set:
\begin{itemize}
\item beam-on: To observe activities on the bunch and just after bunch timing (the timing definition is described later);
\item 20ms-later : To subtract backgrounds from beam-on data\\
We took data with a 20 ms delay after each beam bunch spill.
Because of the PMT gains, efficiency of veto counters and other detector response are exactly the same with the last beam spill,
we can extract the backgrounds from the beam signals with less systematic uncertainties;
\item beam-off: To evaluate truly beam unrelated background;
\item cosmic muons: To calibrate the detector.
\end{itemize}

\subsubsection{Measurement}
\indent

\begin{figure}[hbtp]
	\begin{center}
		\includegraphics[scale=0.55]{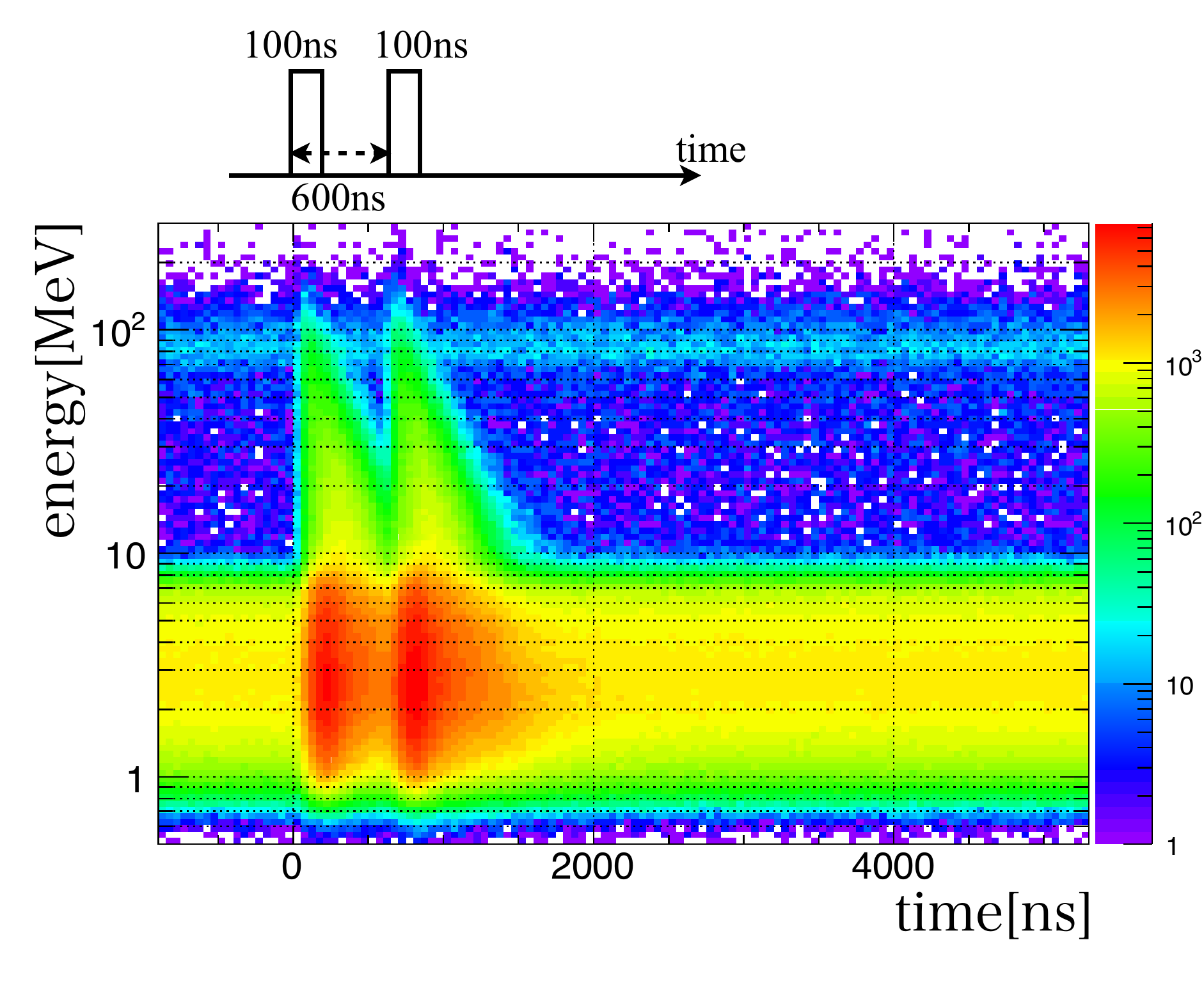}
		\caption{\setlength{\baselineskip}{4mm}Correlation between energy and timing of the events observed at Point 2.}
		\label{500kg_TvsE}
	\end{center}
\end{figure}

The search for beam neutron induced Michel electrons was performed by detecting their prompt signals.
Figure \ref{500kg_TvsE} shows the correlation between the energy and the timing of the events observed at Point 2.
In this measurement the prompt signal is defined by:
\begin{itemize}
\item $20<E[{\rm MeV}]<60$
\item $1.75<t[{\rm \mu s}]<4.65$ from the rising edge of the first beam bunch.
\end{itemize}

\begin{figure}[hbtp]
	\begin{center}
		\includegraphics[scale=0.55]{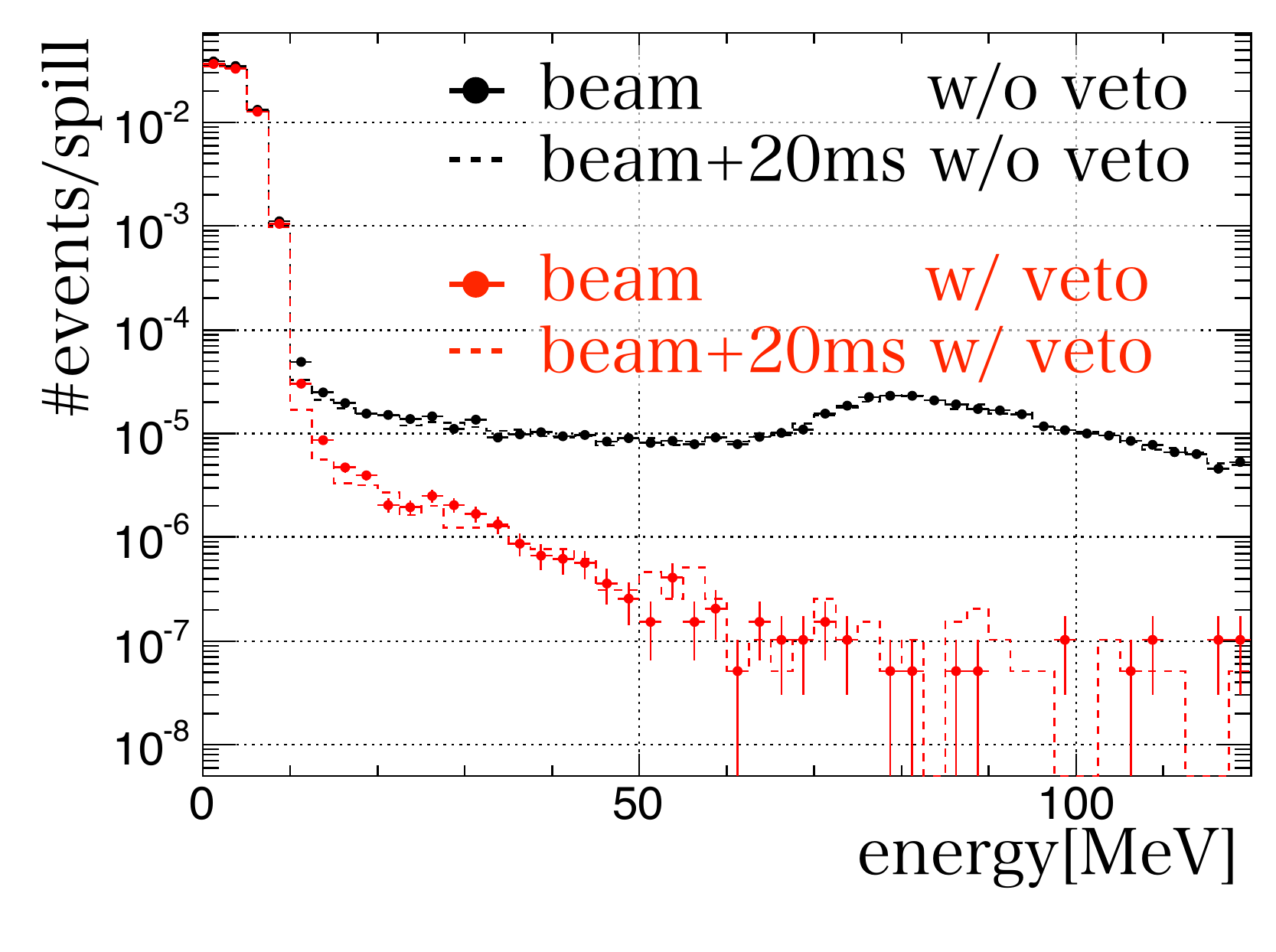}
		\caption{\setlength{\baselineskip}{4mm}Energy distributions of events in the prompt timing window and 20 ms later, before and after applying the charged veto cut.}
		\label{500kg_michele_edep}
	\end{center}
\end{figure}

As described before, we compare the beam-on and also 20ms-later data to subtract other activities with less systematic uncertainties.
A huge number of the clipping muon background is rejected by applying the charged veto cut.
Figure \ref{500kg_michele_edep} shows the energy distributions of events in the prompt timing window, $1.75<t[{\rm \mu s}]<4.65$ from the beam bunches, and 20 ms later, before and after applying the charged veto cut.
Without applying the charged veto cut, we observed $(1.68 \pm 0.03) \times 10^{-4}$/spill for beam data, and $(1.64 \pm 0.03) \times 10^{-4}$/spill for 20ms-later data.
Applying the charged veto cut, we observed $(1.58 \pm 0.09) \times 10^{-5}$/spill for beam data, and $(1.52 \pm 0.09) \times 10^{-5}$/spill for 20ms-later data.
The numbers of events in the prompt energy range are both consistent between beam and 20ms-later data either with or without applying the charged veto cut.

\begin{figure}[hbtp]
	\begin{center}
		\includegraphics[scale=0.55]{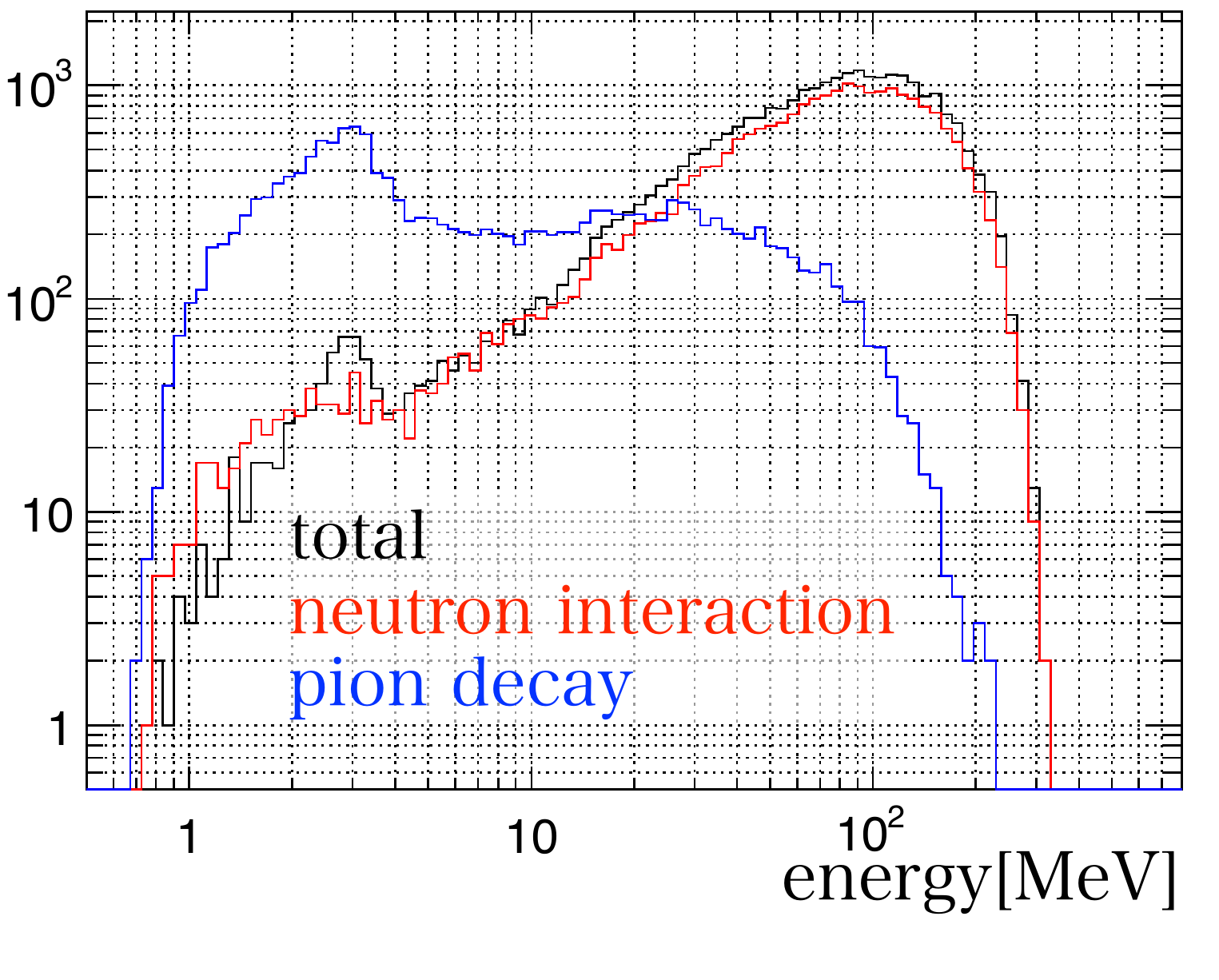}
		\caption{\setlength{\baselineskip}{4mm}Energy deposit on the bunch timing for beam Michel electron with the neutron kinetic energy 300 to 500 MeV (flat). Note this is MC.
		Red, blue and black lines are the energy deposit from neutron interactions, pion decays and their sum in each event.
		A peak around 3 MeV in blue histogram corresponds to the muon kinetic energy, 4.2 MeV, from stopped pion decay including the Birks' quenching.
		Birks' quenching, light attenuation in the scintillator and other detector responses such as resolutions and threshold effects were implemented to the Geant4 based Monte Carlo simulation as described in Appendix~\ref{sc_500kg_mc}.}
		\label{500kg_onbunch_edep}
	\end{center}
\end{figure}

\begin{figure}[hbtp]
	\begin{center}
		\includegraphics[scale=0.55]{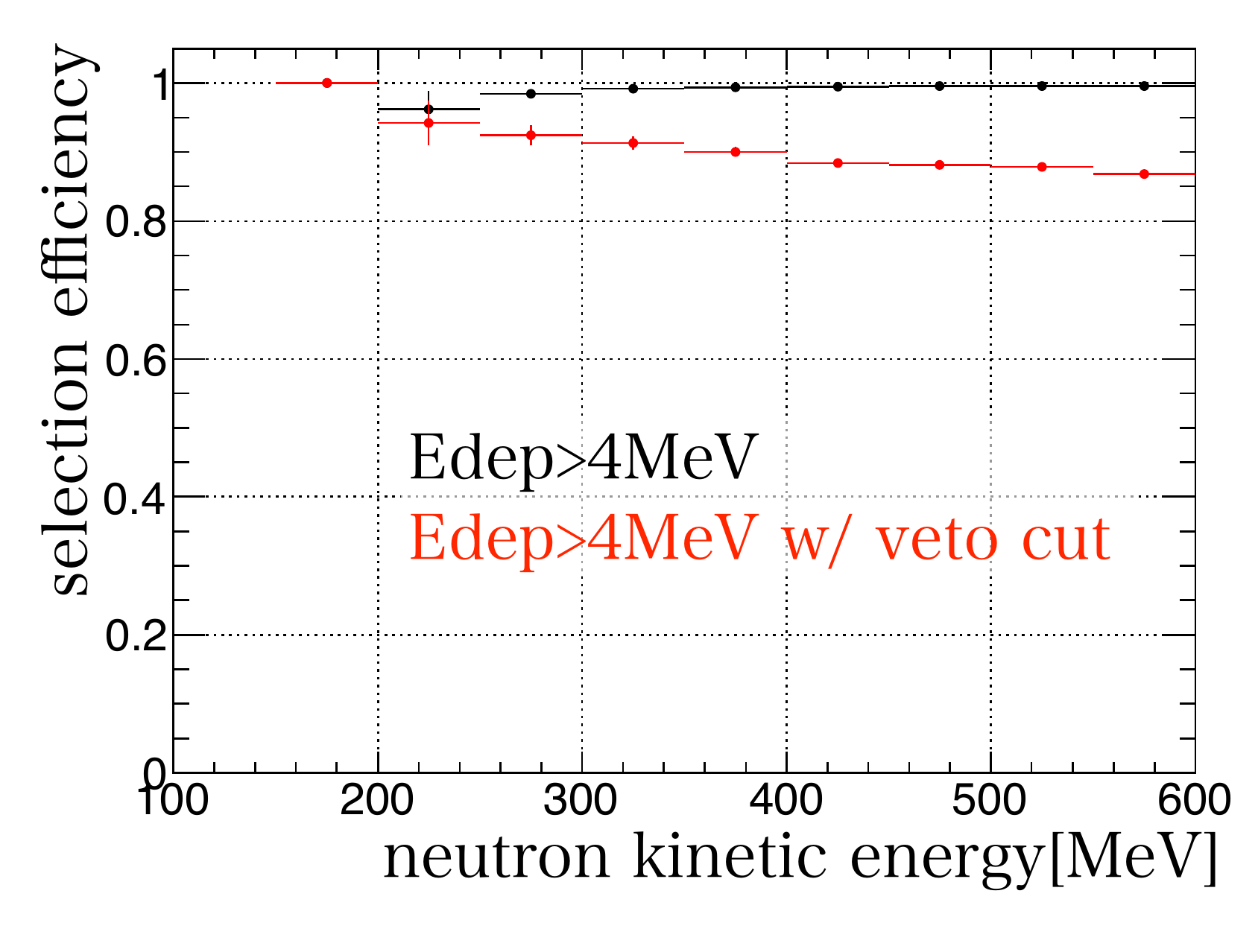}
		\caption{\setlength{\baselineskip}{4mm}Estimated selection efficiency of the on-bunch cut as a function of the incident neutron kinetic energy.}
		\label{500kg_onbunch_eff}
	\end{center}
\end{figure}

To improve the sensitivity, an additional cut was applied before the subtraction.
Figure \ref{500kg_onbunch_edep} shows the energy distribution on the bunch timing when fast neutrons produce charged pions in the 500 kg detector (MC).
We can observe some events on the bunch timing associated with beam Michel electron backgrounds.
On the other hand, as was shown in Fig.~\ref{500kg_measurement},  most of the beam unrelated backgrounds do not have any activities on the bunch timing.
Michel electrons from cosmic muons can have an activity on the bunch timing only when the parent muon comes on the bunch timing accidentally.
However, since the muon is a charged particle, it can be easily rejected by external veto counters.
We can thus strongly suppress backgrounds by requiring on-bunch activities without hits in veto counters.
At least one on-bunch activity (Edep$>$4 MeV) without hits in veto counters was required.
Figure \ref{500kg_onbunch_eff} shows the estimated selection efficiency of this on-bunch cut as a function of the incident neutron kinetic energy based on MC.
Though most of the events have more than 4 MeV energy deposit at on the bunch timing, a part of the events are rejected by self-vetoing.
The selection efficiency has slight dependence on the incident neutron kinetic energy.
Because we do not know the energy spectrum of the incident neutron well, we assumed the selection efficiency, $\epsilon_{\rm on bunch}=0.9$.

\begin{figure}[hbtp]
	\begin{center}
		\includegraphics[scale=0.55]{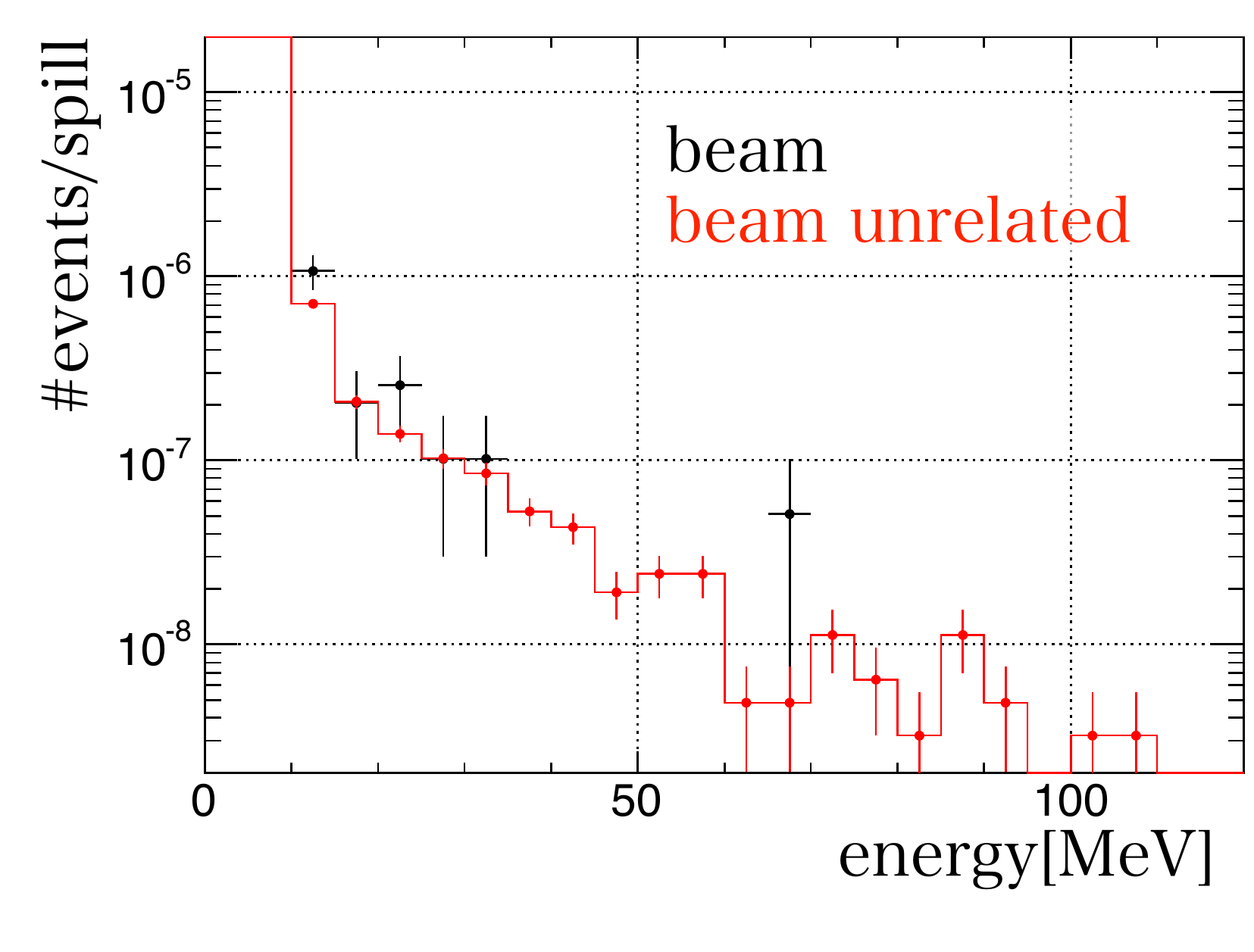}
		\caption{\setlength{\baselineskip}{4mm}Energy distributions after applying the on-bunch cut.
		The beam unrelated distribution was obtained with the accidental coincidence probability, 3.1\%, and the energy distribution of 20ms-later data without applying the on-bunch cut.}
		\label{500kg_michele_result}
	\end{center}
\end{figure}

Figure \ref{500kg_michele_result} shows the energy distributions after applying the cut (the 500 kg detector data).
The beam unrelated distribution was obtained with the energy distribution of 20ms-later data without applying the on-bunch cut, and the accidental coincidence probability: hit rate of neutral activities on the bunch timing.
The mean hit rate without veto activities on the bunch timing was 3.1\%.
The observed event rate during beam-on was $(4.60 \pm 1.53) \times 10^{-7}$/spill,
while the estimated event rate by beam unrelated activities was $(4.91 \pm 0.28) \times 10^{-7}$/spill, and both rates are consistent.
By considering the efficiency of the on-bunch cut, $\epsilon_{\rm on bunch}=0.9$, the upper limit of the event rate of the beam Michel electron is thus $2.5 \times 10^{-7}$/spill (90\% C.L.).
Because we set the timing window, $1.75<t[\mu s]<4.65$ from the beam, for the beam Michel electron,
the obtained upper limit is thus equivalent to $4.7 \times 10^{-7}$/spill (90\% C.L.) for the timing range, $1<t[\mu s]<10$ from the beam.

We then extrapolate the obtained upper limit to the 25 tons detector.
By considering the difference of the beam power (300kW $\to$ 1MW) and the detector acceptance\footnote{\setlength{\baselineskip}{4mm} We assumed fast neutrons, which produce beam Michel electrons, are directly coming from the mercury target.
The lead shield below the 25 tons detector, described in Section \ref{sec:Bkg_delayed_gammaextrapolate}, was also considered.
The scale factor by the detector acceptance is estimated to be 7.6 in average over two detectors.},
the obtained upper limit is equivalent to $1.1 \times 10^{-5}$/spill/detector/MW (90\% C.L.).
The upper limit is about 53 times larger than the event rate assumed in the proposal, and it is simply limited by the target mass and the measurement period.

%% file: 3bkg_prompt_500kg.tex
\subsection{Background from Accidental Coincidence}
\indent

Another important background is due to accidental coincidences.
In order to estimate the accidental background, an absolute rate of the background
measurements on the prompt and the delayed region with the 500 kg detector are
crucial. Measurements with small size detectors are also important to estimate
the contents (PID) of the prompt background and to estimate the rejection
ability of the lead blocks of low energy gammas.

We first explain the background measurements for the prompt signal region,
and then for the the delayed region.

\subsubsection{Background Measurements for Prompt Background Region}
\label{SEC:AP}
\indent

The estimation of the contents and rate of the prompt region
from a small size liquid scintillator and NaI(Tl) detectors is shown first,
then the measurement with the 500 kg detector for the prompt region is also described.
The results of the small size detectors and 500 kg detector are scaled to the
proposed 25 tons detector at the end.

\subsubsection*{PID and Energy Measurements with Small Size Detectors}
\indent

Appendix~\ref{sec:BKG_PID} shows the complete description of the measurements
with the small size detectors at Tohoku University, therefore only 
a summary of the measurements is shown here.  

\begin{figure}
	\centering \includegraphics[width=0.75\textwidth]{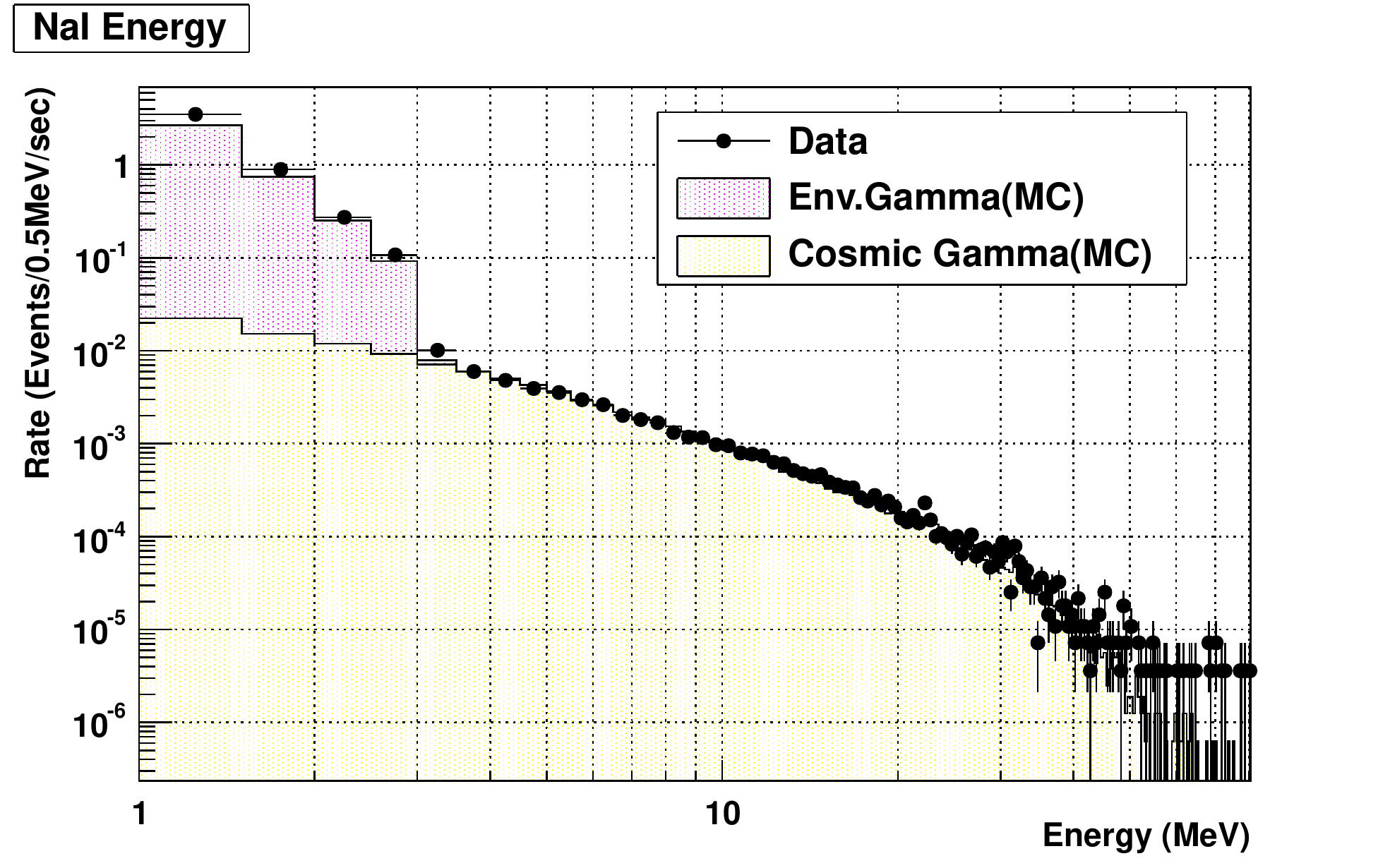}
	\caption{\setlength{\baselineskip}{4mm}
NaI energy spectrum for the remaining events after applying the veto cut.}
	\label{fig:NaI_veto2}
\end{figure}

A Sodium Iodide (NaI) scintillator counter (a cylinder with 2" 
of diameter and 2" height) was first used to measure the 
gamma flux that contributes to the prompt event of 
the accidental background. Cosmic ray muons are rejected using 
plastic scintillators, which surrounds NaI counters, with an efficiency better than
99$\%$. Figure~\ref{fig:NaI_veto2} shows the 
results of the measurement, where the remaining gammas energy spectrum can 
be described by two exponential functions with decay constants 
of 3 and 26~MeV (the gammas for MC were generated following the two exponential functions from the NaI surface assuming an isotropic direction). Comparing data and MC spectra, these components rates were measured to be 150 and 25~Hz/m$^2$, respectively.

\begin{figure}[htpb!]
	\centering \includegraphics[width=.75\textwidth]{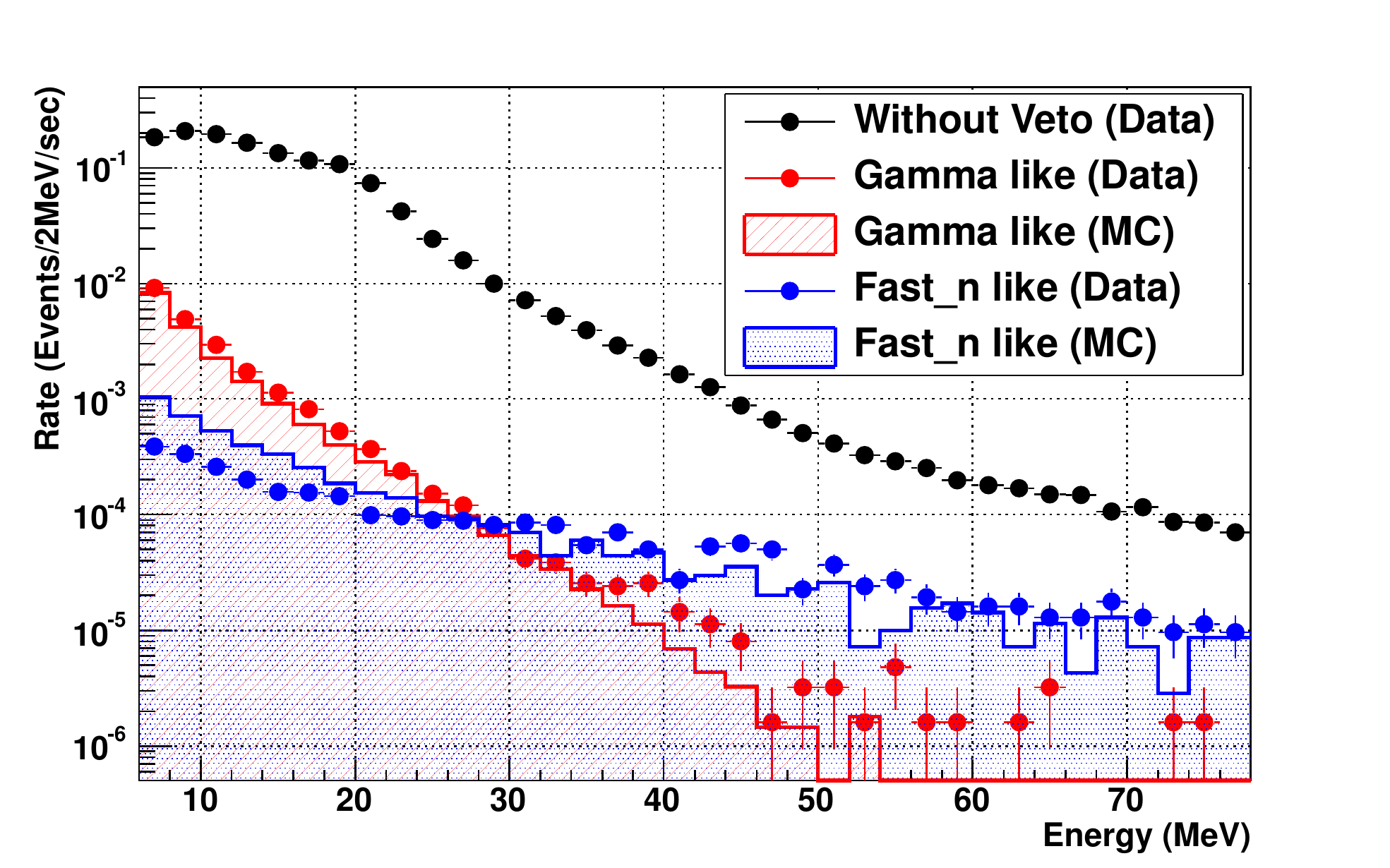}
	\caption{\setlength{\baselineskip}{4mm}
Reconstructed energy distribution for events with low or none energy deposition in the veto. The data and MC components of neutrons and gammas, selected as explained in Section B, are also compared. Data events without veto applied are also shown for comparison.}
	\label{fig:BKG_Meigo_Spec2}
\end{figure}

Another important measurements has been done using a small liquid 
scintillator (NE213)
which can separate neutrons from gammas efficiently.
A cylindrical aluminium housing (5" of diameter and 2" height), with white painted inner walls, 
was filled with NE213, closed with a glass plate 
and attached to a 5"  PMT (R1250-03).
The NE213 detector is also surrounded by the plastic scintillators, and Fig.~\ref{fig:BKG_Meigo_Spec2} shows the results of the measurement.

With this set-up, the fast neutron flux detected above 20~MeV is of ($1.28 \pm 0.05) \times 10^{-3}$~Hz (statistical uncertainty), while the MC gives a rate of $1.12 \times 10^{-3}$~Hz. These numbers are close to the one used in the 
proposal~\cite{CITE:P56Proposal}. 
For gammas, in the same energy range, the measured rate is of ($1.18 \pm 0.04) \times 10^{-3}$~Hz (statistical uncertainty), while the MC gives $0.95 \times 10^{-3}$~Hz. Therefore, the data and MC above 20~MeV for both gammas and neutrons agrees within 20\% of the uncertainty. For the MC presented here and for the one in the first proposal, the same cosmic neutrons generator was used. The generator's flux and spectrum were tuned with the Tohoku University's Reactor Monitor detector, composed of 200~litters of liquid scintillator~\cite{CITE:P56Proposal}.

\subsubsection*{Measurement for Prompt BKG with 500 kg Detector}
\indent

\begin{figure}[hbtp]
	\begin{center}
		\includegraphics[scale=0.55]{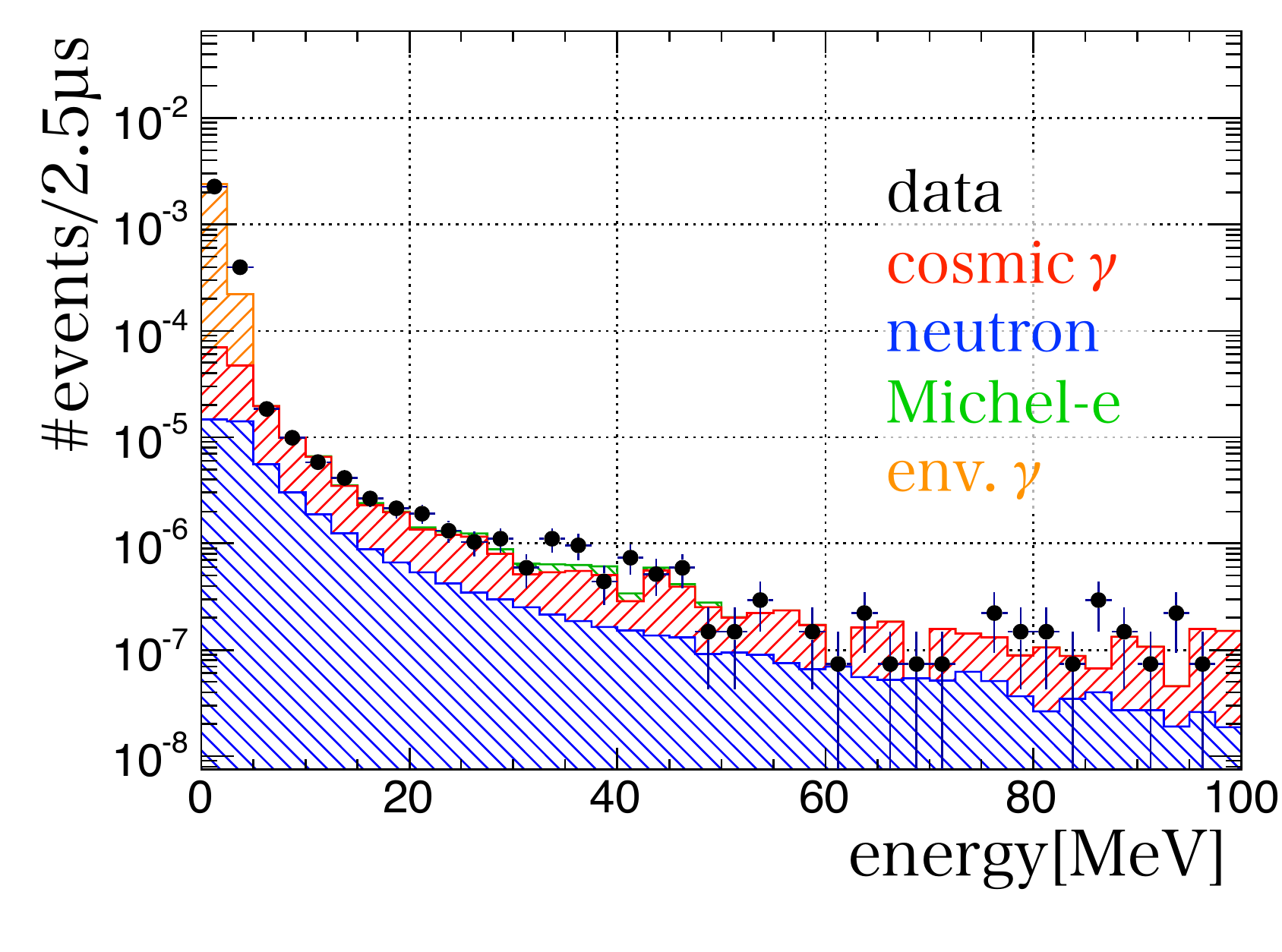}
		\caption{\setlength{\baselineskip}{4mm}Energy distribution of beam-off data after applying charged veto cut and the cosmic Michel electron cut.}
		\label{500kg_prompt_MichelCut}
	\end{center}
\end{figure}

Figure \ref{500kg_prompt_MichelCut} shows the energy distribution of beam-off data after applying charged veto and the Michel electron cuts. And the Monte Carlo estimation with cuts is overlaid.
The cosmic muons are eliminated by applying the charged veto cut because of the high veto efficiency ($>\!99.5$\%) as described in Appendix~\ref{sec:VetoEff}.
The Michel electrons can be rejected by detecting the parent muons coming earlier than the prompt events, and the remaining events are composed of cosmic gammas and neutrons.

The 25 tons detector for the sterile neutrino search will have PID capability, which can reject fast neutrons relative to electrons and gammas by a factor of more than 100\cite{CITE:P56Proposal}.
Neutrons in the remained events are thus not harmful and only gammas can be a prompt background.
The flux ratio and energy distributions of the cosmic gamma and neutron were measured by using small size liquid scintillator and the NaI(Tl) detector at Tohoku university as described in the previous section.
By using the $\gamma/n$ ratio, the number of gammas in the remained events is equivalent to $(6.4 \pm 0.5) \times 10^{-6}$ /2.5$\mu$s.

Note that the predicted rate from the measurement at the RCNS (Tohoku 
University) is consistent with the rate measured by the 500 kg detector at 
the MLF within 6$\%$. 
Thus, a reliable extrapolation to the proposed 25 tons detector is possible, 
since the MC model agrees for different detector types (NaI, NE213, 
Reactor Monitor, and 500~Kg plastic scintillator) at different 
locations (RCNS and MLF).

%% file: 4bkg_prompt_tohoku.tex
\subsubsection*{Background at the proposed P56 25 tons detector}
\label{subs:BKG_spected}
\indent

\begin{figure}[htpb!]
	\centering	\includegraphics[width=.6\textwidth]{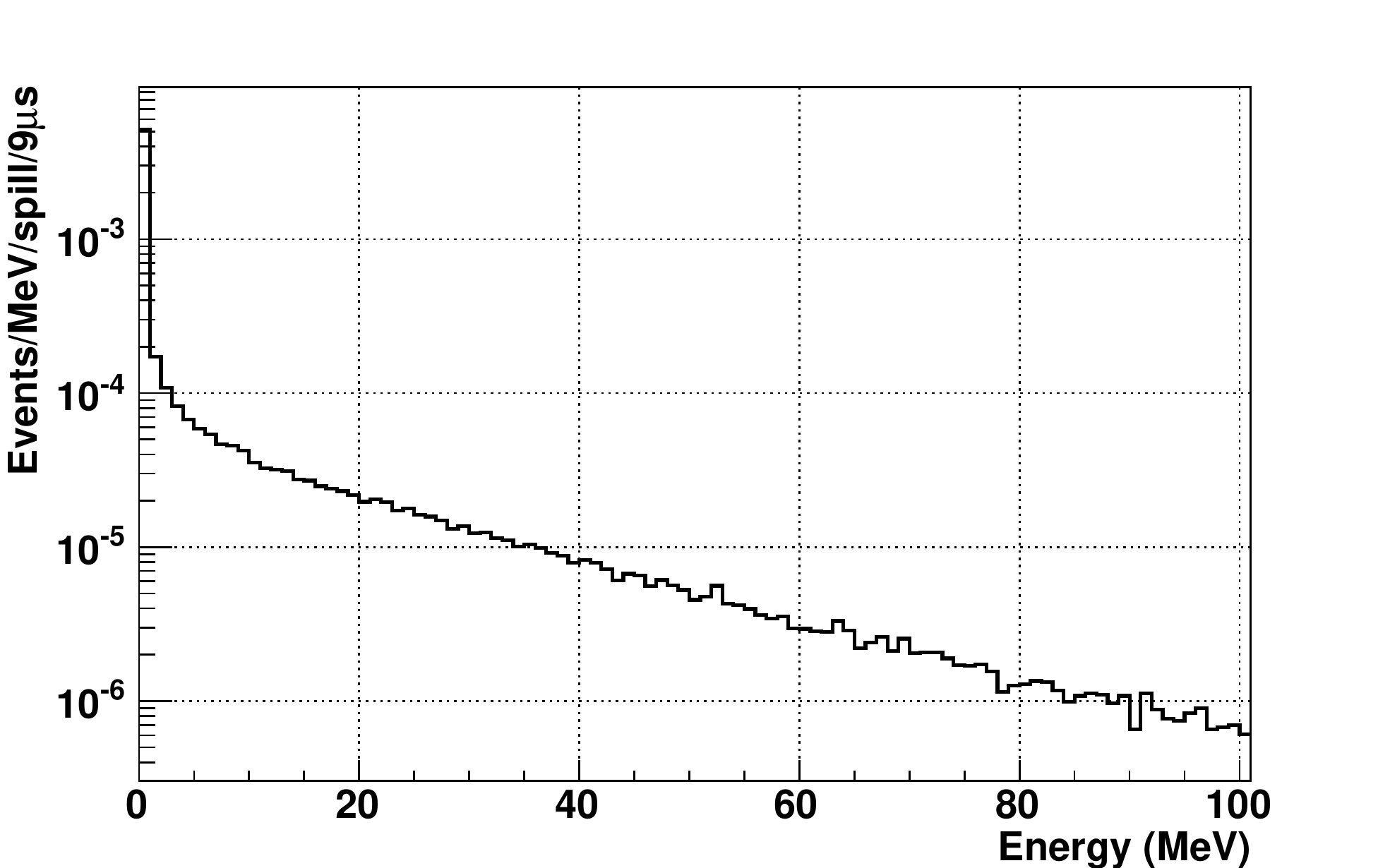}
	\caption{\setlength{\baselineskip}{4mm}
High energy MC gammas energy distribution for the proposed detector.}
	\label{fig:BKG_spec_proposal}
\end{figure}

As a final step, the background measurements explained in the previous sections were scale to this 25 tons proposed detector. When generating high energy gammas, following the measured spectrum and the same MC generators for the 500 kg detector for the proposed detector, the rate of events remaining after the selection criteria is of  $3.8 \times 10^{-4}$~events/spill/9$\mu$s, a value 29 times higher than the last proposal. This is simply because we didn't consider this contribution. Figure~\ref{fig:BKG_spec_proposal} shows the spectrum of such events.
The estimated number of neutrons events, using the same rate as in the previous proposal, is 32.2 in 4~years, considering 4000~hours of operation in each year.

%% file: 5bkg_delayed_500kg.tex
\subsubsection{Background measurements for Delayed Signal}
\label{sec:Bkg_delayed}
\indent
\indent
In this section we report on background measurements and background estimation of the 25 tons detector for delayed signal.
As seen in section~\ref{subs:BKG_spected}, 
the background rate for the prompt signal is about 30 times higher than the last proposal.
Therefore we have to manage the accidental background rate reducing the background rate for the delayed signal to a value more than 30 times smaller than the proposal. 
The schematics of the 25 tons detector is shown in Fig.~\ref{FIG:50ton}.
Outside the acrylic vessel is the buffer region,
which is separated into two layers with a black sheet.
The most outside region is used as a veto cut.

The energy region for the delayed signal is 7 $< E [{\rm MeV}] <$ 12. 
This range is different from the region taken in the previous proposal to reduce the amount of background.
The effect of tightening energy selection is discussed in Section 3.1.1.

Backgrounds for the delayed signal were also measured with the 500 kg scintillator detector at the 3rd floor in MLF.
There are two types of backgrounds for the delayed signal.
\begin{itemize}
\item Gammas by neutron captures in the materials outside of the detector. 
The gammas directly hit the detector from outside and mimic a delayed signal.
\item Neutrons which come into the detector from outside, 
are captured in the detector and emit gammas of which energy is same as delayed signal.
\end{itemize}
Based on the measurement, we estimated MC models of backgrounds for gammas and neutrons.
Using the MC models, we estimated amounts of background in the 25 tons detector.

\subsubsection*{Delayed Background from Beam Gammas}
\label{sec:Bkg_delayed_gamma}
\begin{figure}
\begin{center}
\includegraphics[width=12cm]{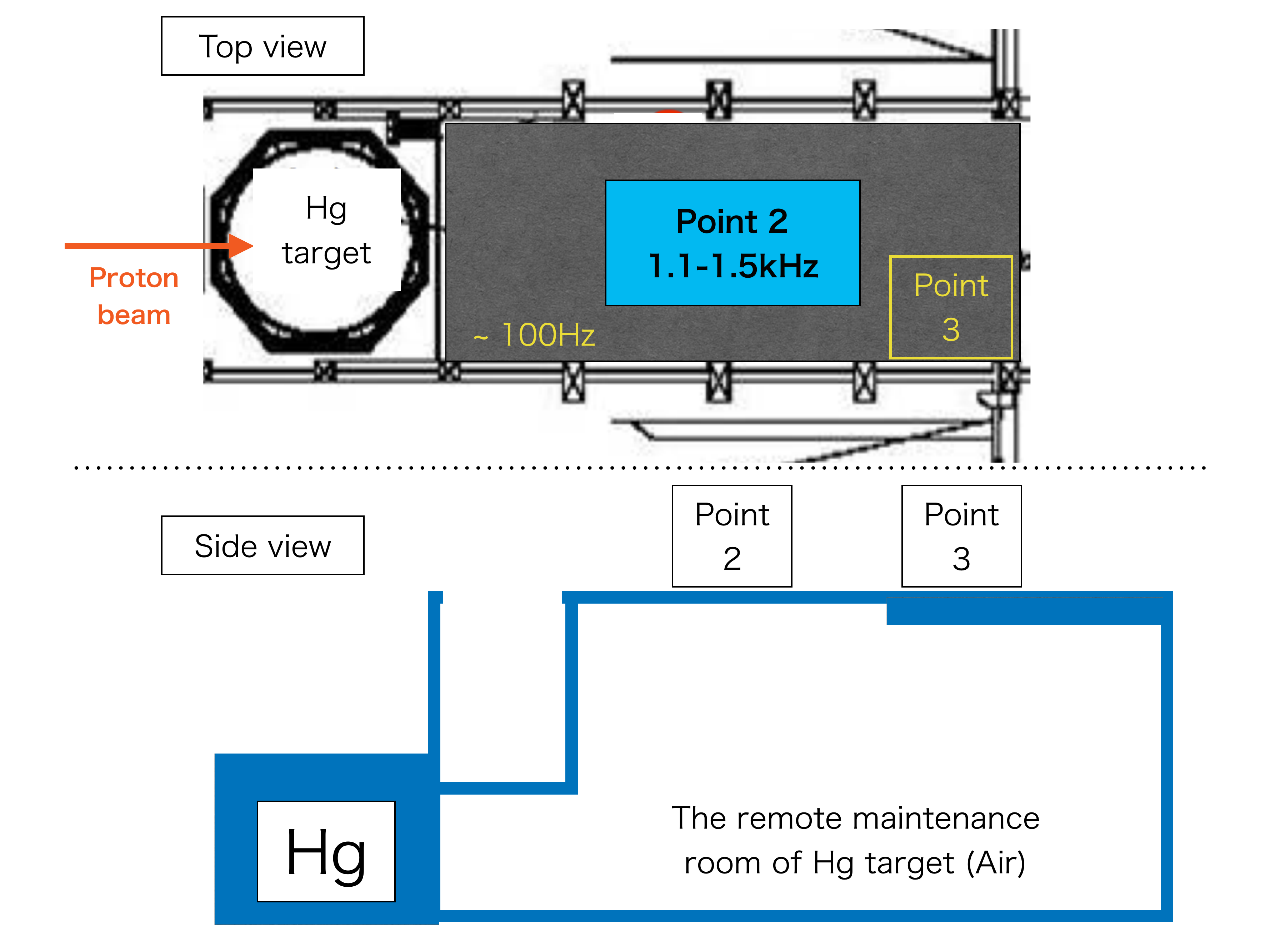}
\caption{\setlength{\baselineskip}{4mm}
Top part of the figure shows the top view of measurement location and the result of preliminary measurement. 
In the cyan area around Point 2, the background rate was more than 1 kHz. 
In the gray area, the background rate was about 100 Hz. 
Bottom part of the figure is the side view of measurement location. 
The floor of Point 2 is made of concrete which has 120 cm thickness and is thinner than that of Point 3 by 30 cm.}
\label{fig:Delayed_premeasurement}
\end{center}
\end{figure}

\indent
\indent
As a preliminary measurement we observed an unexpected background rate around Point 2 and 3 with a small scintillator and an oscilloscope. 
Around Point 2, the background rate was more than 1 kHz with a threshold equivalent to $\sim$ 1 MeV.
This rate was more than 10 times higher than that of other locations including Point 3.
The location and result are shown in the top part of Fig.~\ref{fig:Delayed_premeasurement}.
Because this rate decreased to 200 Hz when 5 cm-thick lead blocks were put below the small scintillator, 
this background can be regarded as gammas.
In addition, this rate decreased to 100 Hz with 10 cm-thick lead blocks under the small scintillator. 

On the other hand, the rate decreased to 150 Hz when 5-cm thick lead blocks surrounded almost all over the small scintillator.
Because lead blocks of bottom side had larger shield effect than those of other side, gamma background was coming from the floor. (See also Appendix C for 
more details.)

Measurements with the 500 kg detector also indicated the background rate at Point 2 was several ten times higher than the rate at Point 3 in the energy region of 7 $< E [{\rm MeV}] <$ 12.
Therefore it is important to understand the source of gamma background at Point 2 and how to suppress it.

\begin{figure}
\begin{center}
\includegraphics[width=10cm]{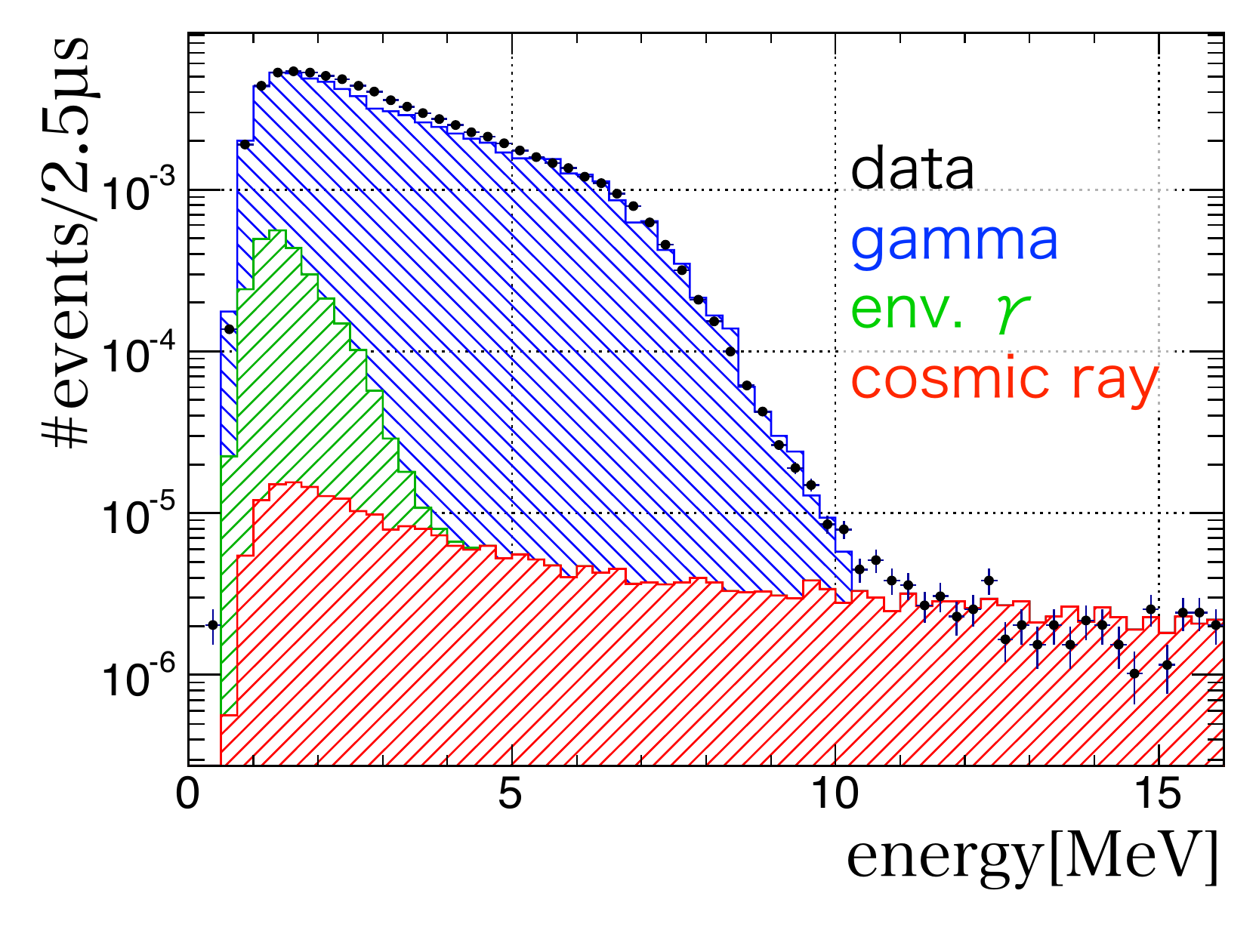}
\caption{\setlength{\baselineskip}{4mm}
Energy distribution of background for delayed signal and other activities measured with the 500 kg detector}
\label{fig:Delayed_gamma_measurement}
\end{center}
\end{figure}

The energy spectrum for the delayed signal at Point 2 is shown in the black dots of Fig.~\ref{fig:Delayed_gamma_measurement}.
In the energy region of 7 $< E [{\rm MeV}] <$ 12, the background rate with the 500 kg detector is $(2.06\pm0.01)\times 10^{-3}$/2.5$\mu$s/0.3MW.
To explain this background, we estimated a MC model.

As shown in Fig.~\ref{500kg_TvsE}, the energy distribution does not change till 40 ms after the beam bunch.
Therefore the background rate for the 2.5 $\mu$s time window is linearly scaled to the rate for the 100 $\mu$s time window.

\subsubsection*{MC model for gamma background}
\label{sec:Bkg_delayed_gammaMC}
\indent
\indent
In order to construct a reliable MC model for gammas production, 
the followings are assumed.  
As shown in the bottom part of Fig.~\ref{fig:Delayed_premeasurement}, the remote maintenance space of the mercury target exists under Point 2 and 3. 
Neutrons generated by the beam are captured at the ceiling of the room which is made of concrete.
Gammas are emitted as a secondary product of the neutron capture by the concrete, and consequently they reach Point 2 and 3.
Because the concrete thickness at Point 2 is about 120 cm, 
which is removable for maintenance and thinner than at Point 3 by 30 cm, 
Point 2 have a higher background rate.

\begin{figure}
\begin{center}
\includegraphics[width=10cm]{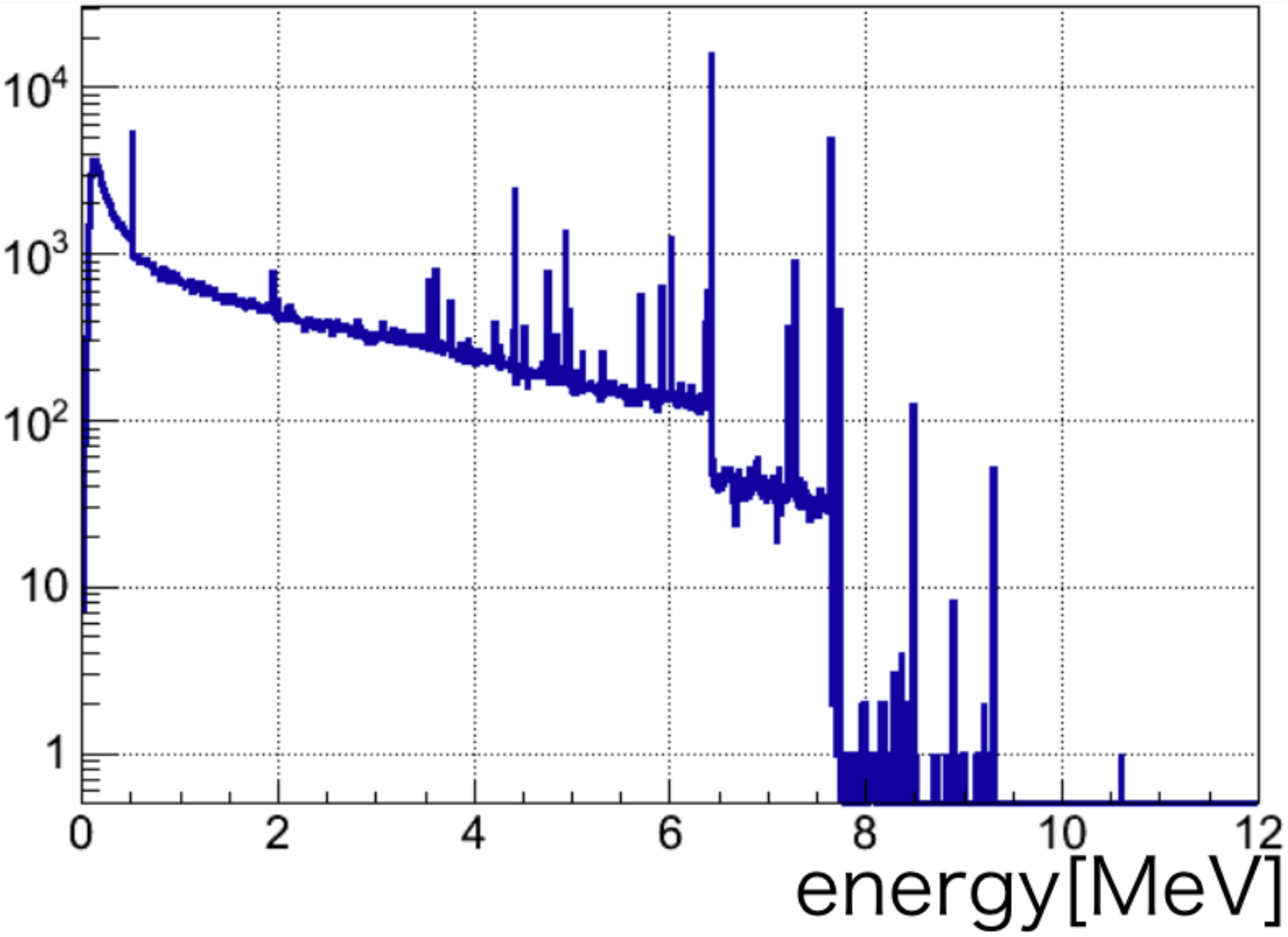}
\caption{\setlength{\baselineskip}{4mm}Energy distribution of gammas in the MC model of Point 2} 
\label{fig:MCgammaE}
\end{center}
\end{figure}

\begin{figure}
\begin{center}
\includegraphics[width=10cm]{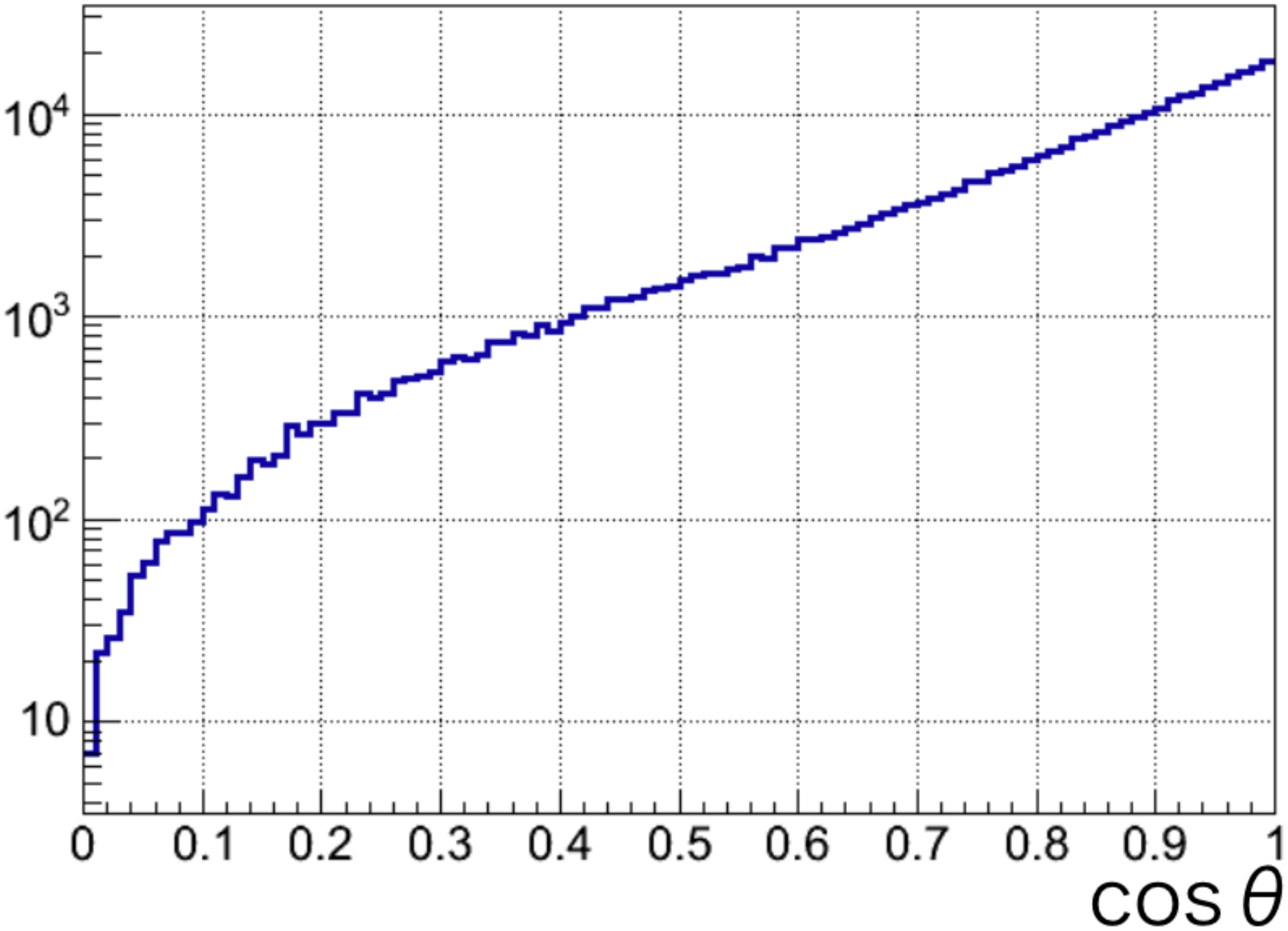}
\caption{\setlength{\baselineskip}{4mm}
Angle distribution of gammas in the MC model of Point 2. $\cos\theta=1$ means that gamma with a vertical direction.} 
\label{fig:MCgammacos}
\end{center}
\end{figure}

Having these assumptions in mind, the energy and angle distributions of gamma of the MC model are shown in Fig.~\ref{fig:MCgammaE} and \ref{fig:MCgammacos}.
Gammas reaching Point 2 have energies reflecting the neutron capture by concrete and 
vertical gammas reach the detector easier than the ones with large angle.

In the MC simulation, these gammas are generated uniformly on a floor around the detector.
We validated this MC model, comparing between measurements and MC with the small plastic scintillator in various shield configurations (see Appendix~\ref{sec:tohoku24_measurement}). 
We also validated the MC model with data taken by the 500 kg detector.
As shown in Fig.~\ref{fig:Delayed_gamma_measurement}, the energy spectrum is well explained by the MC model (blue area), environmental gammas (green area) and cosmic rays (red area).

\subsubsection*{Extrapolation to the 25 tons detector for gamma background}
\label{sec:Bkg_delayed_gammaextrapolate}
\indent

Using the MC model, we estimated the background rate of the 25 tons detector.
The background rate, $N^{25{\rm t}}_{\rm expected}$, is
\begin{equation}
N^{25{\rm t}}_{\rm expected}=\frac{N^{25{\rm t}}_{\rm MC}}{N^{\rm 500kg}_{\rm MC}}N^{\rm 500kg}_{\rm obs}, \label{eq:extrapolate}
\end{equation}
where $N^{25{\rm t}}_{\rm MC}$ is the calculated number of events of the 25 tons detector with MC, 
$N^{\rm 500kg}_{\rm MC}$ is the calculated number of events of the 500kg detector with MC and $N^{\rm 500kg}_{\rm obs}$ is the observed background rate.
Without any shield, $N^{25{\rm t}}_{\rm expected}$ is 1.5 /100$\mu$s/MW/detector in the energy region of 7 $< E [{\rm MeV}] <$ 12. 
This value is 32 times larger than the value of the proposal.
To compensate this increment, at least $\sim$ 6 cm thick lead blocks are 
needed.   
As confirmed in Appendix~\ref{sec:tohoku24_measurement}, adding a lead shield below the detector is effective to reduce this gamma background. Note that we tested the effects of 10 cm thick lead blocks in the configurations. 
In this report, we assume to put 12.5~cm thick lead shield under the 25 tons detector to suppress the background, 
and to compensate the prompt background increase as well. 
Thus, $N^{25{\rm t}}_{\rm expected}$ is reduced to 1.2 $\times 10^{-3}$ /100$\mu$s/MW/detector.~\footnote{
\setlength{\baselineskip}{4mm}
As discussed later, we apply $\Delta VTX_{\rm OB-delayed}$ cut to reduce neutron background. 
The background rate of beam gammas is also affected by the cut and reduced by 2.0\% as well as neutrino signal. 
However, the effect of the reduction can not be seen in this digits.
}

\subsubsection*{Delayed Background from Beam Neutrons}
\label{sec:Bkg_delayed_neutron}
\begin{figure}
\begin{center}
\includegraphics[width=10cm]{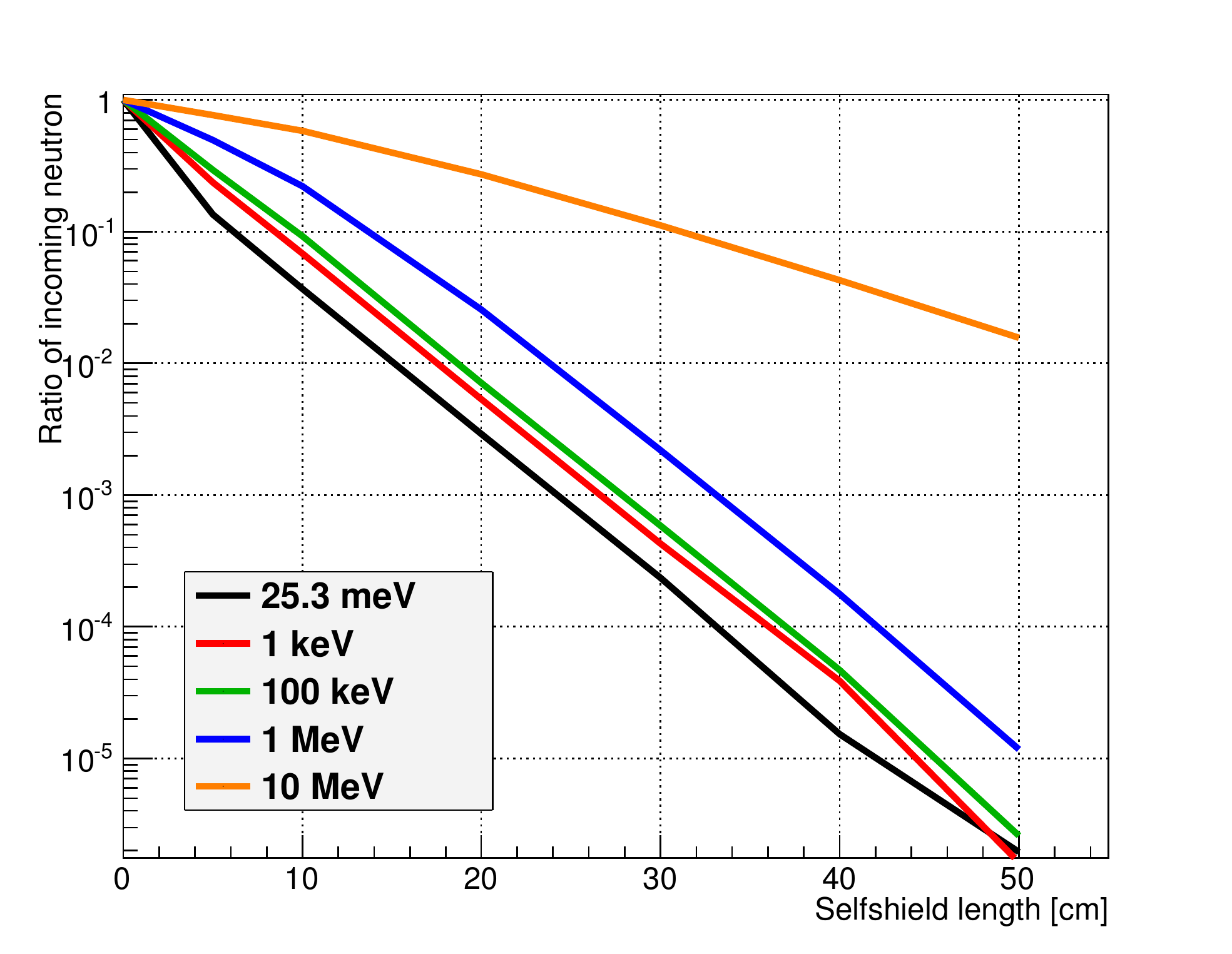}
\caption{\setlength{\baselineskip}{4mm}Self shield effect of liquid scintillator as function of the shield thickness.}
\label{fig:self_shield_n}
\end{center}
\end{figure}

\indent
\indent
As shown in Fig.~\ref{FIG:50ton}, the 25 tons detector have a 50~cm thick buffer region that acts as a self-shield.
This shield effect for the low energy neutrons (less than 10 MeV) is shown in Fig.~\ref{fig:self_shield_n} (See Sec. 5.8 of ~\cite{CITE:P56Proposal}) and 
its neutron rejection power is shown 
in Table~\ref{tab:rejectionpower}, including the energies higher than 10 MeV. 
Low energy neutrons (kinetic energy $E_n\lesssim 10$ MeV) is significantly reduced by self-shield of the 25 tons detector.
On the other hand, high energy ($E_n\gtrsim 10$ MeV) neutrons can reach the fiducial volume and are thermalized and captured by the Gd.
Therefore neutrons with $E_n\gtrsim 10$ MeV become background for delayed signal.

Because such neutrons come at beam on-bunch timing only, 
we estimated the background rate of neutron at the 25 tons detector from on-bunch energy spectrum measured with the 500 kg detector.
We use the energy spectrum above 15 MeV since the gammas are dominant below 15 MeV.
For on-bunch timing, we use 3 $\mu$s time window after the rising edge of the first beam bunch.
In Fig.~\ref{500kg_TvsE}, 2D plot for energy and timing of activities measured with the 500 kg detector is shown.
Energy of on-bunch activities is projected to the black cross in Fig.~\ref{fig:Onbunch_500kg_fit}.
An estimating method is as follows;
\begin{enumerate}
\item We estimated a neutron flux which explains the measurement of the 500 kg detector with MC.
\item Amount of activities in the 25 tons detector was estimated using the flux and MC.
\end{enumerate}
For the following estimation, we assumed that all activities of on-bunch timing in the 500 kg detector are caused by neutrons.
This assumption could be conservative to estimate amount of neutron background because there could be some fraction of gammas as well.
\begin{table}
\begin{center}
\begin{tabular}{|c|c|c|c|c|c|}
\hline
Initial energy [MeV] & 0-20 & 20-40 & 40-60 & 60-80 & 80-100 \\
\hline
rejection power & 75 & 13 & 6.1 & 4.4 & 3.7 \\
\hline
\hline
Initial energy [MeV] & 100-120 & 120-140 & 140-160 & 160-180 & 180-200 \\
\hline
rejection power & 3.3 & 3.1 & 2.9 & 2.7 & 2.6 \\
\hline
\end{tabular}
\caption{\setlength{\baselineskip}{4mm}Rejection powers of selfshield of the 25 tons detector. Rejection power is ratio of the number of incoming neutrons at the surface of the 25 tons detector to the number of neutrons reaching the surface of the fiducial volume.}
\label{tab:rejectionpower}
\end{center}
\end{table}
\begin{figure}
\begin{center}
\includegraphics[width=10cm]{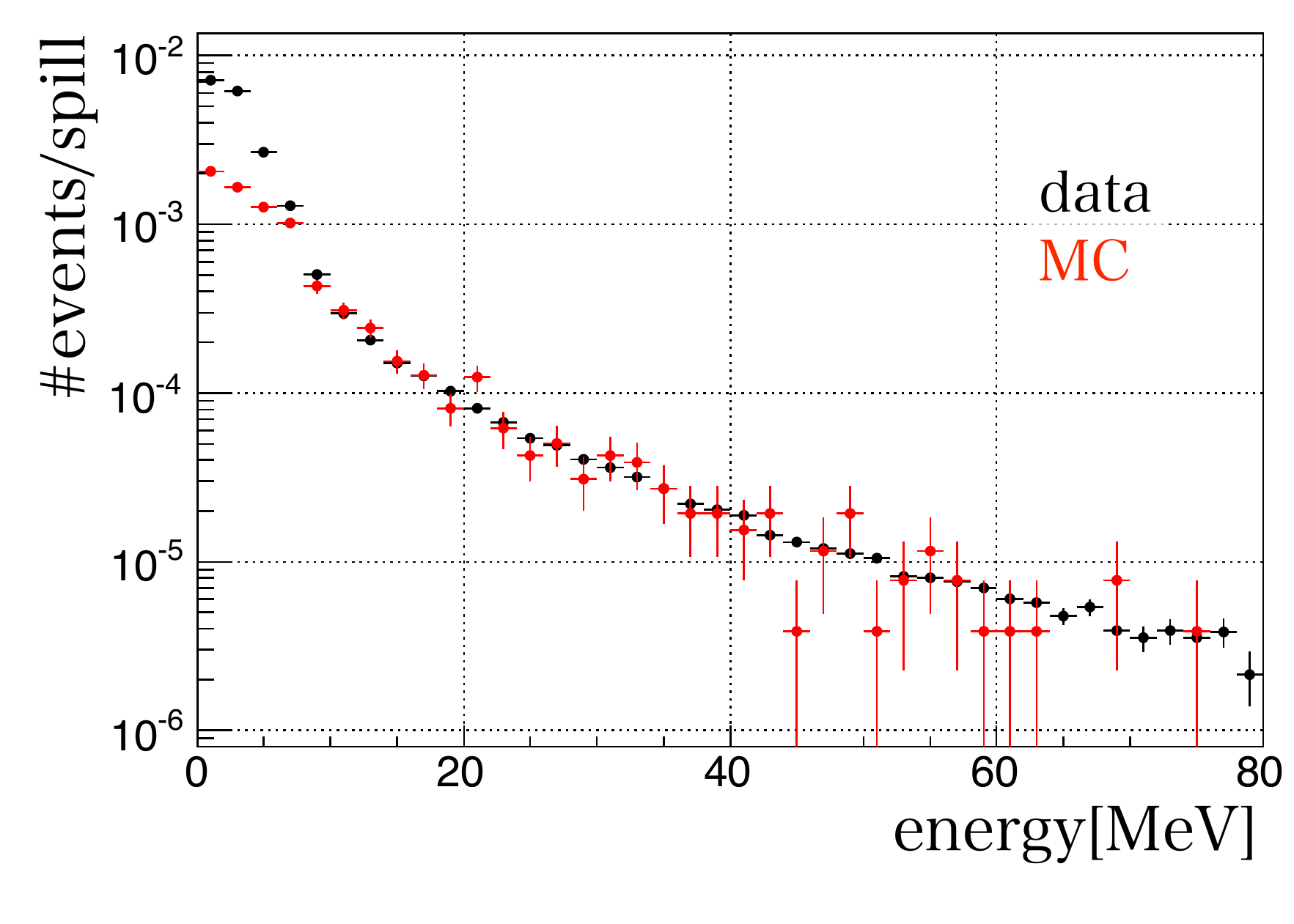}
\caption{\setlength{\baselineskip}{4mm}Energy distribution of on-bunch timing with the 500 kg detector. The black cross is the measured energy spectrum and the red one is a MC fitted to the measured one.}
\label{fig:Onbunch_500kg_fit}
\end{center}
\end{figure}

\subsubsection*{Neutron flux}
\label{sec:Bkg_delayed_neutronflux}
\begin{figure}
\begin{center}
\includegraphics[width=10cm]{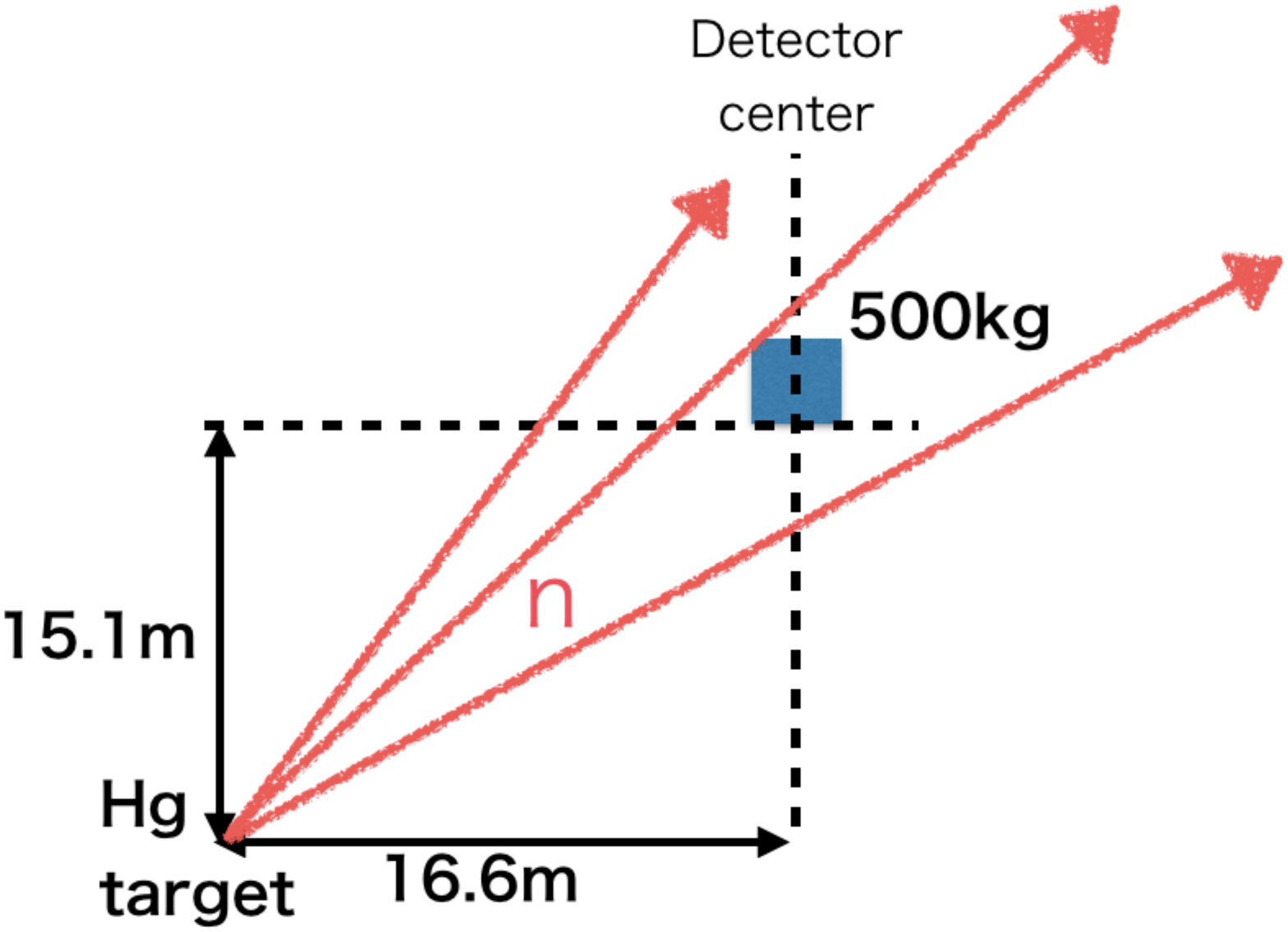}
\caption{\setlength{\baselineskip}{4mm}Geometry of the mercury target and the 500 kg detector}
\label{fig:Nflux_scheme_500kg}
\end{center}
\end{figure}

\indent
\indent
To explain the measured spectrum, we estimated an energy spectrum of the neutron flux.
The geometry used is shown in Fig.~\ref{fig:Nflux_scheme_500kg}.
We assumed that neutrons come directly from the mercury target.
The flux, $\phi(E_{\rm n})$, is expressed as
\begin{equation}
\phi(E_{\rm n}) = \frac{\alpha}{30} \exp(-E_{\rm n}/30{\rm MeV}), \label{eq:beamNflux}
\end{equation}
where $E_{\rm n}$ is kinetic energy of neutron and $\alpha$ is a number of neutrons that are generated at the mercury target in a spill.
$\alpha$ was estimated to be 387/spill as a fit result.
To calculate the visible energy spectrum of the MC, we considered the quenching effect of plastic scintillator using the Birks' law as well as the neutron reactions inside the scintillators in GEANT4.
As shown in Fig.~\ref{fig:Onbunch_500kg_fit}, this flux reproduces well the measurement above 15 MeV.

\subsubsection*{Estimation of neutron background for the 25 tons detector}
\label{sec:Bkg_delayed_neutronestimate}

\subsubsection*{Rate calculation}
\indent

Using the estimated flux, we calculated the neutron rate, which creates the 
fake delayed activity of the 25 tons 
detector\footnote{\setlength{\baselineskip}{4mm}
Note that the Birks' law as well
as the neutron reactions inside the liquid scintillator in GEANT4 are also 
included to calculate this rate.}.
This fake delayed gammas from neutron-captures mimic
the neutrino signals if prompt fake gammas also exist accidentally.
 
\begin{figure}
\begin{center}
\includegraphics[width=10cm]{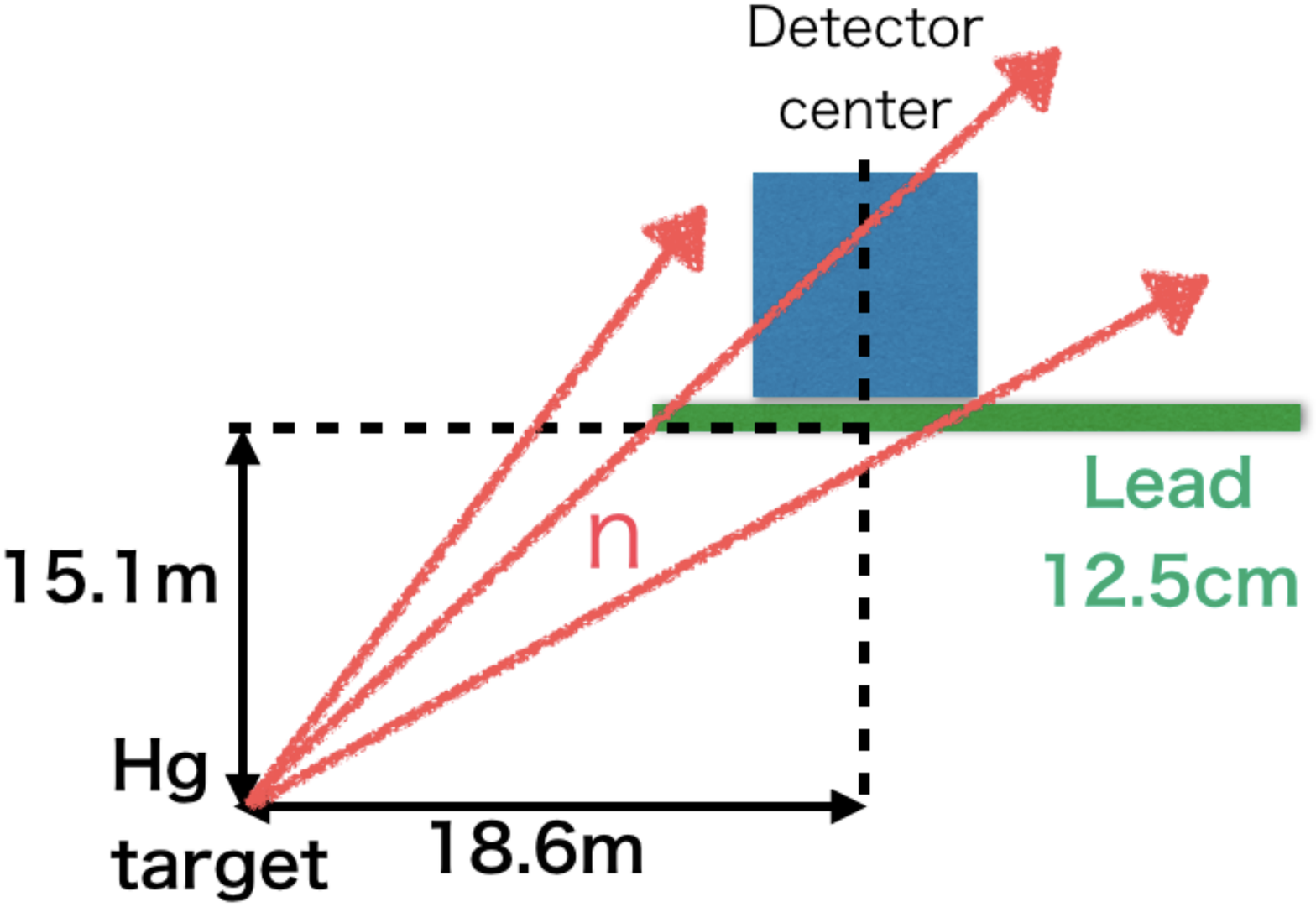}
\caption{\setlength{\baselineskip}{4mm}Geometry of the mercury target and the 25 tons detectors}
\label{fig:Nflux_scheme_25t}
\end{center}
\end{figure}

The geometry used is shown in Fig.~\ref{fig:Nflux_scheme_25t}.
In the MC, the 25 tons detector is located above a 12.5~cm thick lead shield. 
Such a configuration is used for the MC gamma background as well.

As a result of the calculation, we have a rate of the 
background in delayed region of
0.0163/spill/MW/detector.

\subsubsection*{Further Reduction Methods}
\indent

This rate is much larger than the background from beam gammas, thus we have to 
consider further reduction methods.

\begin{figure}
\begin{center}
\includegraphics[width=10cm]{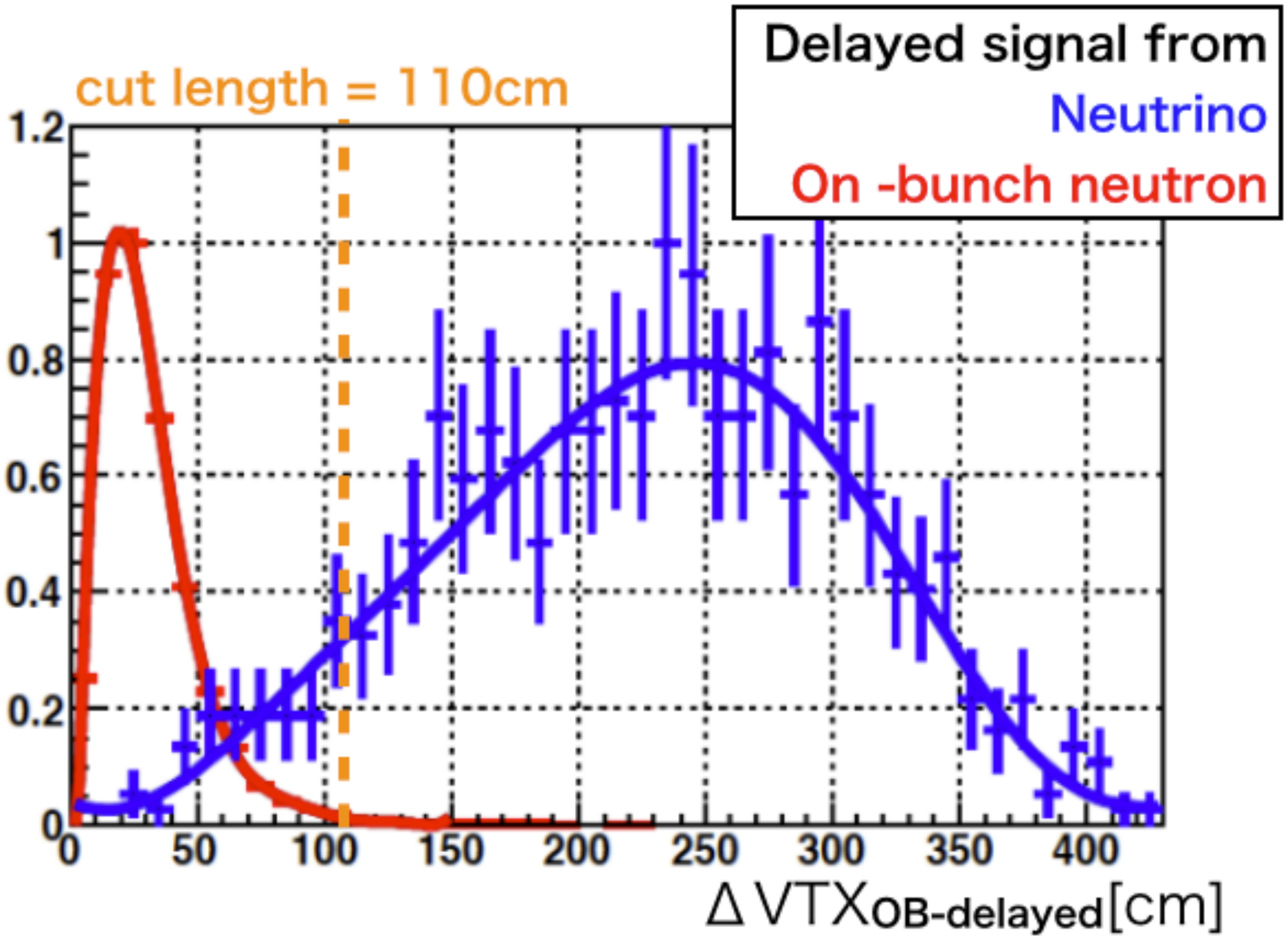}
\caption{\setlength{\baselineskip}{4mm}
$\Delta VTX$ distributions of on-bunch and delayed signals (MC); 
the red line shows the distance between on-bunch hit and delayed hit positions from on-bunch neutrons while the blue line shows the distance between on-bunch and delayed hits from the uncorrelated events.}
\label{fig:DeltaVTXOB}
\end{center}
\end{figure}

Fortunately, 0.0155/spill/MW/detector out of 0.0163
of the background has hits which have more than
0.5 MeV energy inside the black-sheet region in Fig.~\ref{FIG:50ton} on 
the beam bunch timing due to the proton recoils of the beam neutrons.
This recoil proton hit position on the bunch timing and the delayed fake gamma 
position after the neutron thermalization has a strong spatial correlation.  
(The red line in Fig.~\ref{fig:DeltaVTXOB} is the distance between the hit
position of the recoil proton and that of the delayed gamma). 
This correlation (we call this 
variable $\Delta VTX_{\rm OB-delayed}$) reduces the background from 
neutron-capture by 97.4$\%$ when setting 110 cm as 
the cut value. Therefore the remained background out of 
0.0155/spill/MW/detector is 0.0155$\times$0.026 = 0.0004/spill/MW/detector.

There are invisible energy ($E <$ 0.5 MeV) hit events on the beam bunch 
timing inside the black-sheet region (0.00077/spill/MW/detector). However,  
most (97$\%$) of them have hits in the veto (outside the black-sheet) 
region ($E_{VETO} >$ 0.5 MeV). Therefore, we can cut them using the information.
The remained background rate out 
of 0.00077/spill/MW/detector is negligible compared to the former one.

Using these two further reductions, the final delayed background rate from the 
beam neutrons is {\bf 0.0004/spill/MW/detector}. 

%% file: 7bkg_summary_P56.tex
\section{\setlength{\baselineskip}{4mm}
Realistic Rate Estimation for the P56 Experiment using Data
}

\subsection{Additional Selection Cuts from the P56 Proposal}
\label{SEC:additionalcut}

\subsubsection{Tightening Energy Selection for Delayed Activity}
\label{SEC:additionalcut1}
\indent

\begin{figure}[h]
 \centering
 \includegraphics[width=1.1 \textwidth]{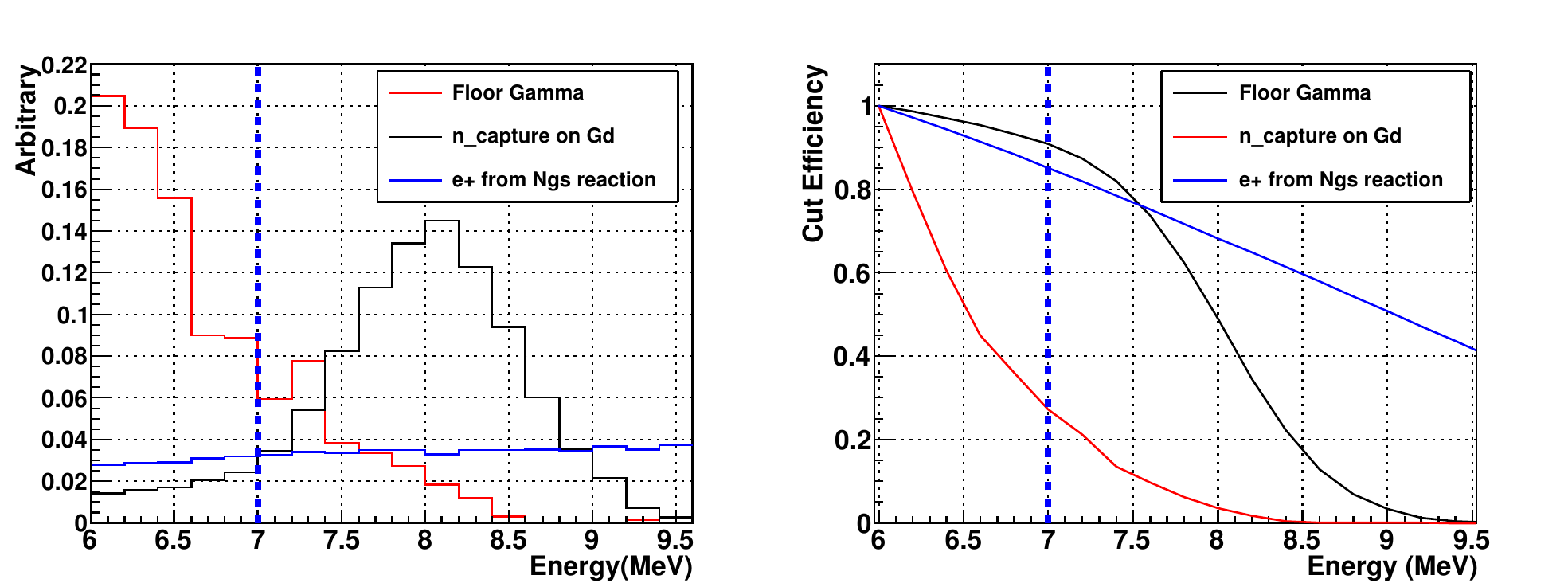}
\caption{\setlength{\baselineskip}{4mm}
 Delayed energy spectra of beam-induced gammas and gammas from thermal neutron capture on Gd(left) and their cut efficiency curves(right). Blue dotted lines show the new lower cut.}
 \label{FIG:EdCut}
\end{figure}

By the measurements at MLF 3rd floor, beam-induced gammas from the floor were observed, and they contribute much to accidental events. To reduce the beam-induced gammas and to improve the S/N ratio, 6 MeV of lower energy cut for delayed events in the proposal is changed to 7 MeV in this analysis.  
Figure.~\ref{FIG:EdCut} shows delayed energy spectra of the beam-induced gammas (red), gammas from thermal neutron capture on Gd (black), and e$^{+}$ from $^{12}C(\nu_{e},e^{-})^{12}N_{g.s.}$ reaction (blue) in the left figure, and their cut efficiency curves in the right plot.
Applying 7~MeV as the lower energy cut, 73\% of the beam-induced gammas can be rejected while the signal efficiency can be kept to be 91~\% of the 6~MeV cut. The signal cut efficiency for delayed events
is changed from 78\% to 71\%. In the case of the $^{12}C(\nu_{e},e^{-})^{12}N_{g.s.}$ reaction, the delayed energy cut efficiency is reduced to 85\% of the 6 MeV lower cut.

\subsubsection{Additional Lifetime Cuts}
\label{SEC:additionalcut2}
\indent

 According to the 2013BU1301 test experiment, the background for 
the prompt signal region is dominated by gammas induced by cosmic rays. 
Therefore, the time structure of the prompt background is
flat (Fig.~\ref{500kg_TvsE}). 
On the other hand, those for the delayed signal region is
dominated by the beam induced gammas and neutrons as described before. 
For gammas, the measured time structure is also flat,
while that for the beam induced neutrons has the same lifetime as neutrino induced
events since beam neutrons come to the detector at the proton beam timing
and are predominantly thermalized. 

 At the P56 proposal~\cite{CITE:P56Proposal}, 
we applied the lifetime cuts for the prompt (1$< \Delta t_{p} < 10 \mu$s)  
and the delayed activities ($\Delta t_{p} < \Delta t_{d} < 100 \mu$s)  
separately. We call this as ``baseline cuts'' hereafter in this subsection. 
The lifetime of the oscillated 
signal hits are shown in Fig.~\ref{FIG:lifetime}(a)(prompt) and (b)(delayed).
However, the tightened cut
can be added to the baseline cuts using two dimensional correlation 
between the prompt and the delayed hits time from the proton beam.  
Table~\ref{TAB:lifetime} and Fig.~\ref{FIG:lifetime}(c),(d),(e) 
show the features of 
the lifetime of events in three categories,
the neutrino signal, the accidental from gammas (prompt) 
+ gammas (delayed) and those from gammas (prompt) + neutrons (delayed).

\begin{table}[htbp]
\begin{center}
\begin{tabular}{|c|c|c|c|}\hline
& Prompt & Delayed & Comments\\\hline
Neutrino Events & Exponential (2.2$\mu$s) & Exponential (30$\mu$s) & \\
Accidental1 & Flat & Flat & $\gamma$+$\gamma$ case \\
Accidental2 & Flat & Exponential (30$\mu$s) & $\gamma$+ n case \\ \hline 
\end{tabular}
\caption{\setlength{\baselineskip}{4mm}
Features of the hit time distribution for three categories. 
}
\label{TAB:lifetime}
\end{center}
\end{table}

\begin{figure}[h]
 \centering
 \includegraphics[width=1.1 \textwidth]{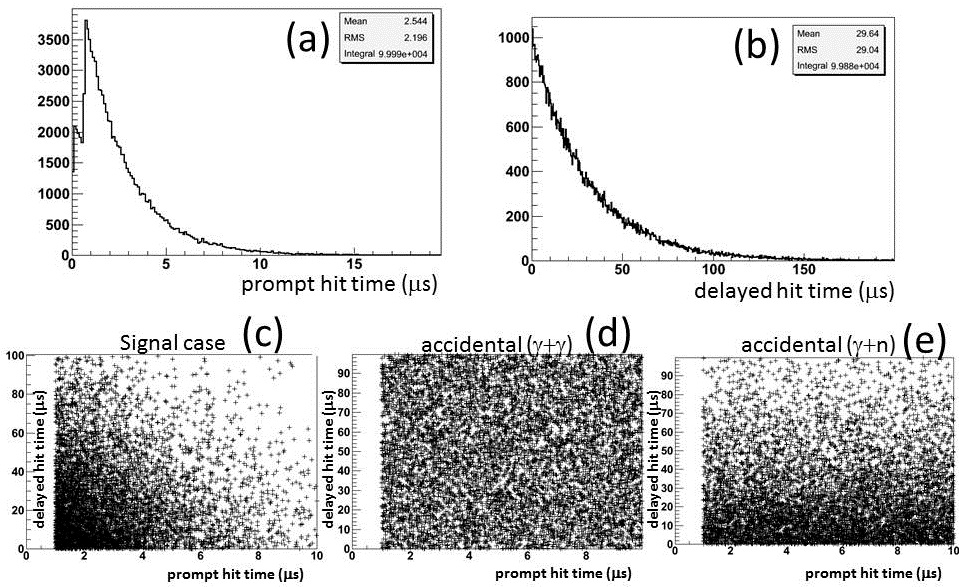}
\caption{\setlength{\baselineskip}{4mm}
(a) Hit time distribution (toy MC) of the oscillated signal (prompt), 
(b) Hit time distribution (toy MC) of the oscillated signal (delayed),
(c) Correlation between prompt and delayed hits for the signal events after the
    baseline cuts,
(d) Correlation between prompt and delayed hits for the accidental background ($\gamma + \gamma$) events,   
(d) Correlation between prompt and delayed hits for the accidental background ($\gamma$ + neutron) events.   
}
 \label{FIG:lifetime}
\end{figure}

In order to maximize the rejection factor of the accidental backgrounds 
and to minimize the loss of the signal efficiency, the likelihood 
function based on the lifetime of the signal hits is created as follows; 

\begin{equation}
\it ln(L) = ln(L_{prompt})+ln(L_{delayed}) 
\end{equation}

This likelihood is calculated in each event using hits time of the
prompt and the delayed activities. Probability Density Functions (PDFs)
to calculate the probability
for the prompt and delayed activities are shown in 
Fig.~\ref{FIG:lifetime}(a) and (b). (note that these distributions
are still not applied for baseline cuts).

\begin{figure}[h]
 \centering
 \includegraphics[width=1.1 \textwidth]{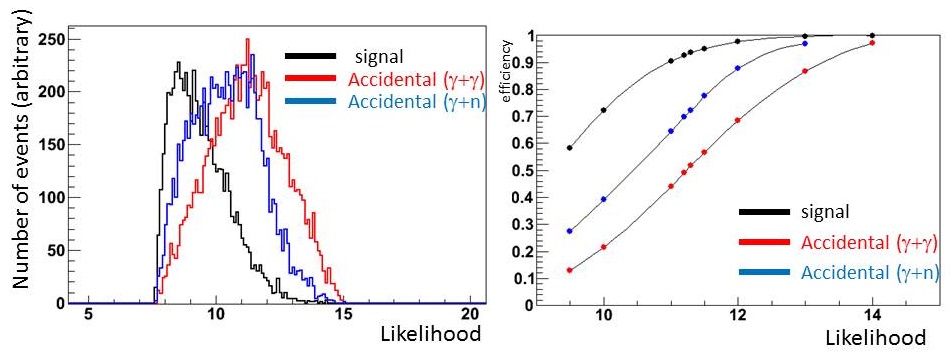}
\caption{\setlength{\baselineskip}{4mm}
 The calculated likelihood distributions (left) and the efficiencies of 
 events in three categories as a function of 
 the likelihood cut value (right).
}
 \label{FIG:likelihood}
\end{figure}

The calculated likelihood distributions, and 
the signal efficiency of these events (three 
categories) are shown in Fig.~\ref{FIG:likelihood}. 
As shown, we can reject more than half of the $\gamma$ + $\gamma$ type 
accidental background, while 91$\%$ of the signal is kept, 
applying a likelihood cut of 11.0. 

We set the cut value of the likelihood at 11.0 in this report.

\subsubsection{$\Delta VTX_{\rm OB-delayed}$ Cut}
\indent

In section~\ref{sec:Bkg_delayed_neutronestimate}, we already discussed the
effects on the reduction of the neutron backgrounds using the
$\Delta VTX_{\rm OB-delayed}$ cut. 
In this subsection, the topic is totally different from that section.
The topic is not the background rate but the signal inefficiency 
for the oscillated neutrino signals due to the $\Delta VTX_{\rm OB-delayed}$ cut. 

Although the $\Delta VTX_{\rm OB-delayed}$ cut is a quite effective to reduce
the neutron background, 
some of the oscillated neutrino events are also cut accidentally. 
In short, if the space in a spill are overlapped 
between oscillated neutrino signal and background neutron events 
accidentally, the signal is cut. 
We have to consider two following cases;

\begin{figure}
\begin{center}
\includegraphics[width=14cm]{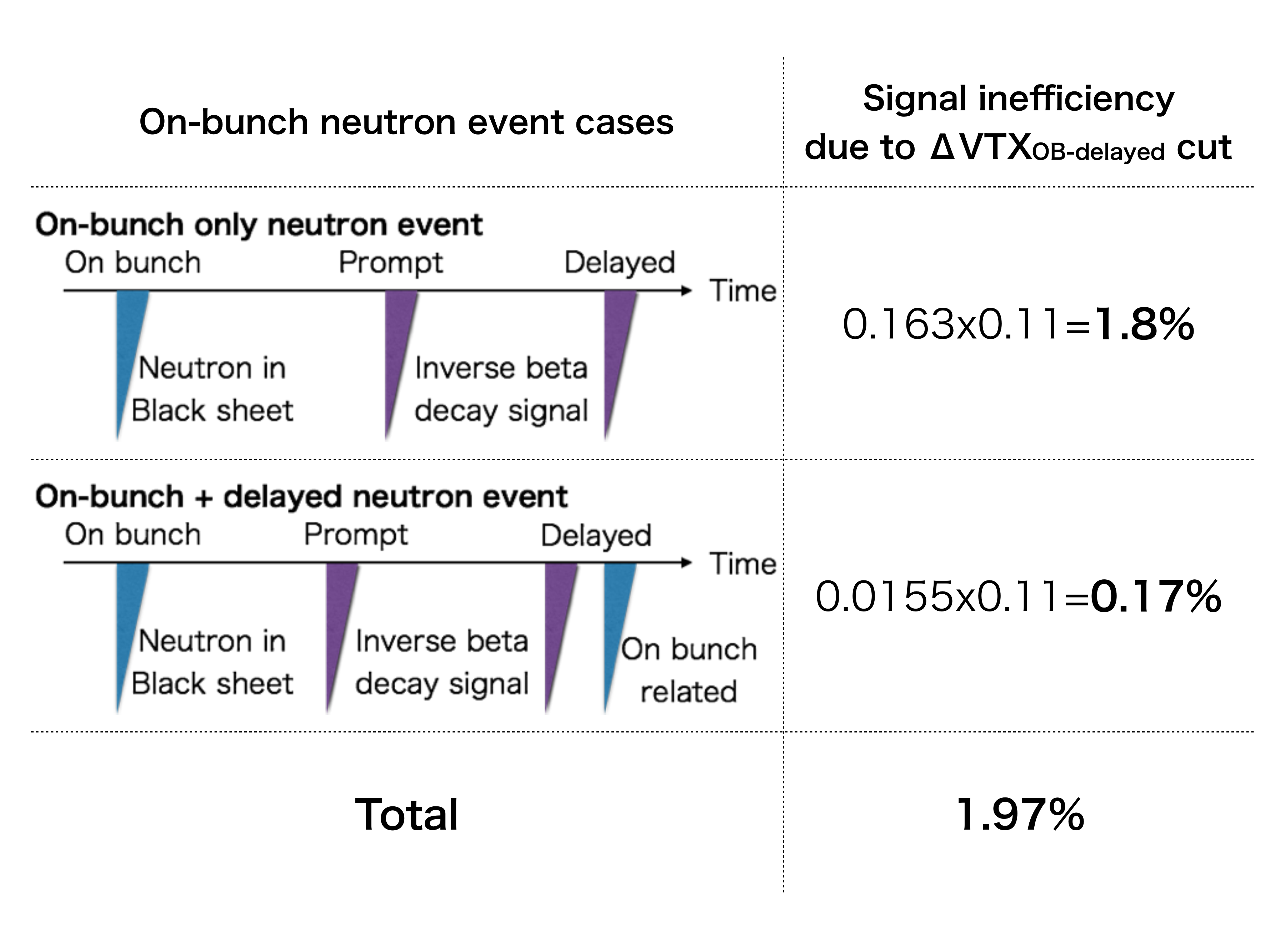}
\caption{\setlength{\baselineskip}{4mm} The cases signal is rejected due to $\Delta VTX_{\rm OB-delayed}$ cut and its inefficiency.}
\label{fig:Delayed_inefficiency}
\end{center}
\end{figure}

\begin{enumerate}
\item ``on-bunch + delayed neutron events'';\\
As mentioned in the section~\ref{sec:Bkg_delayed_neutronestimate}, 
0.0155/spill/MW/detector has both on-bunch and delayed
activities from beam neutrons. 
This means that 1.55$\%$ of the oscillated neutrino signal events have the 
additional on-bunch + delayed neutron activities as shown in the bottom
of Fig.~\ref{fig:Delayed_inefficiency}. If the vertex position of the neutron 
events are overlapped with the signal neutrino events, 
$\Delta VTX_{\rm OB-delayed}$ cut rejects the signal.
The blue line in Fig.~\ref{fig:DeltaVTXOB} shows the 
$\Delta VTX_{\rm OB-delayed}$ distribution for uncorrelated events 
such as this case, and 
an accidental cut efficiency ($\Delta VTX_{\rm OB-delayed} <$ 110 cm)
for the uncorrelated events are 11$\%$.
Thus, the signal inefficiency in the case is 
0.0155 $\times$ 0.11 = 0.17$\%$. 
\item ``on-bunch only neutron events'';\\
In the case that beam neutrons create the hits on the beam bunch timing 
but do not make any delayed gammas (e.g. due to escaping the 
detector region), it also produces the signal inefficiency.
(The top schematic in Fig.~\ref{fig:Delayed_inefficiency}).
According to the rate calculation, 16.3$\%$ of the signal neutrino events  
have such additional ``on-bunch only neutron'', therefore we have a signal
inefficiency of 0.163 $\times$ 0.11 = 1.8$\%$.  
\end{enumerate}
In total, we have 2.0$\%$ for the signal inefficiency.

\subsection{Grand Summary of the Background at P56}
\subsubsection{Summary of Selection Criteria}
\label{SEC:selectsum}
\indent

Table~\ref{TAB:selectionsum} shows a summary of the new selection criteria and the cut efficiency for the oscillated signals with IBD (high $\Delta m^{2}$). The lower energy cut for delayed events is changed, and two new cuts (the $\Delta$VTX$_{OB-delayed}$ cut and life time cut) are added to the selection criteria in the proposal. This is because the current estimation of the accidental prompt events have a 30 times larger background rate than that in the proposal, based on the observation in this background measurement at MLF 3rd floor. Then, the total selection efficiency for the oscillated signal in the high $\Delta m^{2}$ region is reduced to 38\% from 48\% in the proposal (19$\%$ reduction from the proposal). However, the accidental background can be much reduced to an amount comparable with the proposal.

\begin{table}[htbp]
\begin{center}
\begin{tabular}{|c|c|c|}\hline
Cut Condition & Proposal & New\\\hline
1$\le \Delta$t$_{prompt}\le$10$\mu$s&74\% &same\\\hline
6$\le$E$_{delayed}\le$12MeV&78\%&71\%(Ed$\ge$7MeV)\\\hline
20$\le$E$_{prompt}\le$60MeV&92\%&same\\\hline
$\Delta$t$_{delayed}\le$100$\mu$s&93\%&same\\\hline
$\Delta$VTX$_{prompt-delayed}\le$60cm&96\%&same\\\hline
$\Delta$VTX$_{OB-delayed}\ge$110cm&-&98\%\\\hline
Life Time$\le$11&-&91\%\\\hline\hline
Total&48\%&38\% \\\hline 
\end{tabular}
\caption{\setlength{\baselineskip}{4mm}
Summary of selection criteria and efficiency of the oscillated signals for high 
$\Delta m^{2}$ region. 
}
\label{TAB:selectionsum}
\end{center}
\end{table}

\subsubsection{Accidental background Summary}
\indent

Table~\ref{TAB:accsum} shows an accidental background rates summary per detector. Concerning the prompt accidental events, the cosmic-induced gammas were observed and the total prompt accidental rate is 3.9$\times$10$^{-4}$ /spill/detector, which is 30 times of the proposal value. On the other hand, the delayed accidental events rate, of the beam-associated gammas, is reduced to 1.2$\times$10$^{-3}$ /spill/detector, which is 1/39 times of the proposal value. This change is due to the reduction of the gammas from the floor by using a tighter delayed energy cut and putting 12.5cm thick lead shield below the detector. The rate of the beam-associated neutrons decreased to 4$\times$10$^{-4}$ /spill/detector, which is 1/6 times of the proposal value by the on-site measurement. 

\begin{table}[htbp]
\begin{center}
\begin{tabular}{|c|c|c|c|c|}\hline
&Background & Proposal (/spill) & New (/spill)\\\hline
&e$^{-}$ ($^{12}C(\nu_{e},e^{-})^{12}N_{g.s.}$)&8.0$\times$10$^{-6}$&same \\\cline{2-4}
Prompt&e$^{-}$ ($^{12}C(\nu_{e},e^{-})^{12}N^{*}$)&3.8$\times$10$^{-6}$&same \\\cline{2-4}
signals&e$^{+}$ ($\overline{\nu}_{e}$ from $\mu^{-}$)&$<$10$^{-6}$&same \\\cline{2-4}
&Cosmic $\gamma$&-&3.8$\times$10$^{-4}$ \\\cline{2-4}\hline  

&Gamma (Beam associated)&4.7$\times$10$^{-2}$&1.2$\times$10$^{-3}$ \\\cline{2-4}
Delayed&Neutron (Beam associated)&2.4$\times$10$^{-3}$&4$\times$10$^{-4}$\\\cline{2-4}
signals&e$^{+}$ ($^{12}C(\nu_{e},e^{-})^{12}N_{g.s.}$)&1.2$\times$10$^{-5}$&1.0$\times$10$^{-5}$ \\\cline{2-4}
&Spallation products&$\sim$10$^{-4}$&same \\\hline

\end{tabular}
\caption{\setlength{\baselineskip}{4mm}
Accidental background summary (per detector).
}
\label{TAB:accsum}
\end{center}
\end{table}

The total accidental background is estimated as follows:
\begin{equation}
R_{acc}(/spill)=\sum (\sum R^{prompt}_{i} \times R^{delayed}_{i} \times \epsilon^{lifetime}_{i} \times \epsilon^{VTX}_{i}),
\end{equation}
where, $i$ shows some kind of events contributing the accidental backgrounds, R$^{prompt}_{i}$ and R$^{delayed}_{i}$ are accidental prompt and delayed rates for each component in Table~\ref{TAB:accsum}. $\epsilon ^{lifetime}_{i}$ is the lifetime cut (see ~\ref{SEC:additionalcut}). Actually, the $\Delta$VTX cut efficiency depends on each combination of kinds of prompt and delayed like signals, but the $\Delta$VTX cut efficiency for combinations of the $\overline{\nu}_{e}$ prompt signals and delayed like signals of the neutron captured on Gd distributes uniformly in the target volume, and it was used approximately for all components, which is the same condition as the proposal(2.3\%). Finally, total accidental background rate (R$_{acc}$) is 6.3$\times$10$^{-4}$ (/hour/detector).

\subsubsection{Grand Summary}
\label{SEC:GS}
\indent

Table~\ref{TAB:grandsum} shows the grand summary of the number of background events at P56.
The table includes the proposal case with a 17~m baseline and 4000h$\times$4years measurement; the case with new cuts and same operation period but 24~m of the baseline (New1); and the 1.56 times longer period case (5000h$\times$5years), which is the current default condition (New2).
Values in the New1 case for $\overline{\nu}_{e}$ background from $\mu^{-}$(N$_{\overline{\nu}_{e}}$), $^{12}C(\nu_{e},e^{-})^{12}N_{g.s.}$ background(N$_{e^{+}}$) and cosmic-induced fast neutron(N$_{n}$) are estimated as follows:
\begin{equation}
N_{\overline{\nu}_{e}}=377(proposal)\times \epsilon_{BL} \times \epsilon_{ratio},
\end{equation}
\begin{equation}
N_{e^{+}}=38(proposal)\times \epsilon_{BL}\times \epsilon_{Ed} \times \epsilon_{\Delta VTX} \times \epsilon^{e^{+}}_{lifetime},
\end{equation}
\begin{equation}
N_{n}=42(proposal)\times \epsilon^{n}_{liftime} \times \epsilon^{n}_{ratio},
\end{equation}
where, $\epsilon_{BL}$ is the squared baseline ratio(17$^2$/24$^2$), $\epsilon_{ratio}$ is total cut efficiency ratio for the signal between cases of this analysis and the proposal (0.81, see~\ref{SEC:selectsum}). $\epsilon_{Ed}$ and $\epsilon^{e^{+}}_{lifetime}$ are efficiencies of the tightening delayed energy cut (0.85) and the lifetime cut (0.65) for the $^{12}C(\nu_{e},e^{-})^{12}N_{g.s.}$ background, respectively (assuming the same case as the accidental). $\epsilon_{\Delta VTX}$ is $\Delta$VTX cut between on-bunch and delayed events(0.98). $\epsilon^{n}_{lifetime}$ is the lifetime cut for the cosmic-induced neutrons (0.65), $\epsilon^{n}_{ratio}$ is total cut efficiency ratio  for the signal except for the life time cut between cases of this analysis and the proposal (0.88, see~\ref{SEC:selectsum}).

A clear excess of the Michel electron events from beam-associated neutrons comparing the statistical uncertainty was not observed in the on-site measurement. So the only upper limit was estimated from the statistical uncertainty (1.1$\times$10$^{-5}$/spill/9$\mu$s/detector after the on-bunch energy cut).  
The upper limit is 52.5 times of that in the proposal value. Considering the remaining cut efficiencies, except for the on-bunch energy cut and the $\Delta$t$_{prompt}$ cut (total$\sim$0.54), the upper limit with a total of 50 tons (2 detectors) for 4000~hours$\times$4~years measurement is 8.2 comparing with 0.2 after considering the remaining cut efficiencies (total$\sim$0.7) in proposal (the remaining cut efficiencies were not considered for the proposal value of 0.3).

Finally, the dominant background is $\overline{\nu}_{e}$ from $\mu^{-}$ (same as the proposal case), which is 237, while the signal is  342 in case of the best fit values of LSND, with 50~tons detector and current default measurement period(5000~h$\times$5~years).

\begin{table}[htbp]
\begin{center}
\begin{tabular}{|c|c|c|c|c|}\hline
&Contents&Proposal&New1&New2\\
&&(17m)&(24m)&(24m)\\
&&4000h$\times$4y&4000h$\times$4y&5000h$\times$5y\\ \hline
&&&& \\
&$sin^22\theta=3.0\times10^{-3}$&811&307&480\\
&($\Delta m^2$) & ($3.0eV^{2}$) & ($2.5eV^{2}$) & ($2.5eV^{2}$) \\ 
Signal&(Best fit values of MLF)&&&\\ \cline{2-5}
&$sin^22\theta=3.0\times10^{-3}$&&&\\
&$\Delta m^2=1.2eV^{2}$&337&219&342\\
&(Best fit values of LSND)&&&\\\hline\hline
&$\overline{\nu}_{e}$ from $\mu^{-}$&377&152&237\\\cline{2-5}
&$^{12}C(\nu_{e},e^{-})^{12}N_{g.s.}$&38&10&16\\\cline{2-5}
background&beam-associated fast n&0.2(0.3$\times$0.7)&$\le$8&$\le$13\\\cline{2-5}
&Cosmic-induced fast n&42&24&37\\\cline{2-5}
&Total accidental events &37&20&32\\\hline
\end{tabular}
\caption{\setlength{\baselineskip}{4mm}
Grand summary (50~tons in total - 2 detectors). 
}
\label{TAB:grandsum}
\end{center}
\end{table}

%% file: 8sensitivity.tex
\section{Sensitivity using the Latest P56 configuration}
\indent

\subsection{Summary of Changed Points from the P56 proposal}
\indent

The following points are changed from the P56 proposal:
\begin{itemize}
\item The amount of the accidental background and of the background 
     from the beam fast neutrons become realistic due to the MLF 
     2013BU1301 test experiment;
\item The baseline of the detector is longer than that in the proposal
     since the detctor location should have good background rate to 
     perform the mearurement. In this report, we assume 24 meters for the
     baseline;
\item There are some additional lead blocks under the 25 tons detectors and 
     selection cuts to reduce the accidental 
     background while keeping the signal as good as possible;
\item Experimental period is expanded to 5 years from 4 years written 
     in the proposal. Also the default operation time at the MLF was found    
     to be 5000 hours / year during the test experiment, therefore we changed
     it to 5000 hours / year from 4000 hours / year written in the proposal. 
\end{itemize}

For the sensitivity study, the number of events shown in 
Table~\ref{TAB:grandsum}
are used unless it is noted.

\subsection{Methodology of the Fit and Sensitivity}
\indent

The methodology of the fit is exactly the same as that of the P56 proposal.
It uses the profiling method, which treats systematic uncertainties for
the normalization factors of the signal and the backgrounds correctly.

\begin{figure}[h]
 \centering
 \includegraphics[width=1.0 \textwidth]{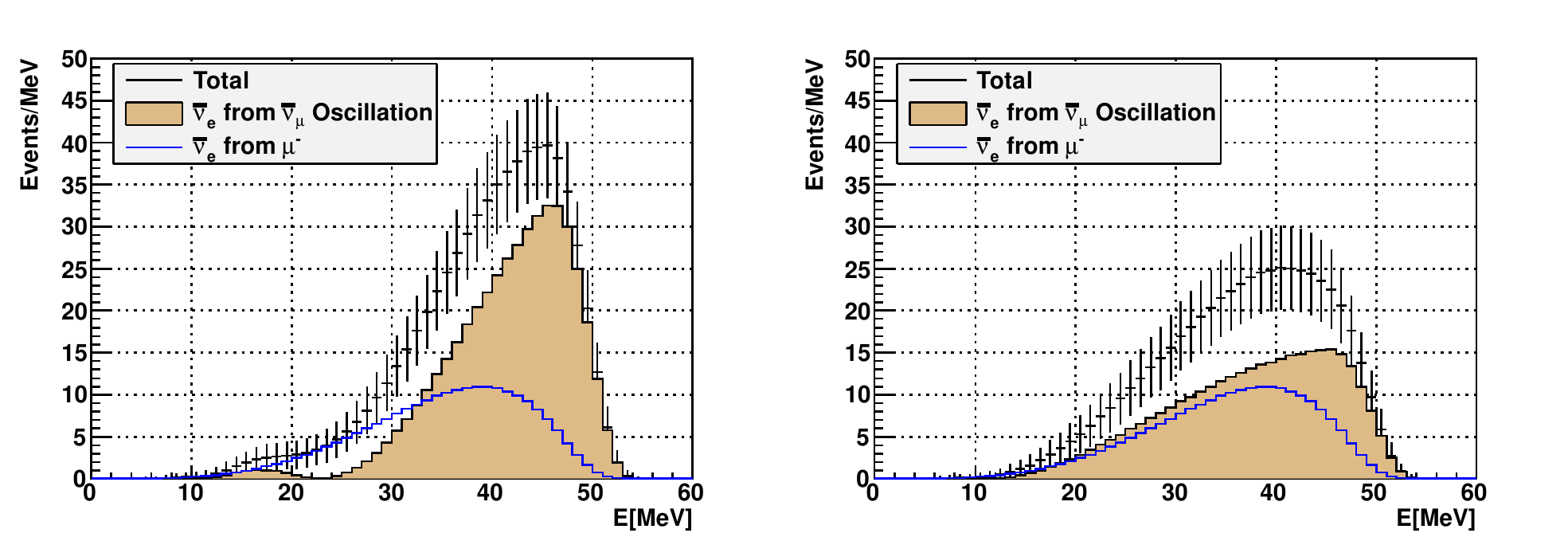}
\caption{\setlength{\baselineskip}{4mm}
The energy spectra with the signal and the dominant backgrund.
Left; oscillatation with ($\Delta m^2$, $\sin^2 2\theta$) = (2.5, 0.003) (MLF best $\Delta m^2$ sensitivity case), right; oscillation with (1.2, 0.003) (LSND best fit case).}
 \label{FIG:ES_24m}
\end{figure}

In this subsection, we neglect the contributions from the accidental
background and the background from the beam fast neutrons. 
The amount of the accidental background is about 13.2$\%$ of the dominant
background ($\bar{\nu_{e}}$ from $\mu^{-}$) and this is smaller than  
that in the proposal. The 90$\%$ C.L. upper limit of the background from 
beam fast neutron is less than 5$\%$ of the dominant background. 
Figure~\ref{FIG:ES_24m} shows the energy spectra after the selection to be used 
for the fitting.

\begin{figure}[h]
 \centering
 \includegraphics[width=0.9 \textwidth]{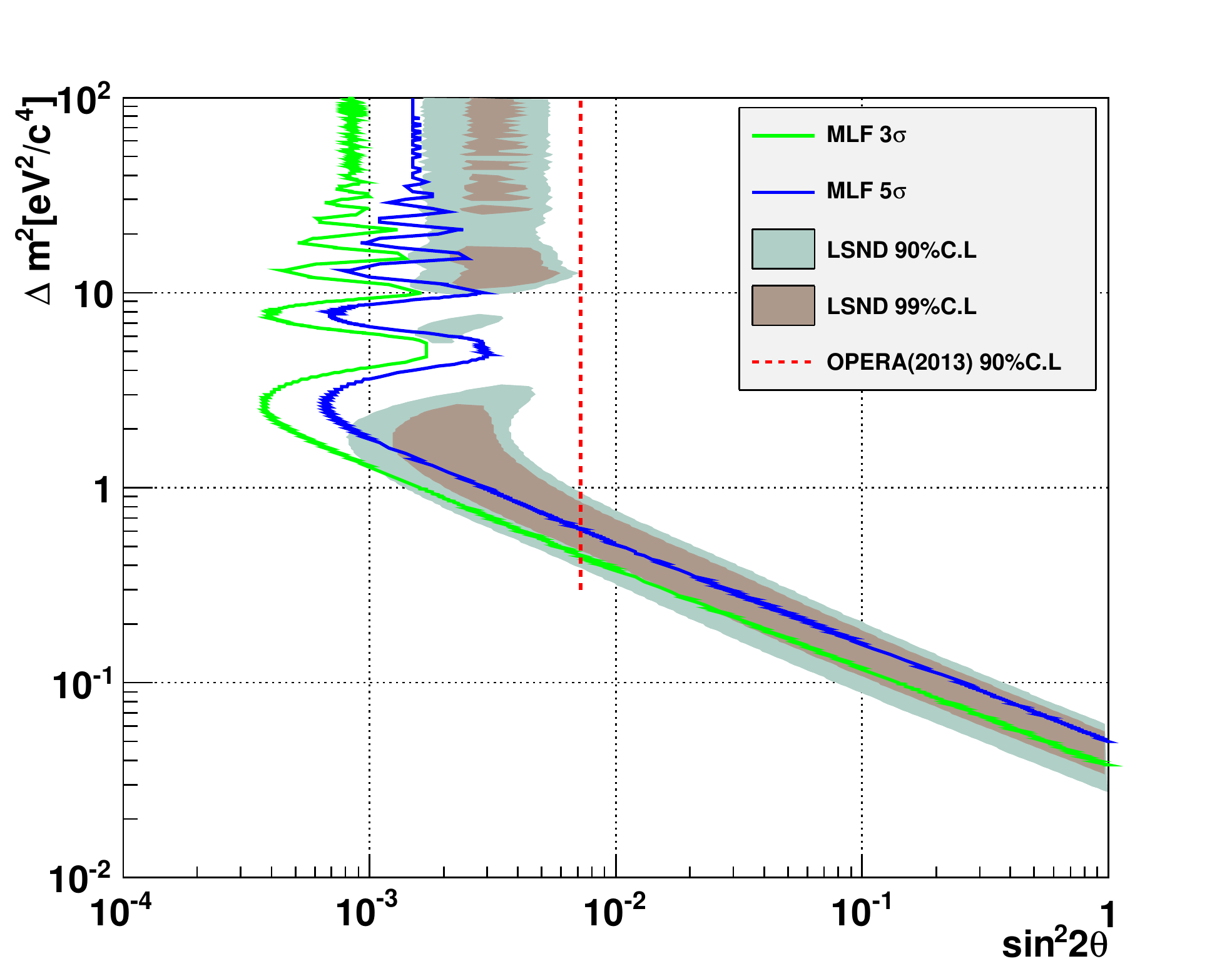}
\caption{\setlength{\baselineskip}{4mm}
Sensitivity of the P56 experiment with the latest configuration 
(MW$\times$5 years).
Blue line shows the 5$\sigma$, while the green one shows 3$\sigma$.  
The exclusion line of the OPERA experiment are also put~\cite{CITE:OPERA}. 
The righthand side region of the line are excluded with 90$\%$ C.L.. 
}
 \label{FIG:senswP56}
\end{figure}

Figure~\ref{FIG:senswP56} shows 
the 3$\sigma$ and 5$\sigma$ sensitivities with this 
condition. We find that;\\
(1) We will confirm or refute the LSND region with 5 $\sigma$ (blue) for 
  $\Delta$m$^{2} > 2 eV^{2}$ region. (same as proposal). \\
(2) The detector baseline is longer than the proposal, thus the coverage region
  of the 3$\sigma$ (green) is larger than that in the proposal in the 
  lower $\Delta$m$^{2}$ region. Most of LSND 90$\%$ C.L region can be
  discussed with 3$\sigma$.\\

\begin{figure}[h]
 \centering
 \includegraphics[width=0.7 \textwidth]{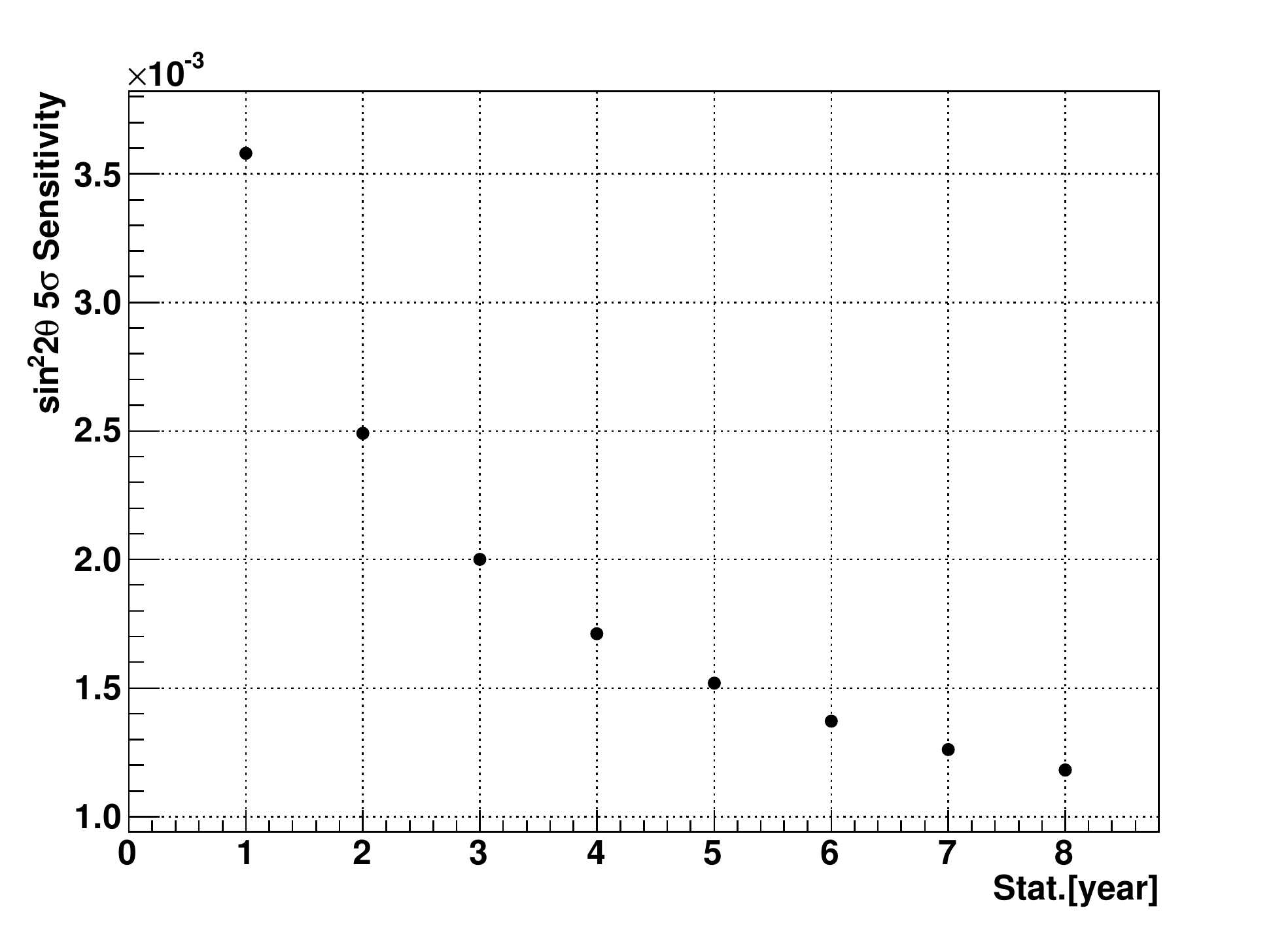}
\caption{\setlength{\baselineskip}{4mm}
The variation of the 5$\sigma$ sensitivity of the P56 experiment as a 
function of operation year (MW $\times$ years).
}
 \label{FIG:sensP56vstime}
\end{figure}

Figure~\ref{FIG:sensP56vstime} shows the variation of 
the 5$\sigma$ sensitivity at high
$\Delta m^{2} (\Delta m^{2} = 100 eV^{2})$ region as a function of the operation 
year (MW $\times$ years). This is a good indicator whether the P56 experiment
can cover whole LSND region above 2 eV$^{2}$ region. 
(at $\Delta m^{2} = 100 eV^{2}$, the left edge of 
the LSND allowed region by 90$\%$C.L is 0.0017). 

With four years of operation, the sensitivity of the P56 starts to cover the
whole LSND region with $\Delta m^{2} > 2.0$~eV$^{2}$.

%% file: 9homework.tex
\section{Milestone for the Experiment}

\subsection{\setlength{\baselineskip}{4mm}
Accidental Background}

\subsubsection{Discussion}
\indent

 In this subsection, we consider the sensitivity variation 
 with changing the amount of the accidental background. This consideration
 reminds us the importance of the background and how to manage them although
 the current best estimation of the background is low enough for
 the experiment.

\begin{figure}[h]
 \centering
 \includegraphics[width=1.0 \textwidth]{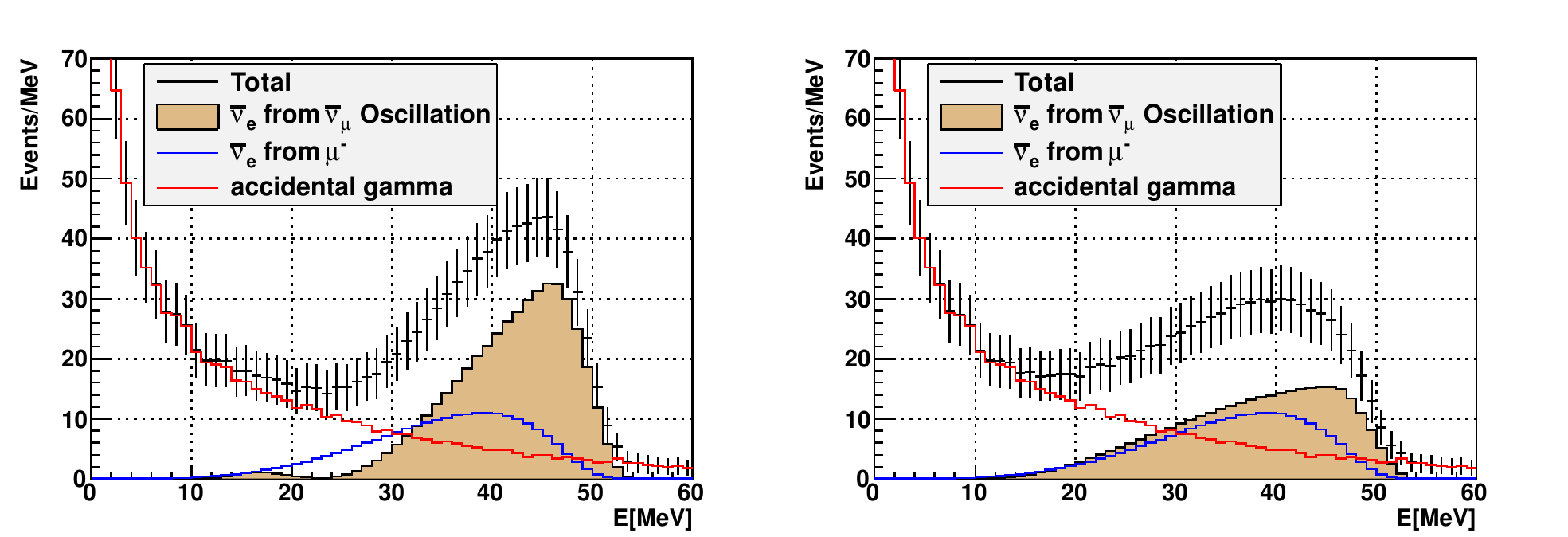}
\caption{\setlength{\baselineskip}{4mm}
Energy spectra with the accidental background with the typical oscillation
signals. Left; oscillatation with 
($\Delta m^2$, $\sin^2 2\theta$) = (2.5, 0.003) 
(MLF best $\Delta m^2$ sensitivity case), 
right; oscillation with (1.2, 0.003) (LSND best fit case).
Crosses correspond to the total,  
the blue histograms correspond to the dominant background
($\bar{\nu_{e}}$ from $\mu^{-}$), and 
red histograms correspond to the accidental backgrounds assuming the 
rate are same as those of the dominant background in $20<E<60 MeV$ region.    
}
 \label{FIG:ESwacc}
\end{figure}

 As explained in the section 2.4.1, the accidental background provides 
 an exponential energy spectrum for the prompt energy. If the total amount 
 of the accidental background were same as the dominant background
 ($\bar{\nu_{e}}$ from $\mu^{-}$), the energy spectrum of the prompt 
 activities are changed to those in Fig.~\ref{FIG:ESwacc}. 

\begin{figure}[h]
 \centering
 \includegraphics[width=0.5 \textwidth]{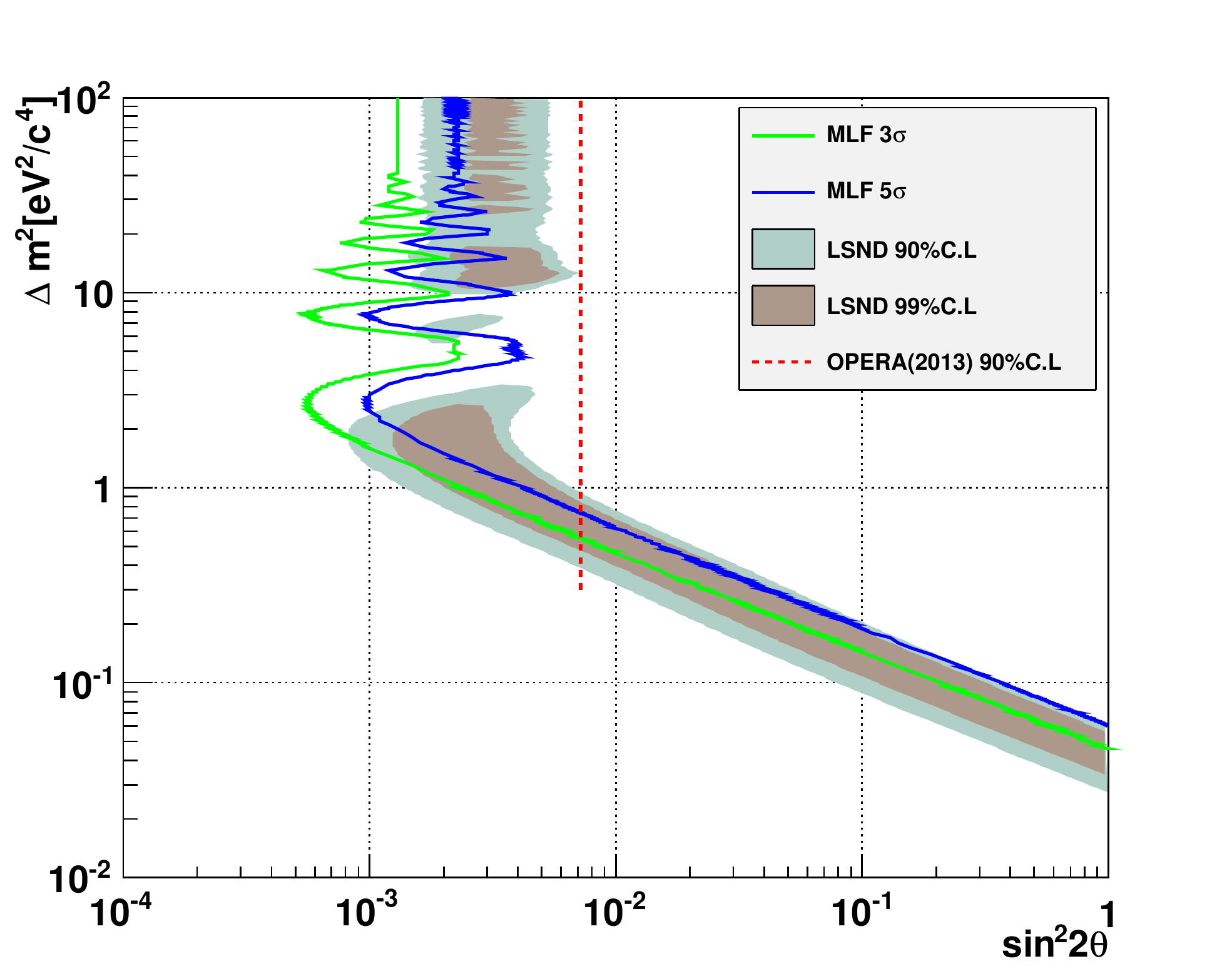}
\caption{\setlength{\baselineskip}{4mm}
Sensitivity of the P56 with the accidental background (in the the case 
that there were same amount as those of intrinsic $\bar{\nu_{e}}$.)  
}
 \label{FIG:senswacc}
\end{figure}

This has a large impact for the sensitivity since the separation power
between the oscillated signals and the dominant background is coming mainly 
from the difference of the energy spectra around 50 MeV.    
The statistical fluctuation of the accidental background reduces the
separation power even though the energy shape and the rate is well determined
by beam-off data (less than 1$\%$).
Fig.~\ref{FIG:senswacc} shows the sensitivity at the case that 
there were same accidental background rate as the dominant background
in $20<E<60$ MeV region. The 5$\sigma$ and 3$\sigma$ sensitivities are worsened
by about 50$\%$.

\subsubsection{Conclusion}
\indent

The energy shape and rate of the accidental background is found 
with the 2013BU1013
test experiment for the first time. Thus, {\bf taking real data 
(not only MC) is crucial} to know the properties of the background deeper.
Also {\bf the reduction of the background using the hardware and the software 
including the analysis improvement} is essential for 
the experiment if there are larger
amount of the accidental background than expected.
    
During the detector construction phase, we build the two detectors
sequentially (not in parallel). Thus, we can check the background rate 
and the energy shape with one detector at first. During this phase,
the best hardware configuration, e.g. shielding,  can be also examined.  
This provides good feed-backs to the other detector configuration
and the whole experiment within the short time scale.

\subsection{Experimental Feasibility}
\indent

The design of the detector tank is already discussed in the last part of this
documentation. It provides not only the static strength of the tank, but also
the endurance of the tank against earthquakes.

We will also prepare a document for other
hardware feasibilities. PMTs, electronics, maintenance of the detector 
will be discussed in the document in detail although part of them were  
already discussed in the proposal. The detector design is similar
to the reactor experiments (well established detector), 
therefore the crucial point to discuss is the location of the detector
and the maintenance of the detector. Note that we have to bring the 
detector out from the MLF building once per year for the
MLF maintenance at least. 

One detector scheme during the construction phase also helps the realization
of the maintenance of the detector.

\section{Requests to PAC}
\indent

We request the following to PAC; \\
(1) Stage-1 approval \\

%% file: 10Acknowledgement.tex
\section{Acknowledgements}
\indent

We warmly thank the MLF people, especially, Prof. Masatoshi Arai, MLF Division leader, 
the neutron source group, muon group and user facility group for the 
various kinds of supports.
The LEPS2 experiment has provided us with good quality scintillators, electronics, cables, PMTs, and we deeply appreciate it.
We borrow cables, electronics, scintillators from J-PARC Hadron 
group, University of Kyoto, JAEA, KEK online group, Belle II 
experiment and T2K experiment and we would like to express appriciation 
for their kindness.

Finally, we thank the support from J-PARC and KEK.

\section{Executive Summary of Background}
\indent

In order to understand the backgound rates for the real experimental 
condition at a glance, a summary of the selection critera and their
efficiencies at the 25 tons detector is shown in this section.
Table~\ref{TAB:ES} shows the executive summary. 
Each number or selection cut are discribed in the noted sections, and if 
there are no notes in the table, they are described in the proposal. 
(Ref~\cite{CITE:P56Proposal})

\begin{table}[htb]
\begin{center}
\begin{tabular}{|c|c|c|c|c|} \hline
Selection & Signal & Beam Fast n & Accidental & Accidental \\ 
          &        &             & ($\gamma$+$\gamma$) & ($\gamma$+n) \\ \hline 
1$<\Delta t_{p}<10\mu$s & 74$\%$ & $<$1.1$\times 10^{-5}$ & 3.9$\times 10^{-4}$ & 3.9$\times 10^{-4}$ \\ 
&& /9$\mu$s/MW/25t  & /9$\mu$s/MW/25t & /9$\mu$s/MW/25t  \\ \cline{1-2}
20$<E_{p}<$60MeV & 92$\%$ & (90$\%$CL) & (sec.~\ref{SEC:AP}) & (sec.~\ref{SEC:AP}) \\ 
&&(sec.~\ref{SEC:ME})&& \\ \cline{1-5}  
$\Delta t_{p}< \Delta t_{d} < 100 \mu$s & 93$\%$ & 93$\%$ & 1.2$\times 10^{-3}$ & 1.6$\times 10^{-2}$  \\  \cline{1-3}
7$<E_{d}<$12MeV & 71$\%$ & 71$\%$ & /spill/MW/25t  & /spill/MW/25t  \\ 
(sec.~\ref{SEC:additionalcut1}) & & & (sec.~\ref{sec:Bkg_delayed_gamma}) & (sec.~\ref{sec:Bkg_delayed_neutron})  \\ \cline{1-5}
$\Delta$VTX$_{p-d}<$60cm & 96$\%$ & 96$\%$ & 2.3$\%$ & 2.3$\%$  \\ \hline \hline
$\Delta$VTX$_{OB-d}<$110cm& 98$\%$ & 98$\%$ & 98$\%$ & 2.6$\%$  \\ 
$\&\&$ $E_{VETO} <$0.5MeV &&&& \\ 
(sec.~\ref{sec:Bkg_delayed_neutron}) &&&& \\ \hline
lifetime (L$<$11) & 91$\%$ & 91$\%$ & 44$\%$ & 65$\%$  \\ 
(sec.~\ref{SEC:additionalcut2}) &&&& \\ \hline\hline
Additional & -- & 1/2000 & -- &  --  \\ 
(described in && OB-d spatial(0.01) & & \\
Ref.~\cite{CITE:P56Proposal}) && $\#$of delayed (0.05) & & \\ \hline \hline
Total & 38$\%$ & 3.1$\times$10$^{-9}$ & 4.6$\times$10$^{-9}$ & 2.4$\times$10$^{-9}$ \\ 
(1 detector)&& /spill/MW/25t & /spill/MW/25t & /spill/MW/25t  \\ \hline \hline  
$\#$Events/5years/50t & & $<$ 13 & 21 & 11 \\ 
(2 detectors)&&&&\\
(sec.~\ref{SEC:GS})&&&&\\ \hline
\end{tabular}
\caption{\setlength{\baselineskip}{4mm}
An executive summary of the selection criteria, their efficiencies  
and predicted background rates of the 25 tons detector(s). 
(The final row shows the number of the events in 2 detectors as shown in
Section.~\ref{SEC:GS}. Except for the line, all rates corresponds to 
one detector) 
For the accidental 
backgrounds, events in the prompt region are mainly coming from gammas 
induced by cosmic rays, while the beam creates events for delayed region.  
Note; ``OB'' means ``on-bunch'', ``p'' corresponds to ``prompt'' and ``d'' 
corresponds to the ``delayed''.  Unless noted, rates and efficiencies are 
discussed in the proposal~\cite{CITE:P56Proposal}.
}
\label{TAB:ES}
\end{center}
\end{table}

\begin{figure}[h]
 \centering
 \includegraphics[width=0.7 \textwidth]{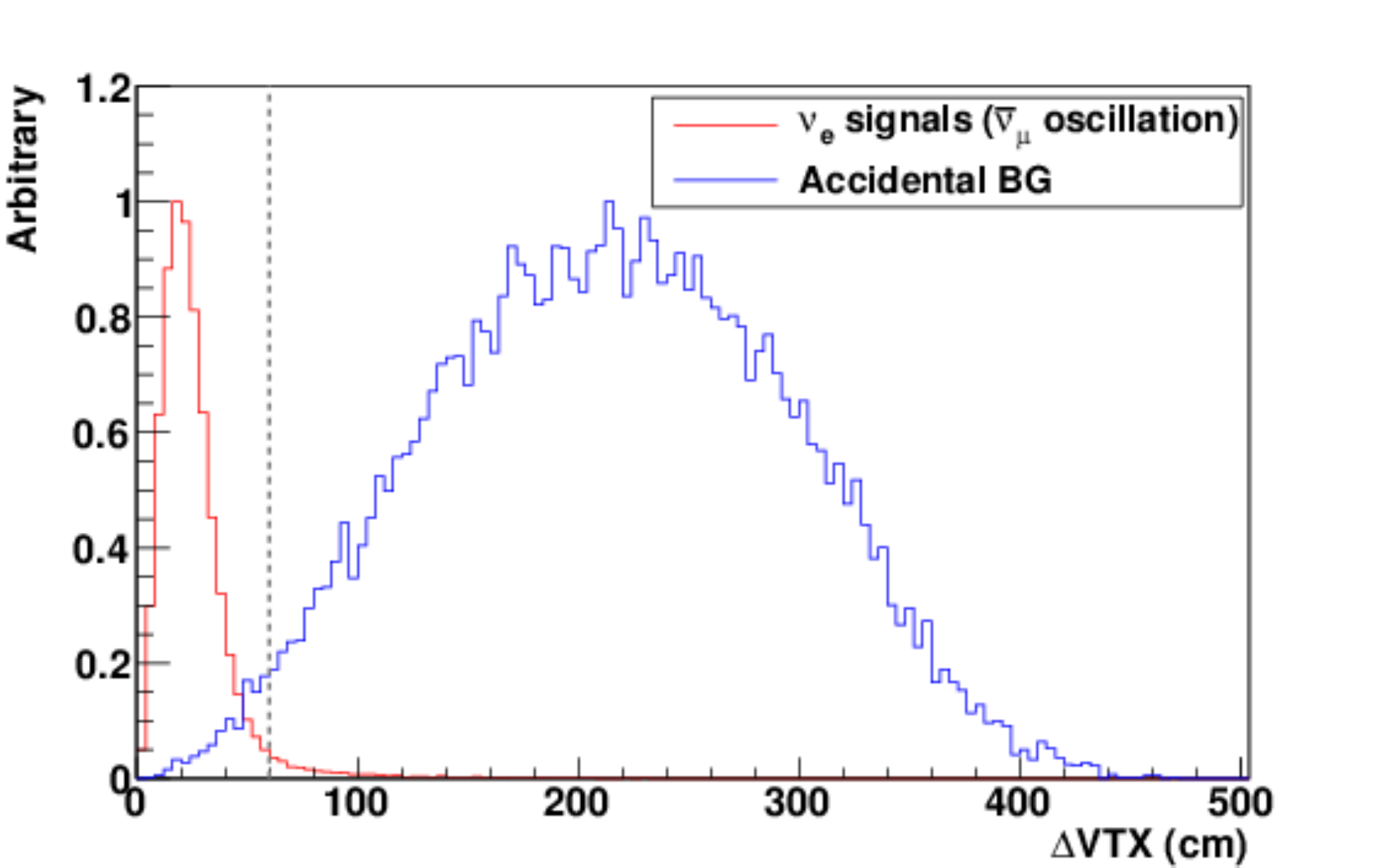}
\caption{\setlength{\baselineskip}{4mm}
$\Delta$VTX$_{p-d}$ distributions of 
$\bar{\nu_{\mu}} \to \bar{\nu_{e}}$ oscillation
events (red line) and the accidental events (blue line). The gray 
doted line shows the $\Delta$VTX$_{p-d}$ cut condition (60 cm).
}
 \label{FIG:DVTXN}
\end{figure}

Note that the accidental background rate is estimated with the 
following equation;
\begin{equation}
R_{acc}(/spill)=R^{prompt} \times R^{delayed} \times \epsilon^{VTX_{p-d}} \times \epsilon^{new},
\end{equation}
where $R$ is the probability ($\#$backgrounds/spill), and  
$\epsilon^{VTX_{p-d}}$ and $\epsilon^{new}$ are the efficiencies of the 
the $\Delta$VTX$_{p-d}$ cut and new additional cuts (the $\Delta$VTX$_{OB-d}$
and the lifetime correlation cuts), respectively. 
Figure~\ref{FIG:DVTXN} shows the 
distance between vertex of the prompt hit and that of delayed hit 
($\Delta$VTX$_{p-d}$), which is the reprint from the proposal. 
Using this simple equation, the accidental background rate is
calculated and is smaller than the intrinsic $\bar{\nu_{e}}$
background. 

%% file: 11detector_calib_500kg.tex
\section{Setup and Calibration of the 500 kg detector}
\indent

\subsection{Setup}
\label{appendix_500kg_setup}
\indent

\begin{figure}[hbtp]
	\begin{center}
		\includegraphics[scale=0.42]{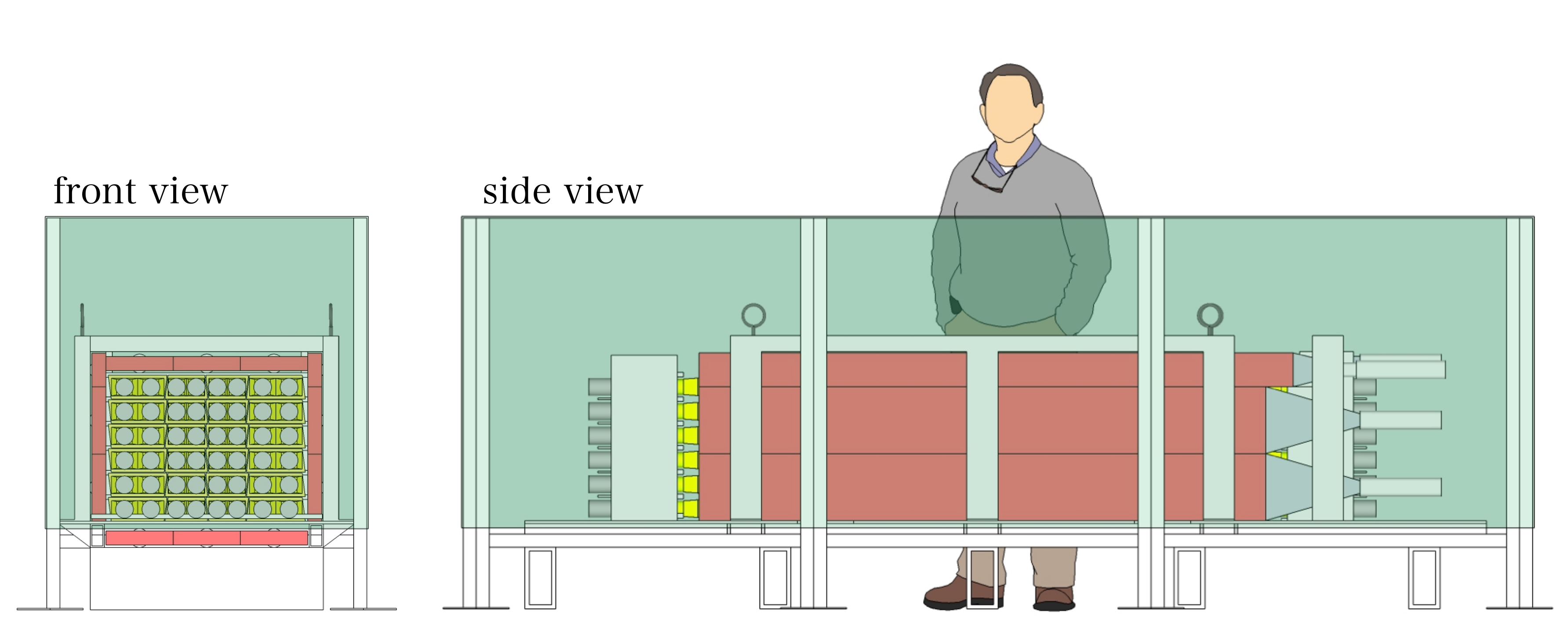}
		\caption{\setlength{\baselineskip}{4mm}Schematic view of the 500 kg plastic scintillator detector at the third floor of MLF(left: front view, right: side view).
		The 500 kg plastic scintillators (yellow) were surrounded by two layers of charged vetoing system, Inner veto(red) and Outer veto(green).}
		\label{500kg_setup}
	\end{center}
\end{figure}

Figure \ref{500kg_setup} shows a schematic view of the 500 kg plastic scintillator counters placed at the third floor of MLF.
It consists of two types of scintillators: 12 pieces of $11.7/13.7$ (trapezoid) $\times 7.6 \times 182$~cm$^3$ scintillator (1D)
and 12 pieces of $16.9/18.8$ (trapezoid) $\times 7.6 \times\ 182$~cm$^3$ scintillator (3D).
Each end of the scintillators was viewed by two PMTs. Signals from each PMT were recorded by FINESSE 500 MHz FADC.
The 500 kg detector was surrounded by two layers of charged vetoing system, the Inner and Outer vetoes.
The Inner veto covers the surfaces of the top, bottom and both sides of the 500 kg detector.
The thickness of the plastic scintillator for Inner veto is 4.5 cm.
The Outer veto surrounds the 500 kg detector and Inner veto, with mostly 2 layers of 6-8mm thick plastic scintillators.
PMT signals from the veto counters were recorded by FINESSE 65 MHz FADC with 50 ns RC-filter.

\subsection{Calibration}
\indent

\begin{figure}[hbtp]
	\begin{center}
		\includegraphics[scale=0.6]{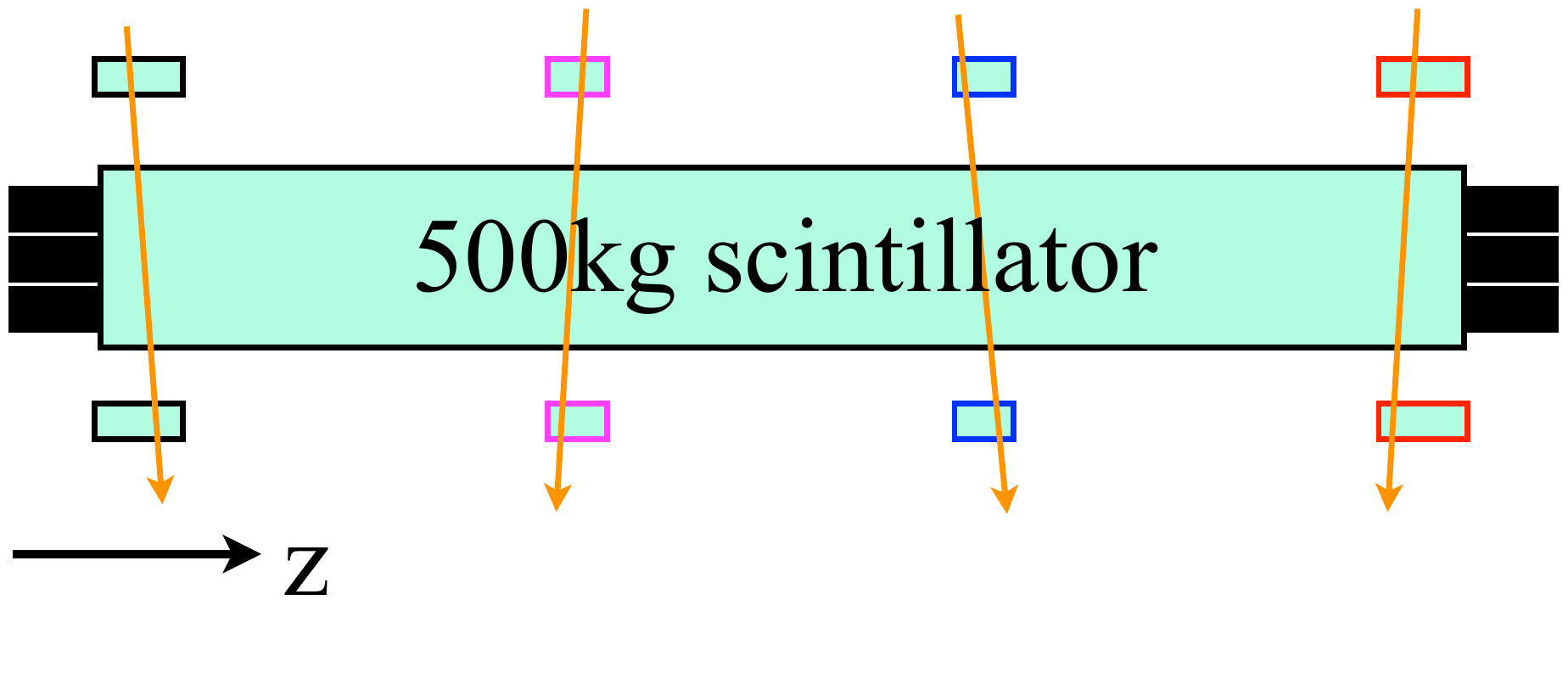}
		\caption{\setlength{\baselineskip}{4mm}Schematic view of the cosmic muon trigger counters.
		We prepared 4 pairs of plastic scintillators to trigger cosmic muons.}
		\label{500kg_setup2}
	\end{center}
\end{figure}

We used cosmic muons to measure the attenuation length of the scintillator, to calibrate the energy and timing.
We prepared 4 pairs of scintillator counters to trigger the cosmic rays.
Figure \ref{500kg_setup2} shows a schematic view of the cosmic muon trigger counters.

\begin{figure}[hbtp]
	\begin{center}
		\includegraphics[scale=0.4]{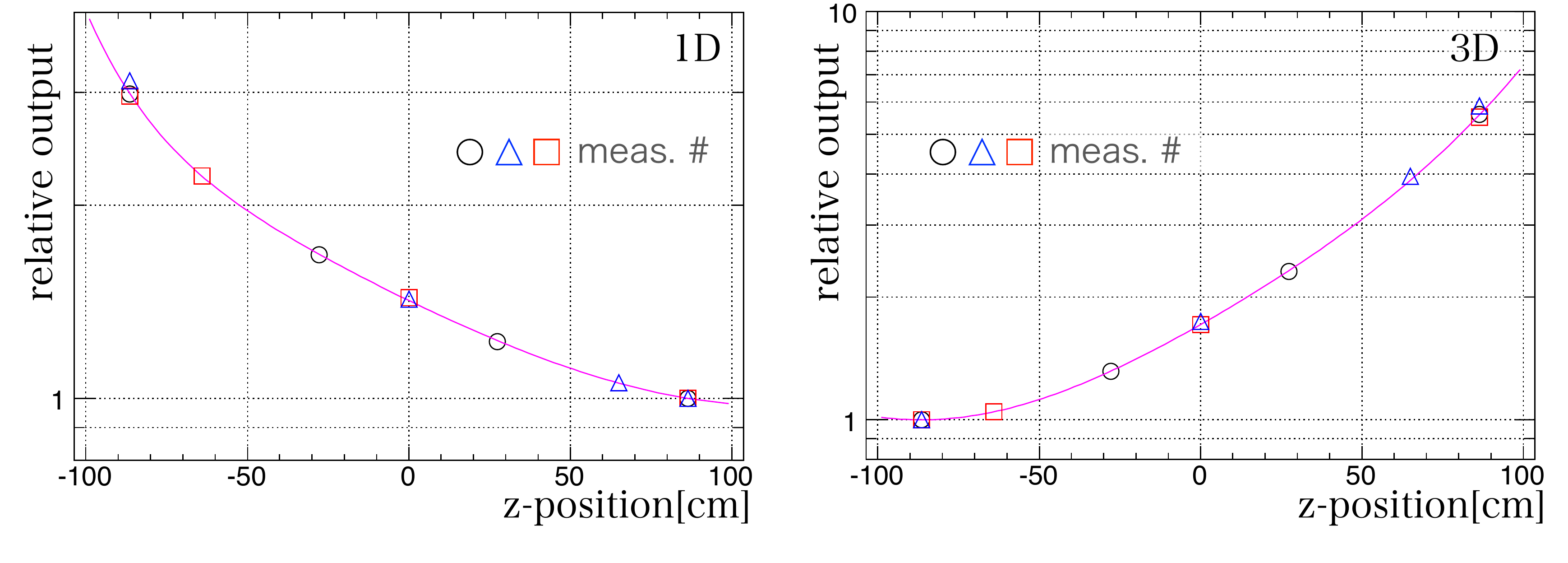}
		\caption{\setlength{\baselineskip}{4mm}Typical measured attenuation curve for each scintillator type (1D and 3D).
		We measured the curves and parameterized one for each scintillator (magenta line).}
		\label{500kg_attenuation}
	\end{center}
\end{figure}

\begin{figure}[hbtp]
	\begin{center}
		\includegraphics[scale=0.6]{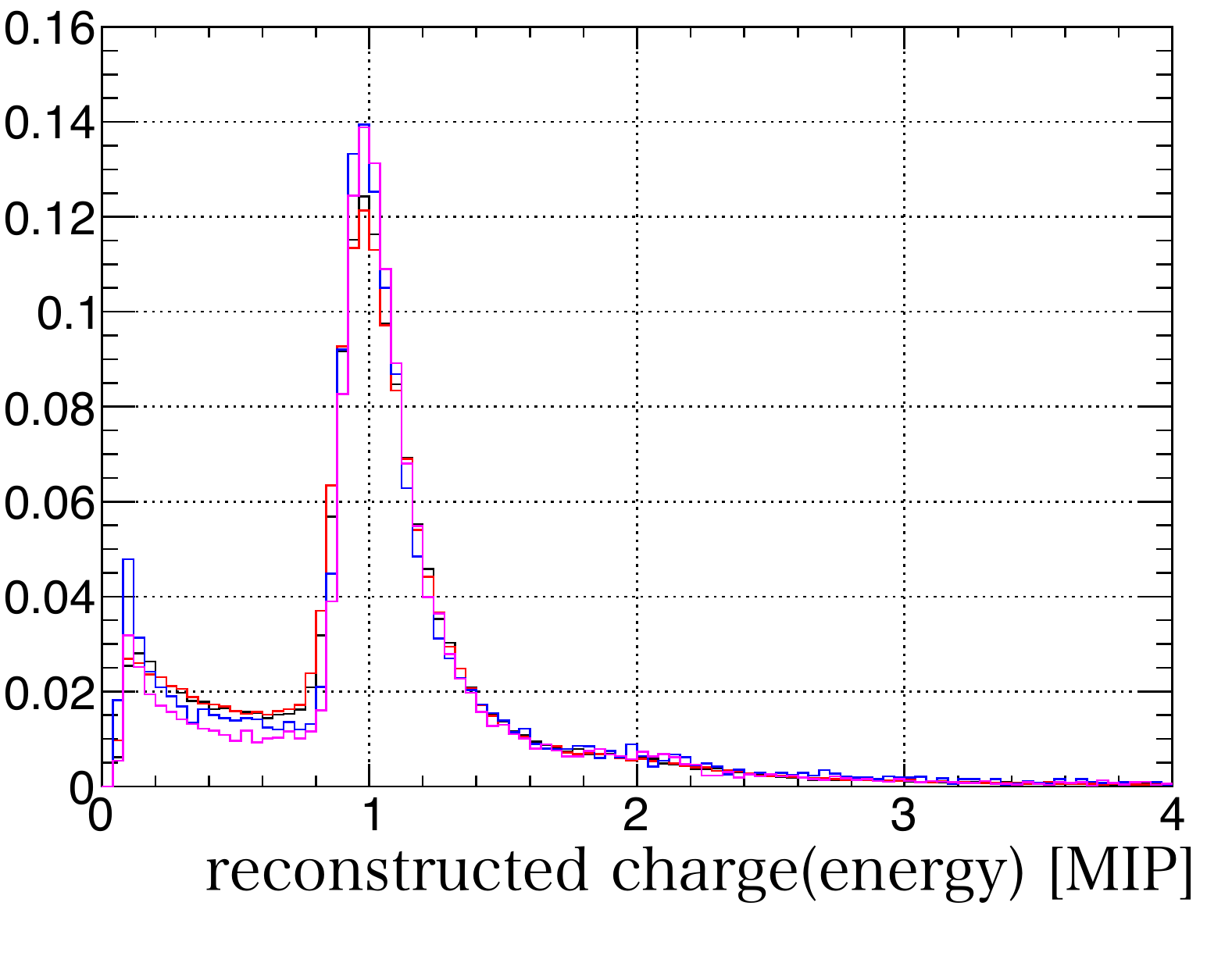}
		\caption{\setlength{\baselineskip}{4mm}Reconstructed charge for different incident positions by considering the measured attenuation length.
		The colors correspond to the events triggered by the cosmic muon counters shown in Fig.~\ref{500kg_setup2}.}
		\label{500kg_reccharge}
	\end{center}
\end{figure}

%%% attenuation
To measure the attenuation length of the scintillators, we made some dedicated runs in which we changed the position of the trigger counters.
Figure \ref{500kg_attenuation} show the measured typical attenuation curves for each scintillator type.
We measured the curve and parameterized one for each scintillator.
By considering the attenuation length, the reconstructed charge was independent from the incident position as shown in Fig.~\ref{500kg_reccharge}.

\begin{figure}[hbtp]
	\begin{center}
		\includegraphics[scale=0.4]{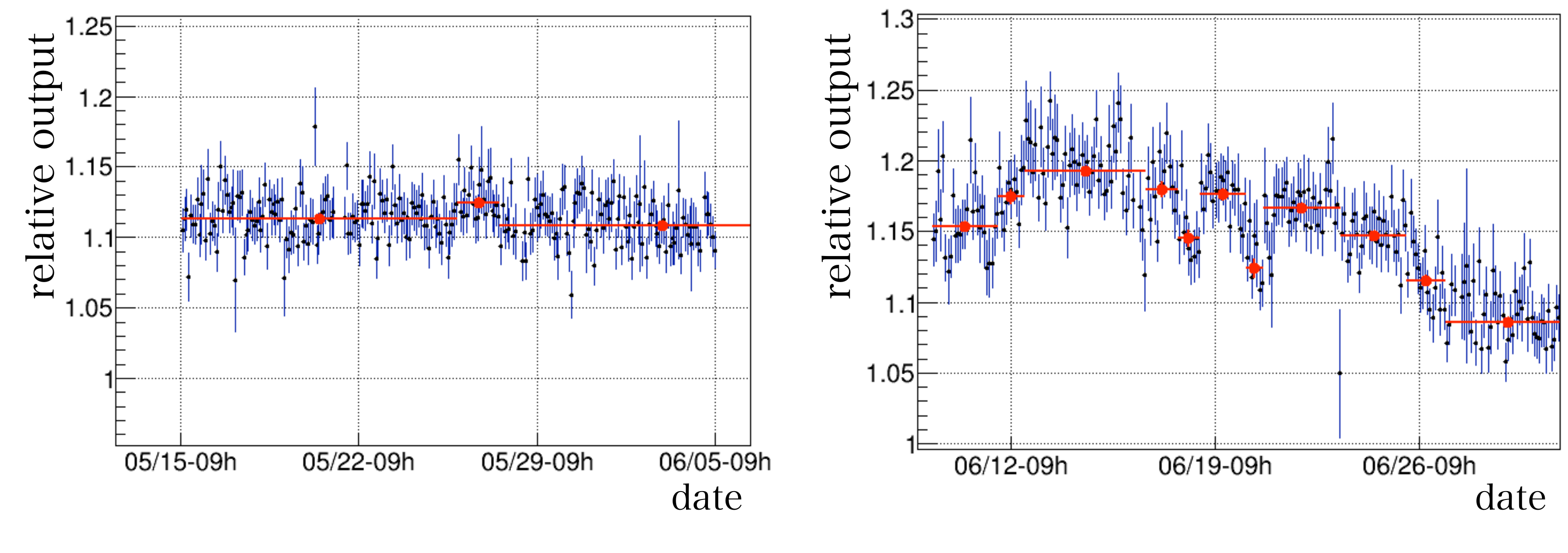}
		\caption{\setlength{\baselineskip}{4mm}Examples of gain histories for relatively stable(left) and unstable(right) PMTs.
		The black dots show the relative gains for each period, and the red dots show the averaged values over stable periods to be used for the gain correction. }
		\label{500kg_gainHistory}
	\end{center}
\end{figure}

%%% gain
The gain of each PMT was also calibrated by using cosmic ray events.
Figure~\ref{500kg_gainHistory} shows the relative gains of certain PMTs as a function of date.
Figure \ref{500kg_gainCorrection} shows the relative energy of a certain scintillator\footnote{\setlength{\baselineskip}{4mm} One scintillator is viewed by four PMTs in total as described in Appendix~\ref{appendix_500kg_setup}.} with the gain correction as a function of date for muon triggered events which are different from the events used for gain correction.
Output from each scintillator is stable within 1-2\% with the gain correction.
\begin{figure}[hbtp]
	\begin{center}
		\includegraphics[scale=0.6]{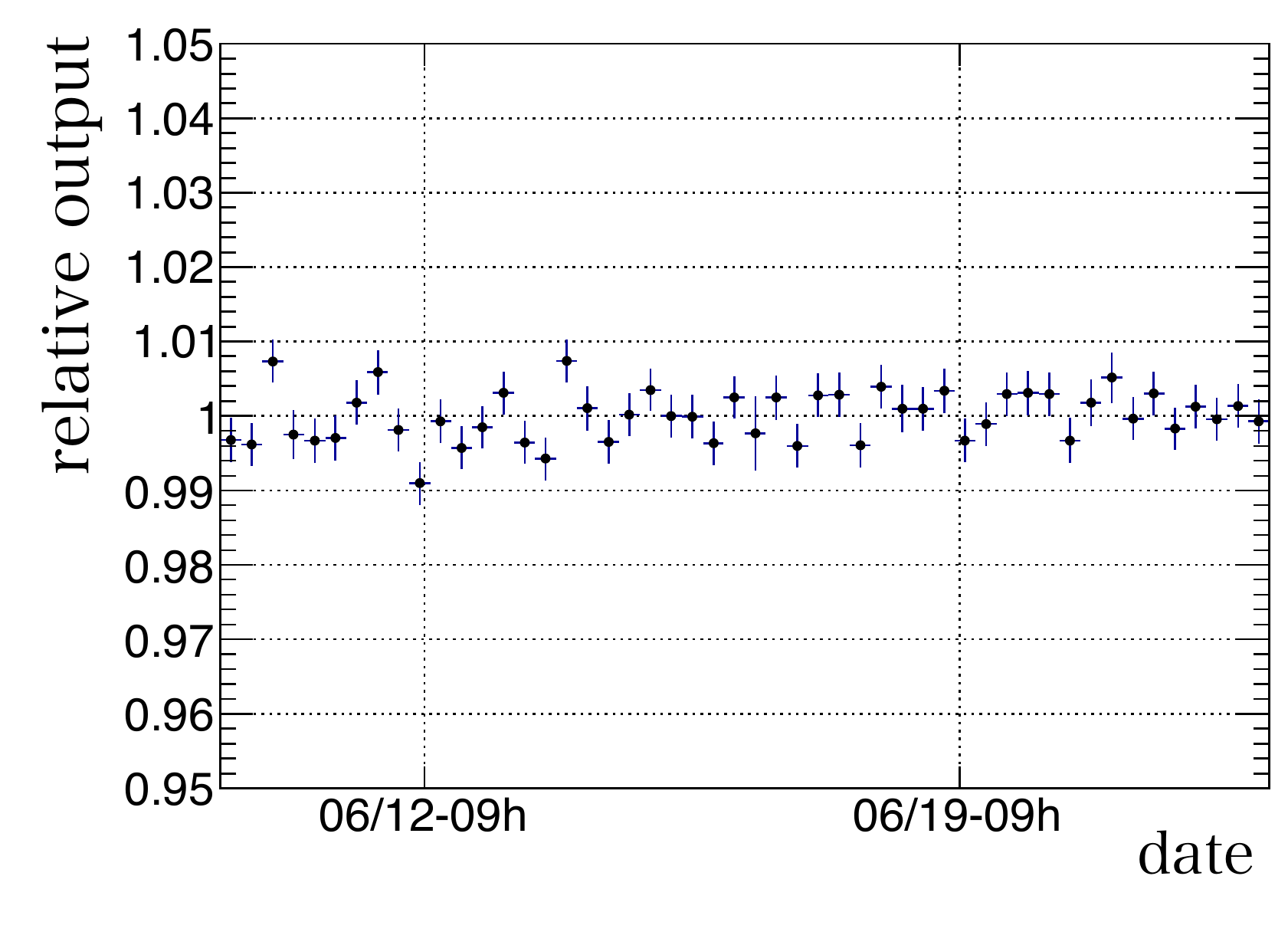}
		\caption{\setlength{\baselineskip}{4mm} Example of the relative energy of a certain scintillator with the gain correction as a function of date.}
		\label{500kg_gainCorrection}
	\end{center}
\end{figure}

\begin{figure}[hbtp]
	\begin{center}
		\includegraphics[scale=0.6]{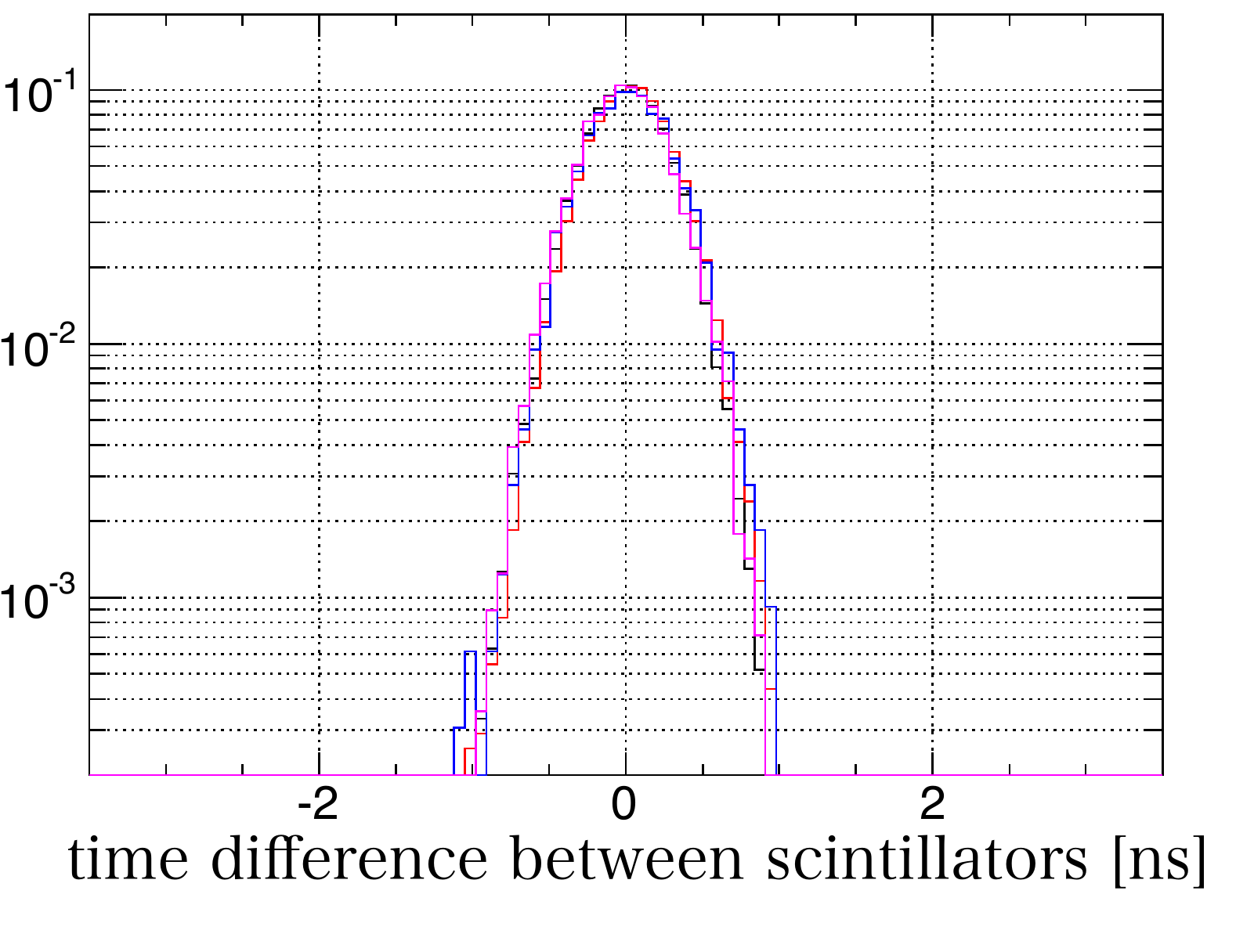}
		\caption{\setlength{\baselineskip}{4mm}Time differences between scintillators of cosmic ray events for different incident positions.
		The colors correspond to the events triggered by the cosmic muon counters shown in Fig.~\ref{500kg_setup2}.}
		\label{500kg_deltat}
	\end{center}
\end{figure}

%%% timing
We also aligned the timing of each PMT.
Time offsets were determined to minimize the time difference between each pair of PMTs on each end.
The velocity of cosmic muons passing through the detector was considered.
Figure \ref{500kg_deltat} shows the time difference between scintillators for cosmic ray events.
The timing of each PMT and scintillator was well calibrated.

\subsection{Resolutions}
\indent

%%% timing resolution
We evaluated the timing, position and energy resolutions of the 500 kg detector.
We first evaluated the timing resolution at each side of the scintillators as a function of the equivalent charge.
The timing resolution was evaluated by using the timing difference between two vertically neighboring scintillators.
The total charge difference between the two scintillators was required to be less than 20 \% of itself,
and we can thus simply divide the width of the timing difference between those two scintillators by $\sqrt{2}$ to get the timing resolution.
Figure \ref{500kg_tres} show the typical timing resolutions as a function of equivalent charge for two types of scintillators.
The obtained timing resolution was parameterized with the function:
\begin{equation}
	\label{func_tres}
	\sigma_t = \frac{p_0}{\sqrt{C}} \oplus p_1,
\end{equation}
where $C$ is the normalized charge with the MIP energy at the far end from the side.
We obtained $p_0=0.173$ ns, $p_1=0.246$ ns for 1D, and $p_0=0.164$ ns, $p_1=0.418$ ns for 3D scintillator (typical values).
The obtained timing resolutions are mainly limited by jitters between channels\footnote{\setlength{\baselineskip}{4mm} Because we used its internal clock for each FADC, each FINESSE card worked asynchronously.
We added a reference signal at each waveforms end for each channel every event to align and synchronize the FADC cards.}.
The timing of the scintillator was defined by averaging the hit time at both sides of the scintillator.
The typical timing resolutions of the scintillators are 0.19 ns for 1D and 0.26 ns for 3D at the middle of the scintillators for MIP energy.
\begin{figure}[hbtp]
	\begin{center}
		\includegraphics[scale=0.45]{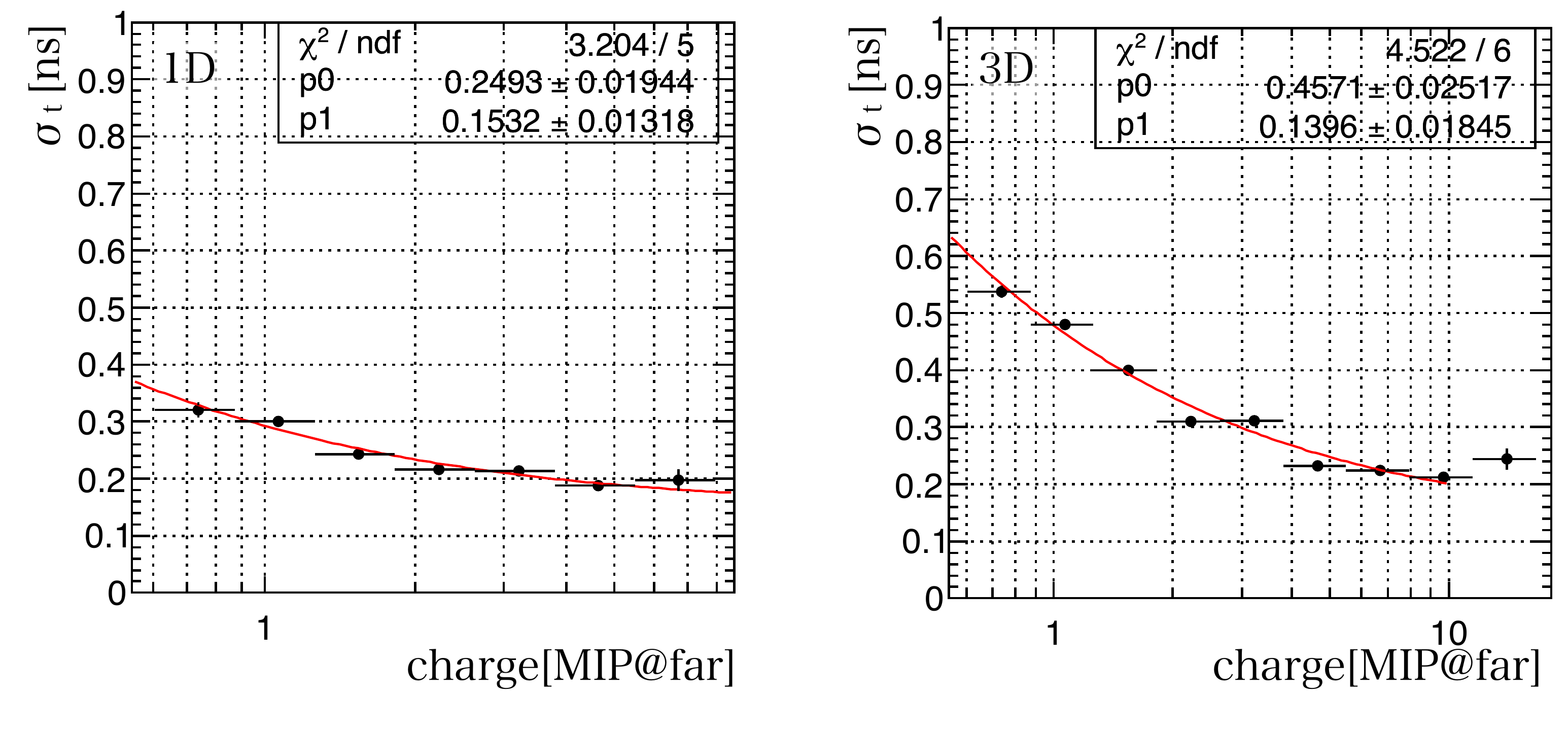}
		\caption{\setlength{\baselineskip}{4mm}Obtained timing resolutions as a function of equivalent charge for two types of scintillators.
		The obtained timing resolution was parameterized with Function \ref{func_tres}.}
		\label{500kg_tres}
	\end{center}
\end{figure}

\begin{figure}[hbtp]
	\begin{center}
		\includegraphics[scale=0.6]{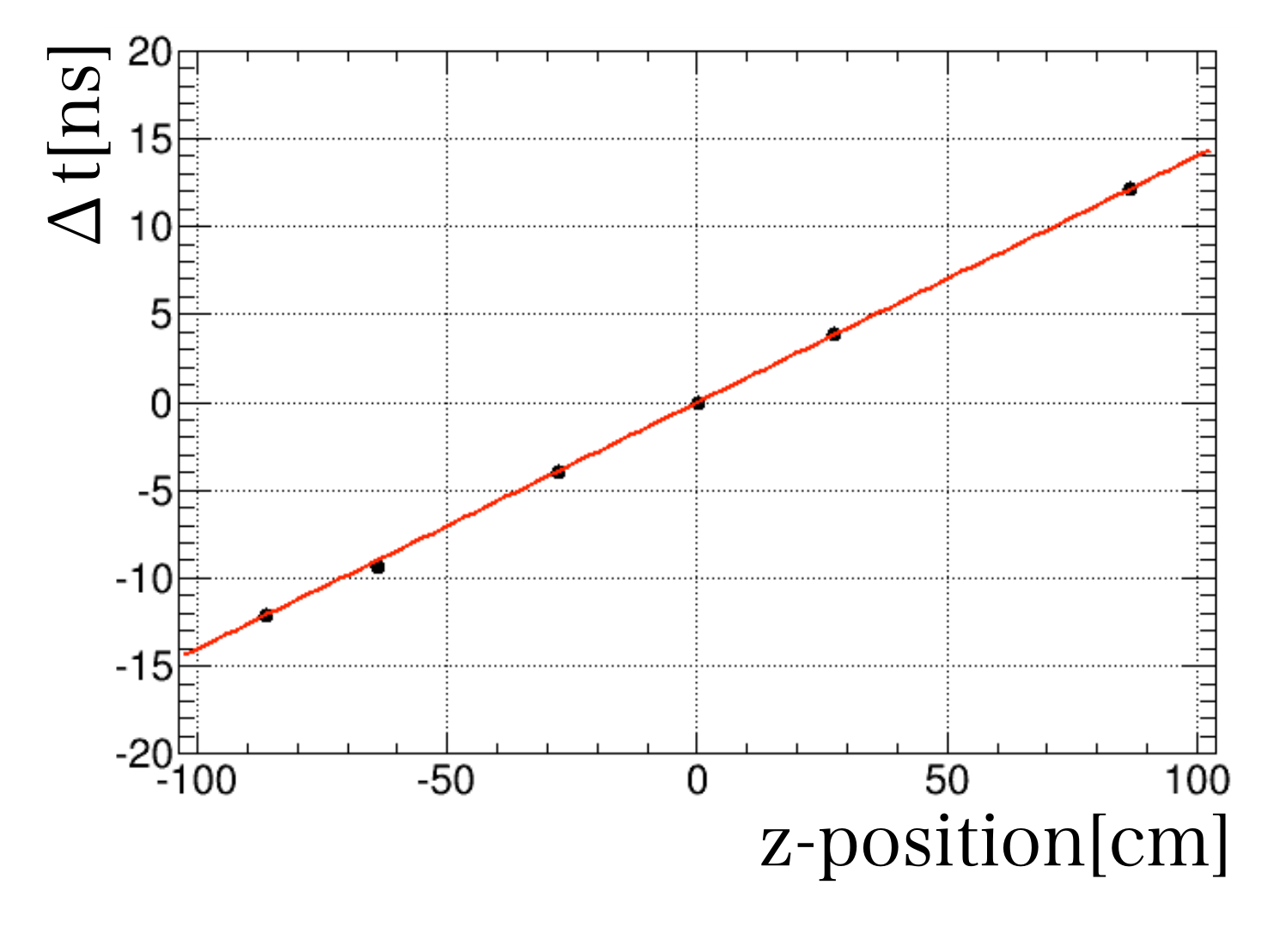}
		\caption{\setlength{\baselineskip}{4mm}Correlation between $z$-position and the time difference between both ends of PMTs.
		The red line is the fitted function and the slope corresponds to the light velocity in the scintillator, 14.3 cm/ns.}
		\label{500kg_velocity}
	\end{center}
\end{figure}

\begin{figure}[hbtp]
	\begin{center}
		\includegraphics[scale=0.6]{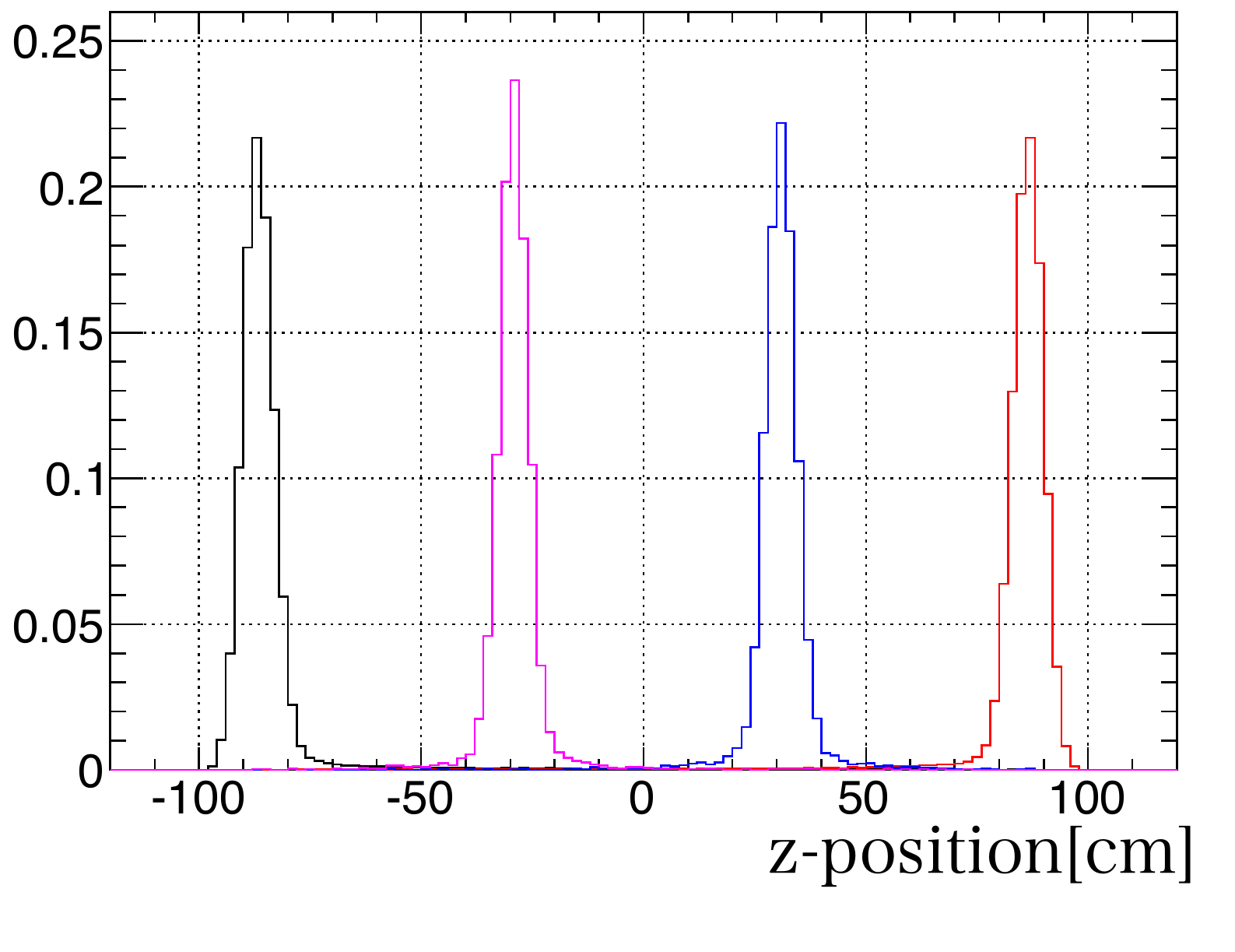}
		\caption{\setlength{\baselineskip}{4mm}Reconstructed $z$-position for different cosmic triggers.
		The colors correspond to the events triggered by the cosmic muon counters shown in Fig.~\ref{500kg_setup2}.}
		\label{500kg_recz}
	\end{center}
\end{figure}

\begin{figure}[hbtp]
	\begin{center}
		\includegraphics[scale=0.6]{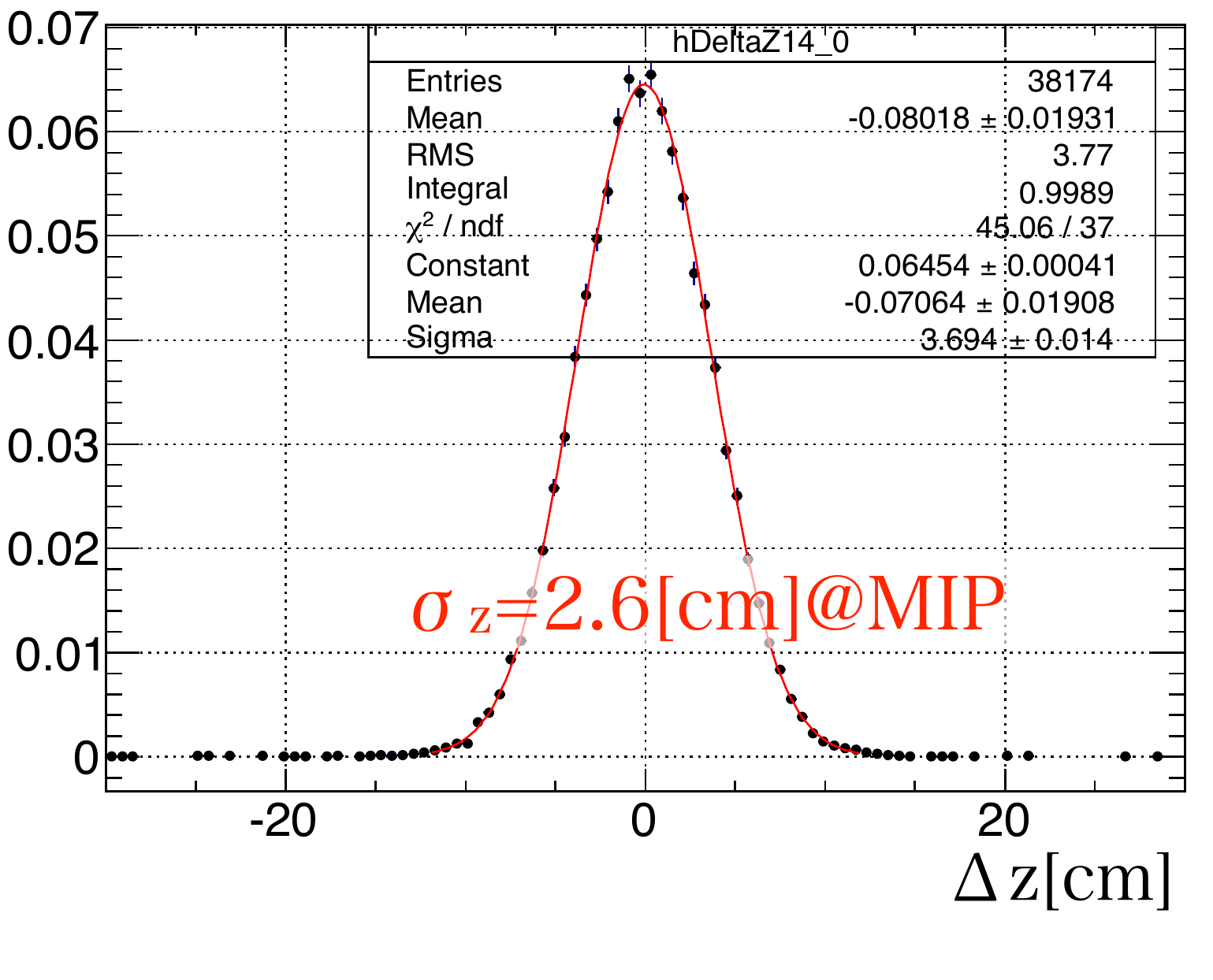}
		\caption{\setlength{\baselineskip}{4mm}Difference of the reconstructed $z$-position between vertically neighboring scintillators for MIP energy.
		By dividing the width by $\sqrt{2}$, we obtained the position resolution $\sigma_z=2.6$ cm for MIP energy.}
		\label{500kg_zres}
	\end{center}
\end{figure}

%%% position resolution
We also evaluated the obtained position resolution.
The hit position in a scintillator along $z$-direction was calculated by using the time difference between PMTs on both ends of the scintillator.
The light velocity in each scintillator was measured by using different cosmic trigger events.
Figure \ref{500kg_velocity} shows the time difference as a function of the hit position.
The typical velocity was 14.3 cm/ns.
Figure \ref{500kg_recz} shows the reconstructed $z$-position for different cosmic trigger events,
and Fig.~\ref{500kg_zres} shows the difference of the reconstructed $z$-position between vertically neighboring scintillators for MIP energy.
By dividing the width by $\sqrt{2}$, we obtained the position resolution $\sigma_z=2.6$ cm for MIP energy.

\begin{figure}[hbtp]
	\begin{center}
		\includegraphics[scale=0.4]{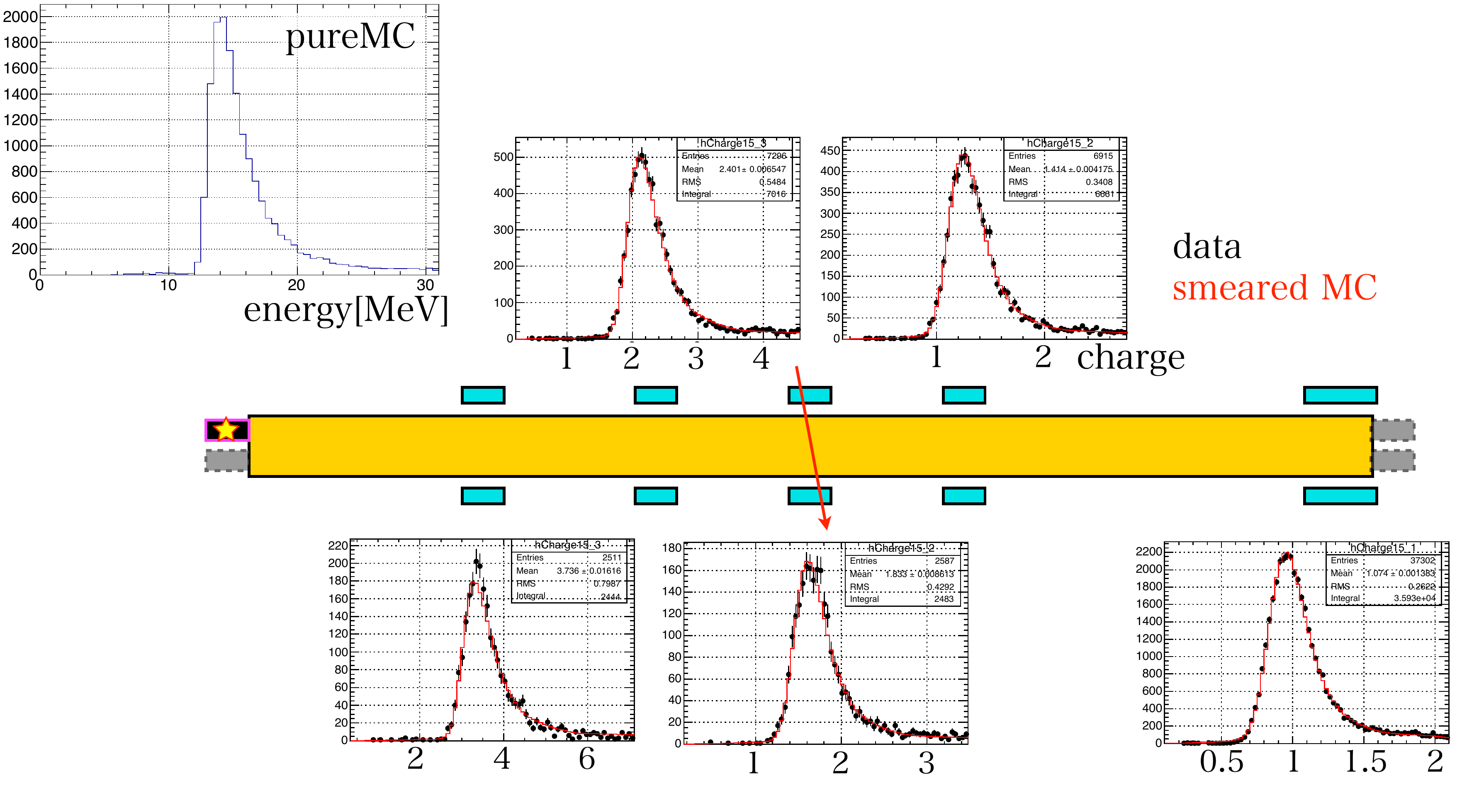}
		\caption{\setlength{\baselineskip}{4mm}Schematic view of the estimation procedure of the charge(energy) resolution.
		The charge resolution was parameterized with Function \ref{func_cres}, and five distributions were fitted with the function at once.}
		\label{500kg_eres_concept}
	\end{center}
\end{figure}

%%% energy resolution
We evaluated the charge(energy) resolution by smearing the output of the pure Monte Carlo simulation, and comparing it with data.
Figure \ref{500kg_eres_concept} shows a schematic view of the estimation procedure of the charge resolution.
We parameterized the charge resolution as follows:
\begin{equation}
	\frac{\sigma_C}{C} = \frac{p_0}{\sqrt{C}} \oplus \frac{p_1}{C} \oplus p_2,
	\label{func_cres}
\end{equation}
where $C$ is the normalized charge with the MIP energy at the far end from the side,
$p_0$ represents the photo-statistics term, $p_1$ represents the noise contribution and $p_2$ represents the calibration precision.
The charge distribution at 5 different positions along the $z$-axis were compared for each PMT.
Because only one PMT was in interest at a time, we set $p_2=0$.
We fitted the charge distributions by changing the remaining of parameters at the same time.
Light attenuation in the scintillator was also considered.
The typical energy resolutions of the scintillators are 3.3\% for 1D and 4.5\% for 3D at the middle of the scintillators for MIP energy.

\subsection{Detector simulation}
\label{sc_500kg_mc}
\begin{figure}[hbtp]
	\begin{center}
		\includegraphics[scale=0.6]{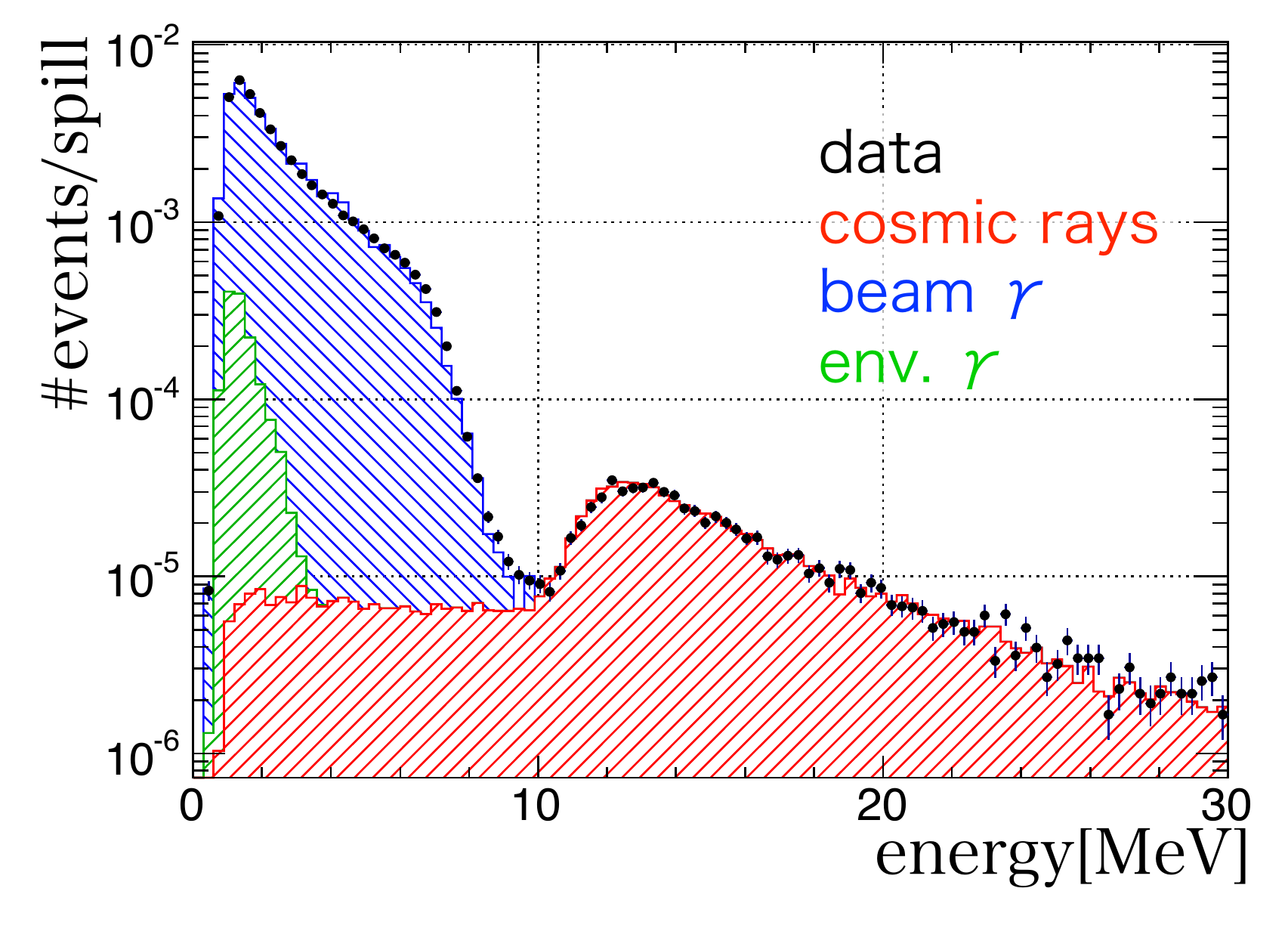}
		\caption{\setlength{\baselineskip}{4mm}Energy distribution for a scintillator during beam period with the Monte Carlo estimation overlaid.}
		\label{500kg_mc}
	\end{center}
\end{figure}

The resolutions described above and other detector responses, such as Birks' quenching, light attenuation and the threshold effect, were implemented to Geant4 based Monte Carlo simulation.
Figure \ref{500kg_mc} shows the deposited energy distribution for a scintillator during a beam period with the Monte Carlo estimation overlaid.
The measured distribution is well reproduced by the Monte Carlo simulation.

\indent

%% file: VetoEff.tex
\subsection{Veto Efficiency}
\indent
\label{sec:VetoEff}

In order to check the effect of having the Inner Veto (IV) and Outer Veto (OV) systems on the MLF third floor background measurement, their particle tagging efficiency was measured. The veto efficiency ($\varepsilon$) is defined as follows:

\begin{eqnarray}
\varepsilon_{\rm IV} & = & \frac{\rm Nb\ of\ coincident\ event\ OV,\ IV\ and\ Target\ (Triple\ Coincidence)}{\rm Nb\ of\ coincident\ event\ OV\ and\ Target\ (Double\ Coincidence)} \\
\nonumber \\
\varepsilon_{\rm OV} & = & \frac{\rm Nb\ of\ coincident\ event\ OV,\ IV\ and\ Target\ (Triple\ Coincidence)}{\rm Nb\ of\ coincident\ event\ IV\ and\ Target\ (Double\ Coincidence)}
\end{eqnarray}

\begin{figure}[htpb!]
	\centering \includegraphics[width=0.75\textwidth]{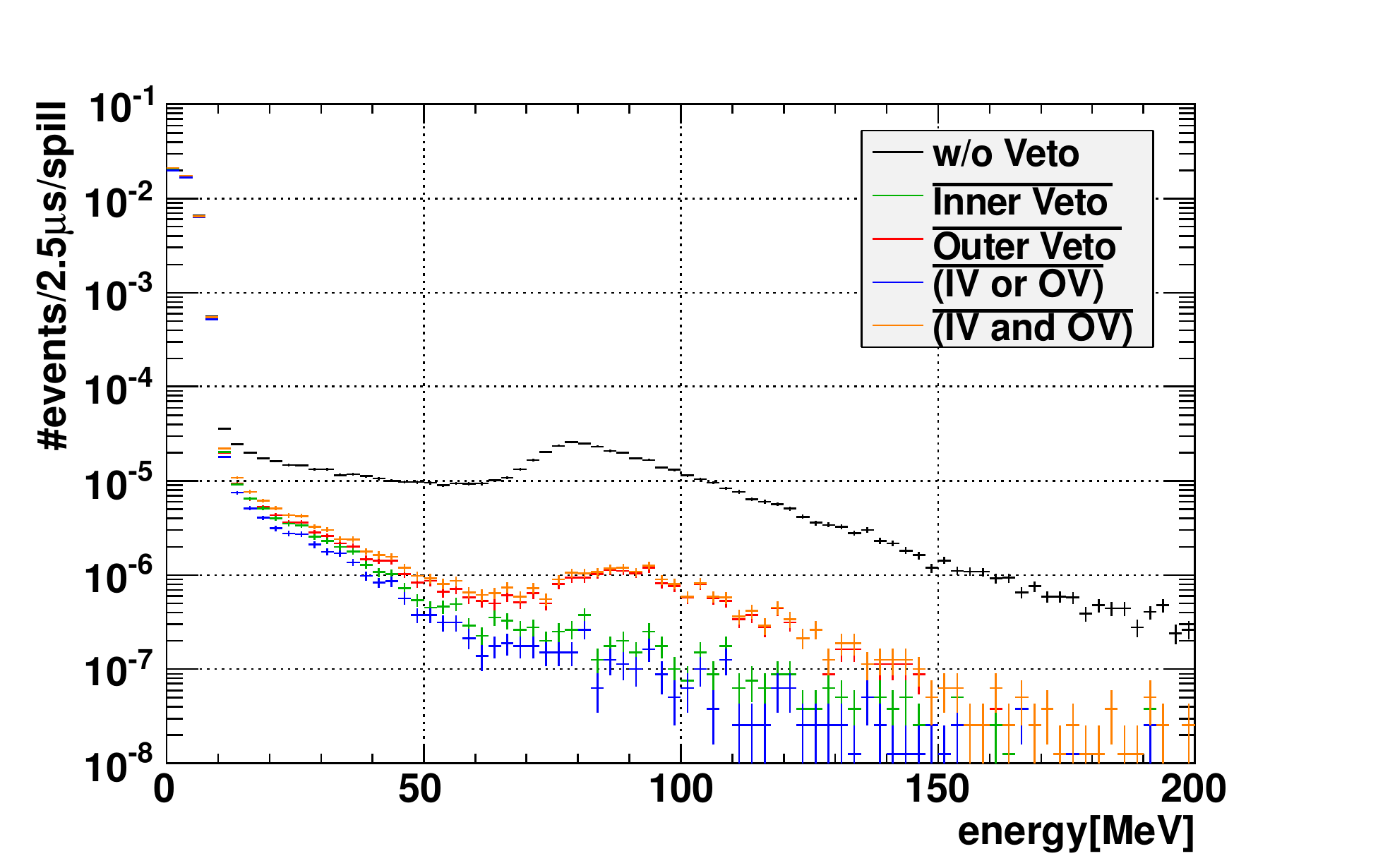}
	\caption{Target energy spectra for different veto conditions.}
	\label{fig:VetoEff}
\end{figure}

The time window selection for the events was $\rm 2.6 \mu s < $ Hit Time $ < \rm 5.3 \mu  s$, and the energy spectra for different veto condition are displayed in figure~\ref{fig:VetoEff}.
Finally, the veto efficiency for different energy ranges is summarized in table~\ref{tab:VetoEff_sum}, where it can be seen that a total veto efficiency is bigger than 99.5\%. In these table and figure, it is also possible to see that muons will deposit an amount of energy larger than 60~MeV and almost of the totality of them will thus be vetoed.
\begin{table}[htpb!]
	\caption{Summary of IV and OV efficiency for different energy ranges.}
	\begin{center}
		\begin{tabular}{cccc}
			\hline \hline
			Energy Range [MeV]	&	$\rm \varepsilon_{IV}$	&	$\rm \varepsilon_{OV}$	&	(IV or OV)		\\
			\hline
			20 $<$ E $<$ 60		&	96.8$\pm$0.2\%			&	94.1$\pm$0.2\%			&	$\sim$99.8\%		\\
			60 $<$ E $<$ 100		&	99.5$\pm$0.04\%			&	96.2$\pm$0.1\%			&	$\sim$99.9\%		\\
			100 $<$ E $<$140		&	99.6$\pm$0.07\%			&	95.1$\pm$0.3\%			&	$\sim$99.9\%		\\
			\hline \hline
		\end{tabular}
	\end{center}
	\label{tab:VetoEff_sum}
\end{table}

%% file: 12detector_Meigo.tex
\section{\setlength{\baselineskip}{4mm}
Small size detectors to measure the PID, Rate and Energy of Prompt BKG}
\indent
\label{sec:BKG_PID}

In a neutrino oscillation experiment, such as one looking for flavour appearance, it is important to understand the backgrounds that can falsify a true neutrino interaction. Even though the selection cuts are applied to maximize the Inverse Beta Decay (IBD) sample in the data, there are accidental coincidences that can pass these cuts. Moreover, an energetic neutron entering the detector, hereafter called fast-n, can create a correlated prompt and delayed event pair when it scatters a proton and it is thermalized by the liquid scintillator, finally being captured by a Gadolinium nucleus, mimicking an IBD event. Natural occurring high energy gammas and neutrons are secondary products of the cosmic rays interactions in the upper atmosphere, from their hadronic and electromagnetic components. In addition, a fast-n can interact with atomic nucleus of the materials in the detector surroundings, leaving the nucleus in an excited state that will produce a high energy gamma during its de-excitation; or the nucleus can absorb the neutron, which will also emit gammas in the process. Therefore, the sources of accidentals and correlated background must be understood and their rates predicted. This section covers the use of small detectors to study such backgrounds. 

\subsection{Measurements at Tohoku University}
\label{subs:RCNS}

In the laboratories of the Research Center for Neutrino Science of Tohoku University (RCNS), several measurements were performed and their data compared with Monte Carlo simulations (MC) in order to understand those backgrounds. Later, as it is described in the next section, these measurements results were compared with the ones performed at the third floor of MLF facility and extrapolated to the detector type and size being proposed in this document.

\subsubsection{NaI}
\label{subsub:NaI}

\begin{figure}[htpb!]
	\centering \includegraphics[width=0.75\textwidth]{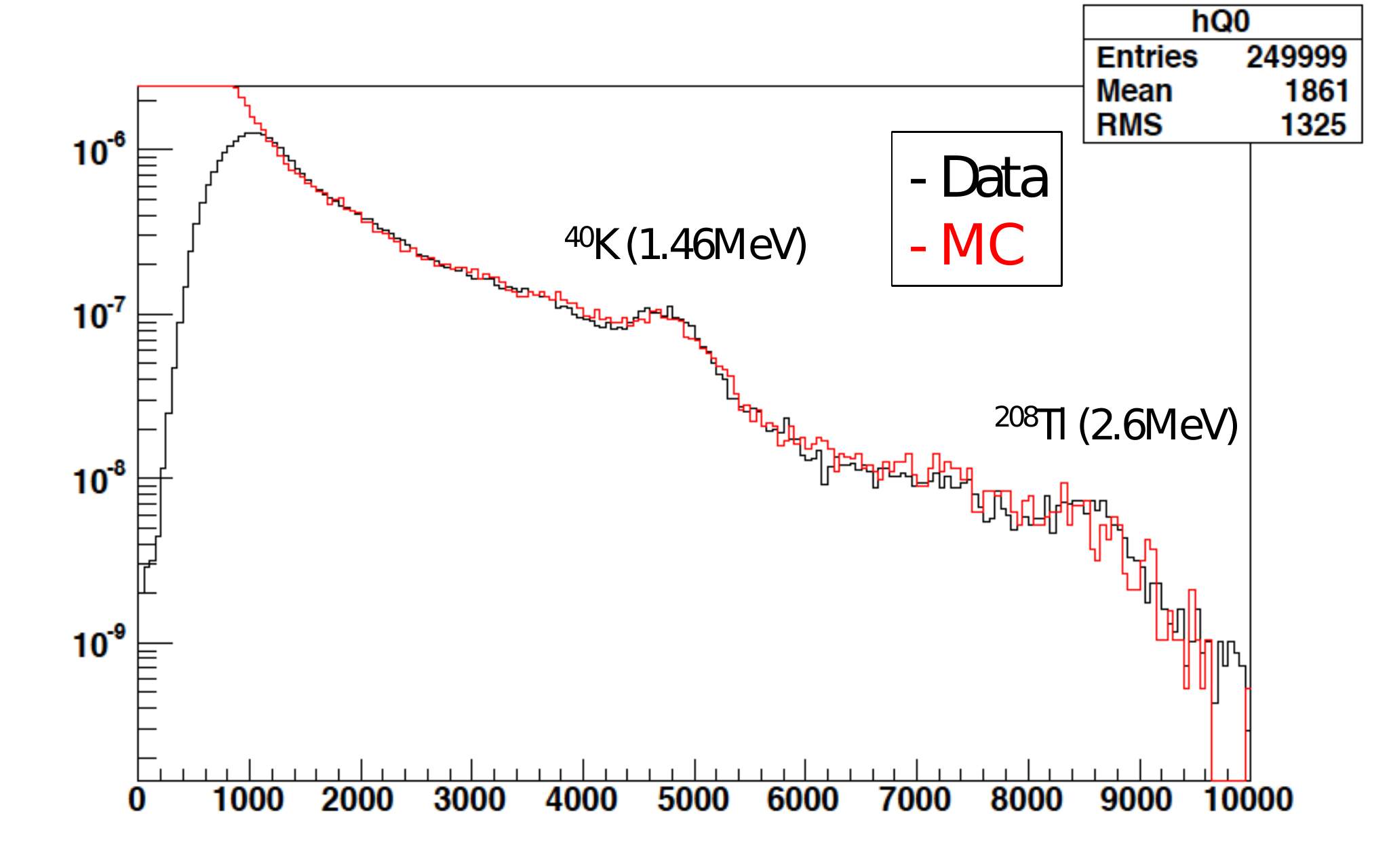}
	\caption{\setlength{\baselineskip}{4mm}
Energy spectra of the low energy ($< 3$~MeV) environmental gammas using a NaI counter, where the potassium and thallium peaks were used for the MC calibration.}
	\label{fig:NaI_calib}
\end{figure}

\begin{figure}[htpb!]
	\includegraphics[width=\textwidth]{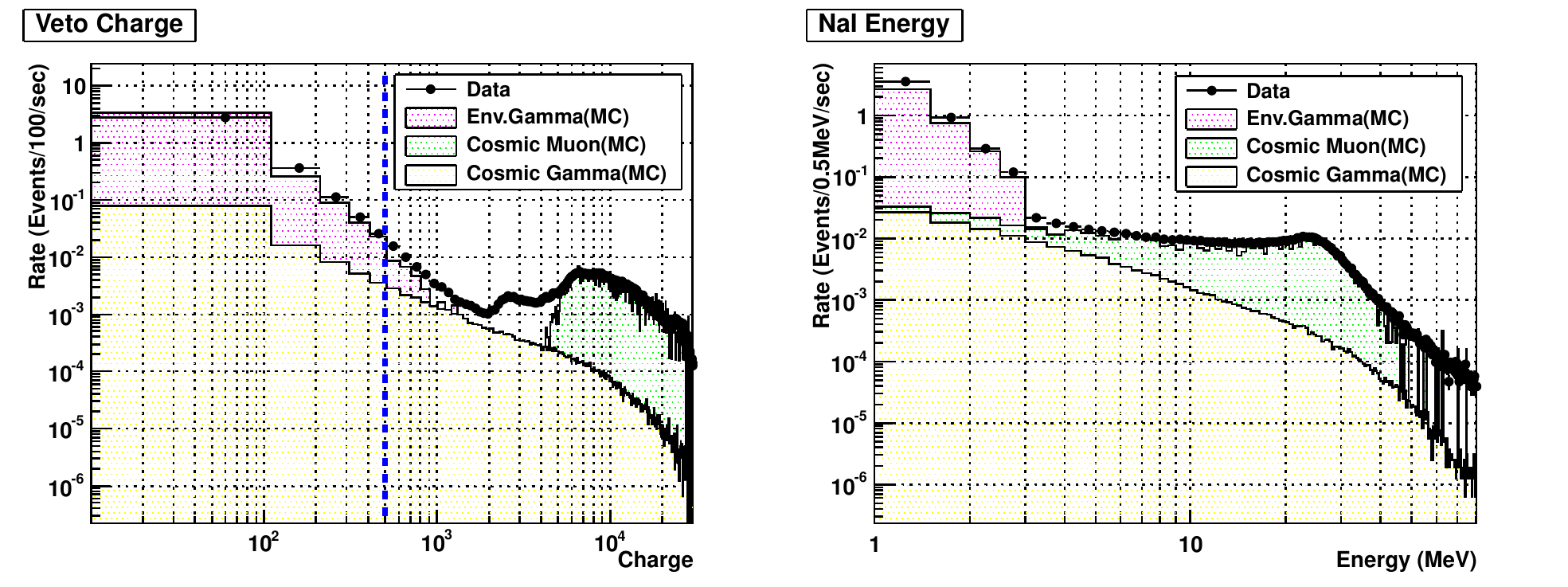}
	\caption{\setlength{\baselineskip}{4mm}
Full energy range spectra for NaI counter and its veto system. The left plot shows the charge distribution for data and MC of the veto system where the doted line represents the veto cut value, while the right plot shows the full NaI energy spectrum where the MC is decomposed in its components: two exponentials with decay constants of 3 and 26~MeV for gammas and cosmic muons.}
	\label{fig:NaI_wo_veto}
\end{figure}

A Sodium Iodide (NaI) scintillator counter (a cylinder with 2" of diameter and 2" height) was first used to measure the high energy gamma ray flux that contributes to the prompt event of the accidental background. The NaI was fully surrounded by plastic scintillator counters (six identical rectangular cuboids of $20\times90\times5$~cm) acting as a veto, in order to ensure the purity of this neutral component. Figure~\ref{fig:NaI_calib} shows the data used to calibrate the MC for this set-up, while the right plot of figure~\ref{fig:NaI_wo_veto} shows the full energy range and the MC with its components. After the environmental gammas, cosmic muons and neutrons were generated, a missing component of the spectrum was identified to be high energy gammas, which energy spectrum can be described by two exponential functions with decay constants of 3 and 26~MeV. These components rates was measured to be 150 and 25~Hz/m$^2$, respectively.

\begin{figure}
	\centering \includegraphics[width=0.75\textwidth]{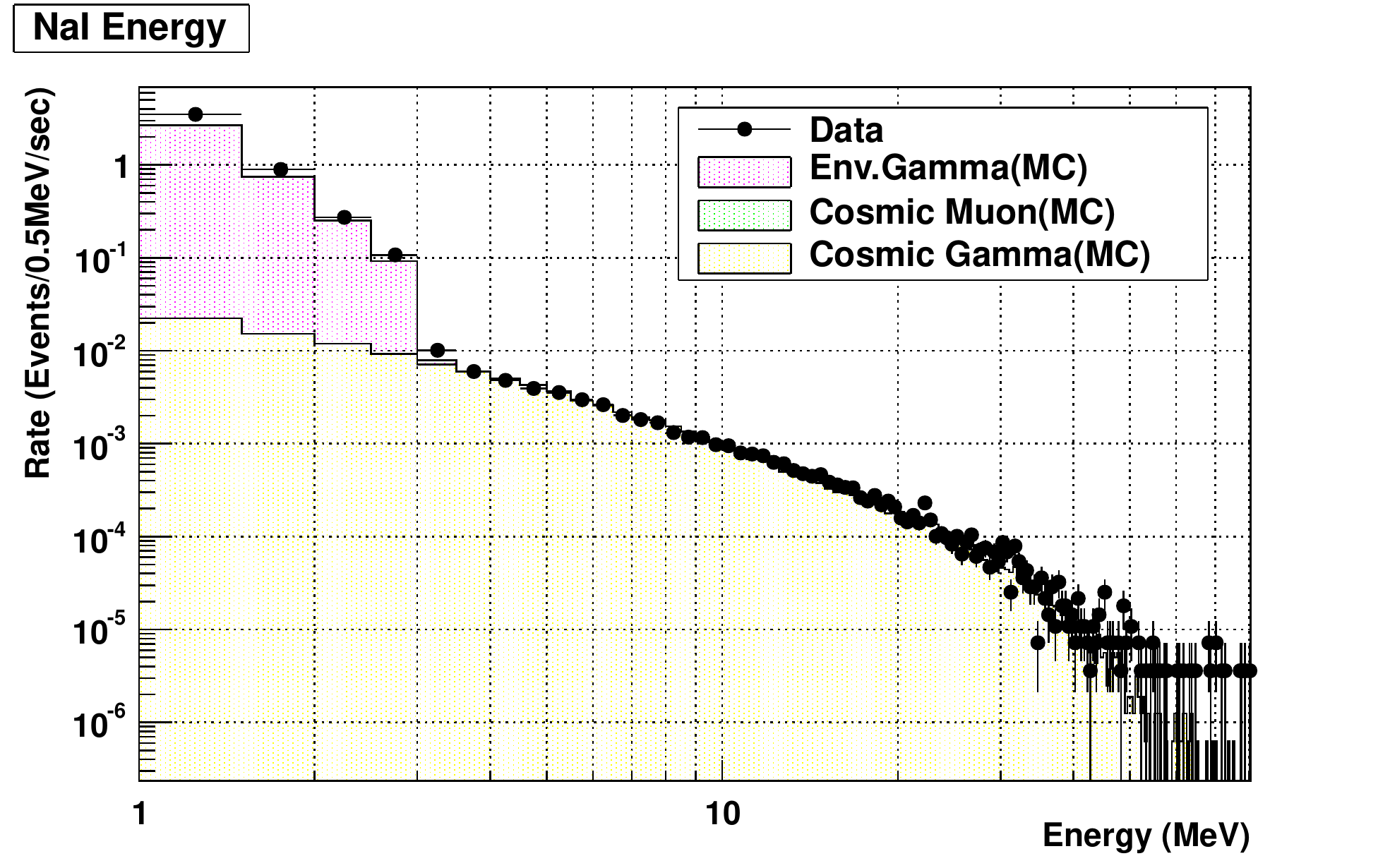}
	\caption{\setlength{\baselineskip}{4mm}
NaI energy spectrum for the remaining events after applying the veto cut.}
	\label{fig:NaI_veto}
\end{figure}

Finally, the effect of adding the veto system was studied. The left plot of figure~\ref{fig:NaI_wo_veto} shows the data veto spectrum compared with total the MC and its components. %
The doted line represents the cut value used to reject events depositing considerable (bigger than 500 default units of charge [DUQ]) amounts of energy in any part of the veto. This plot also shows that there is an excess when comparing the MC and data histograms between the cut value and 5000~DUQ. One possible explanation for this excess is that it is caused by Michel-$e^-$, and that the simulation does not apply well the low energy muons that decay and produce such events. Thus, the 500 low value cut was used, to avoid this unknown region. The remaining events, after applying the veto, are shown in figure~\ref{fig:NaI_veto}, where it can be seen that on the prompt energy window (20 to 60~MeV) the events are mainly gammas.

\subsubsection{NE213}
\label{subsub:NE213}
\begin{figure}[htpb!]
	\includegraphics[width=0.75\textwidth]{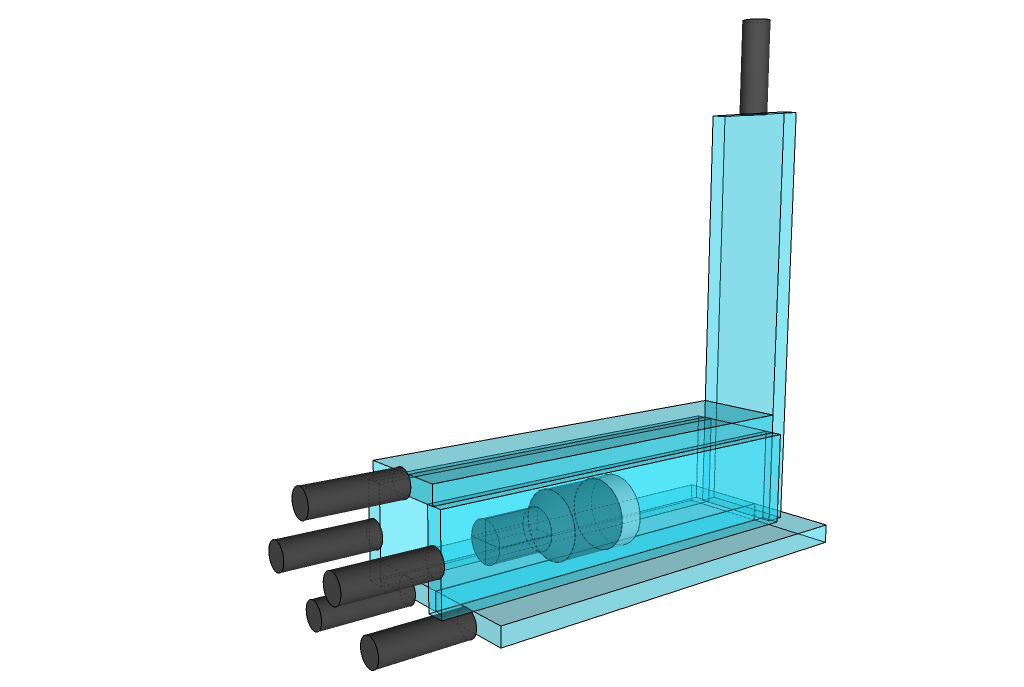}
	\caption{\setlength{\baselineskip}{4mm}
Scheme of the NE213 detector surrounded by the veto system formed by plastic scintillator blocks.}
	\label{fig:NE213_scheme}
\end{figure}

The next important component of the natural background that was studied is the fast-n. Thus, a liquid scitillator with Particle Identification (PID) capabilities was chosen. A cylindrical  aluminium housing, with white painted inner walls, of 613.6 cm$^3$ was filled with NE213, closed with a glass plate and attached to a 5 inches PMT (R1250-03). This detector where surrounded by the same plastic scintillators, described in the previous subsection, making the veto system. The detector and its veto are shown in figure~\ref{fig:NE213_scheme}.

The NE213 is a well known scintillator for Pulse Shape Discrimination (PSD). The ratio of its fast and slow time decay constants is dependent on the type of the incident particle. Therefore, it was used to separate the neutron (recoiled proton) from gamma (mainly Compton electrons and positrons from pair production) signals.

\begin{figure}[htpb!]
	\includegraphics[width=0.5\textwidth]{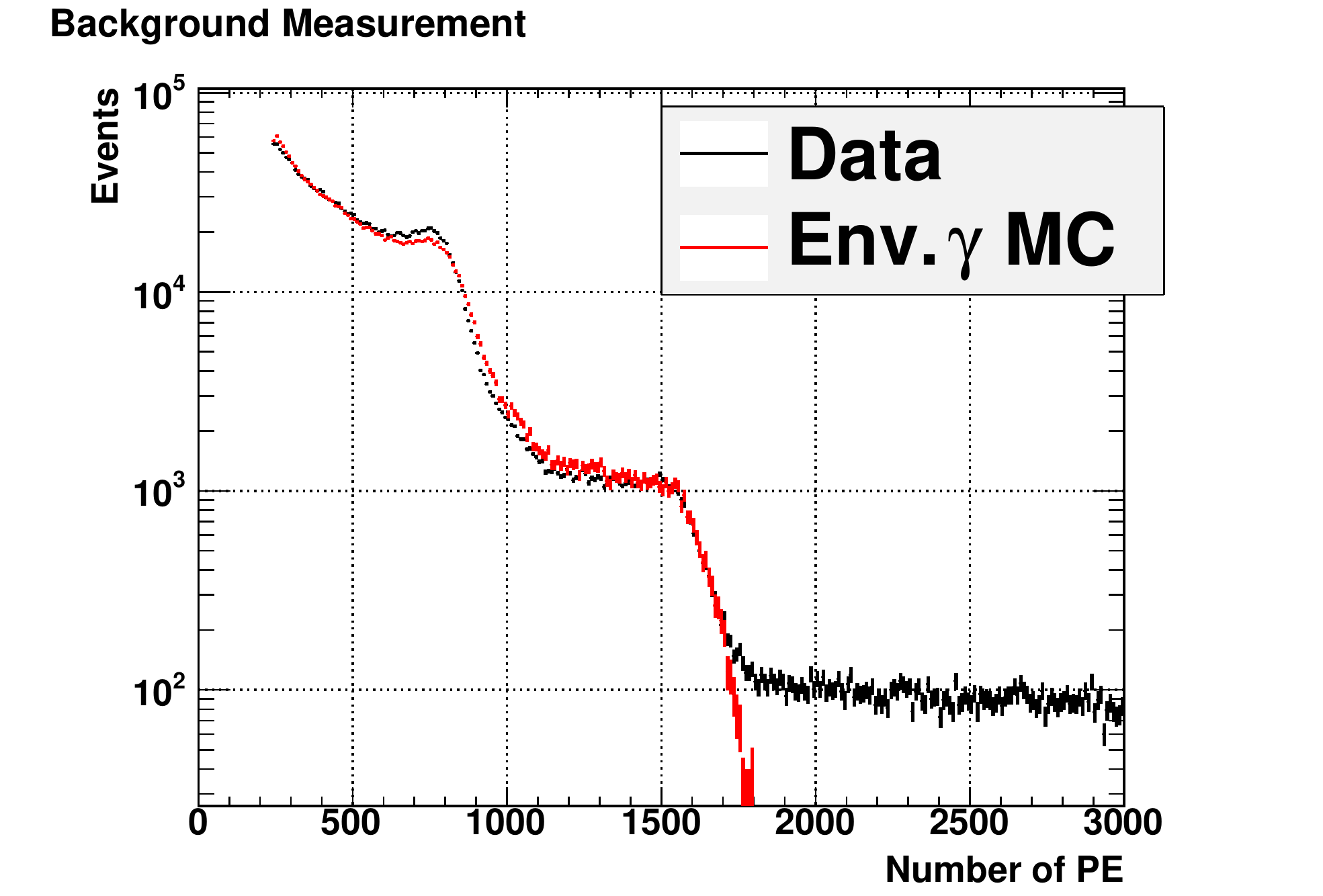}
	\includegraphics[width=0.5\textwidth]{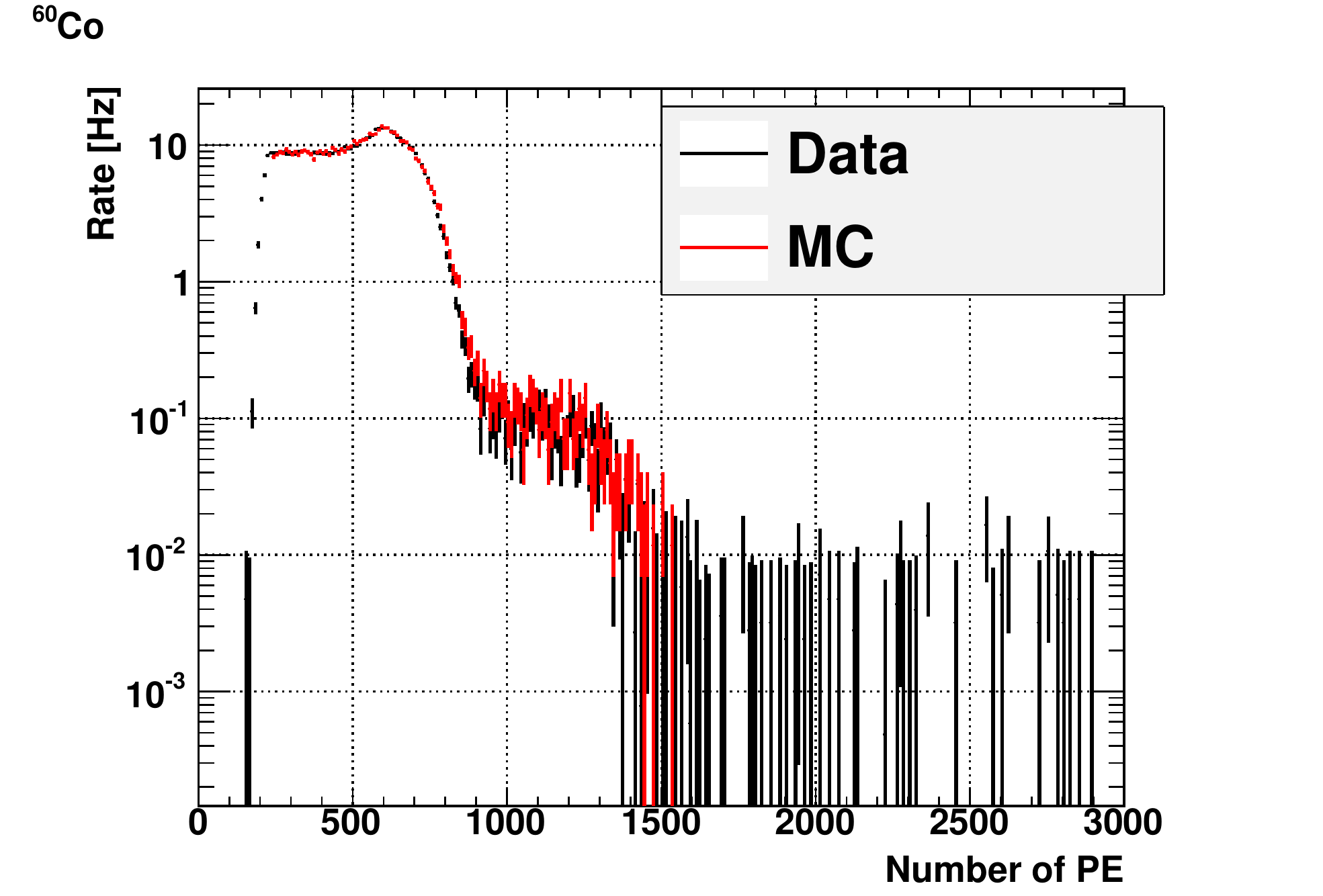}
	\caption{\setlength{\baselineskip}{4mm}
Energy spectra of the environmental gammas and $^{60}$Co using the NE213 counter.}
	\label{fig:NE213_calib}
\end{figure}

As for the NaI case, the NE213 set-up was compared with MC for tuning its parameters. On figure~\ref{fig:NE213_calib} the photon-electron (PE) spectra of data and MC for environmental gammas and a Cobalt-60 ($^{60}$Co) source are presented. %
While the nominal value is 13565 PE/MeV the plots shows that a tuned MC agrees well for a value of 650 PE/MeV. This difference is mainly due the Quantum Efficiency (QE) of the PMT, the reflection inefficiency of the walls and leakage through the optical contacts.

\begin{figure}[htpb!]
	\includegraphics[width=0.5\textwidth]{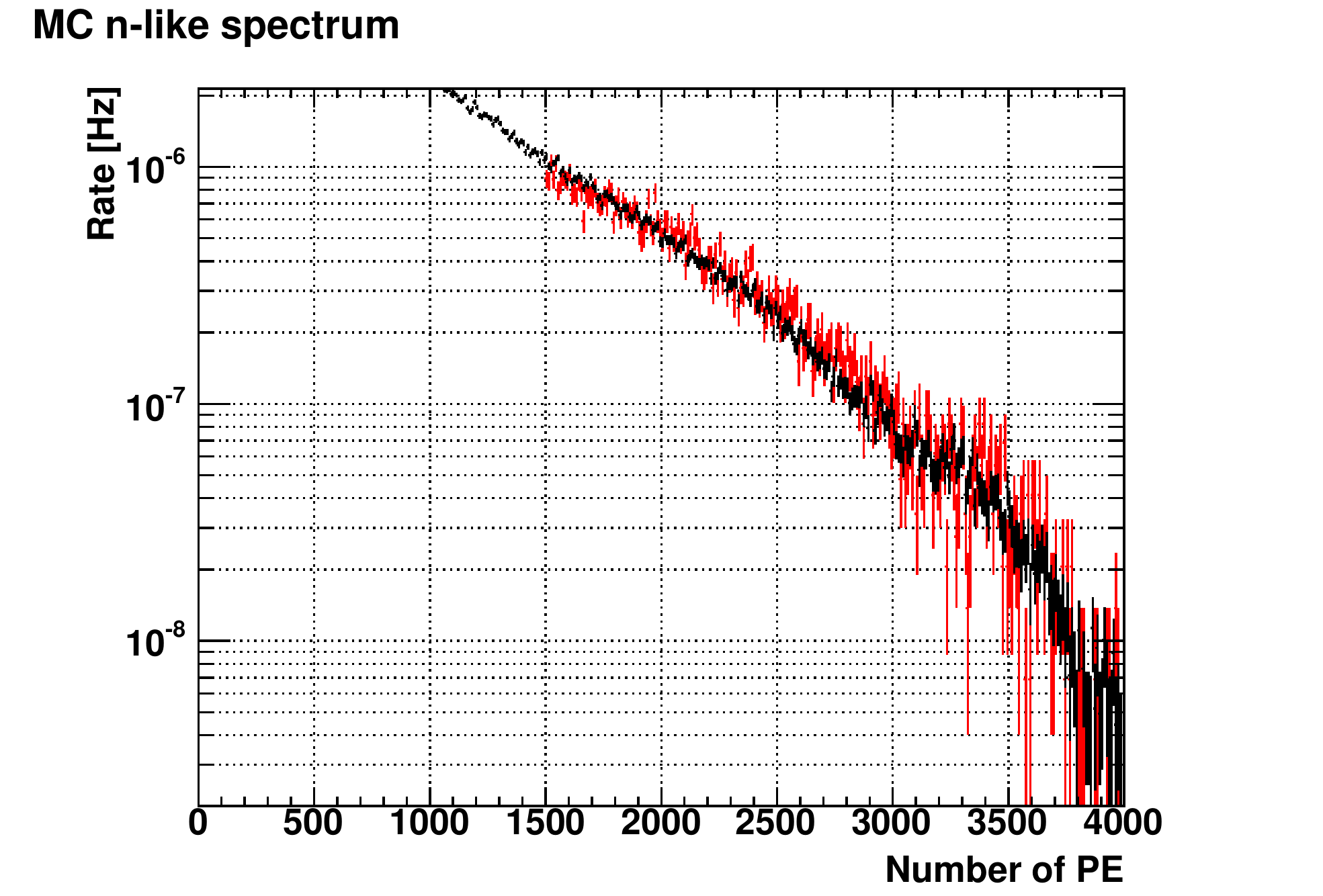}
	\includegraphics[width=0.5\textwidth]{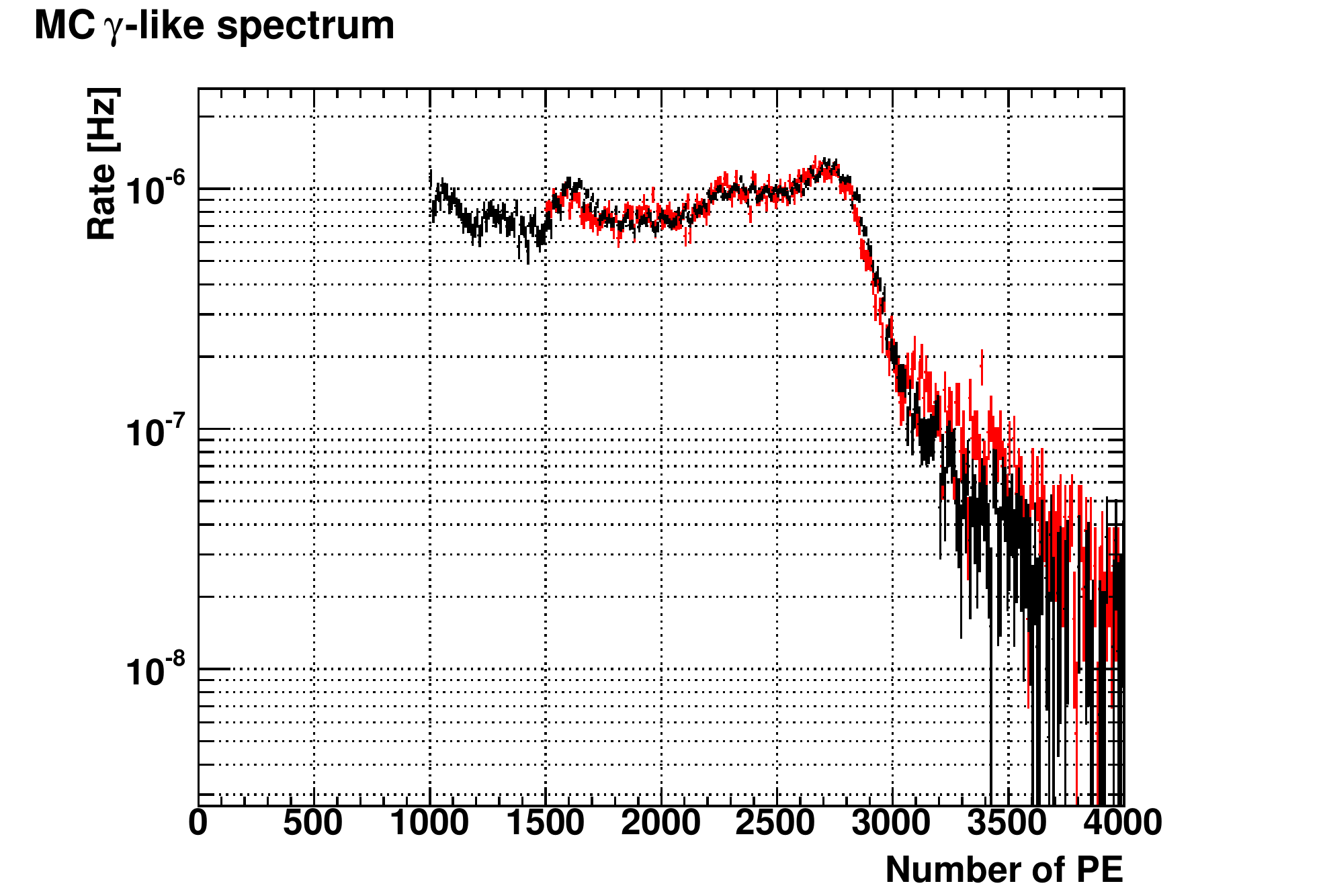}
	\caption{\setlength{\baselineskip}{4mm}
AmBe data and MC comparison. The left plot is the events tagged as neutrons and the right plot for the ones tagged as gammas. The agreement between the spectra for the neutrons shows that a good value for the Birks constant was chosen.}
	\label{fig:AmBe_calib}
\end{figure}

Another important parameter tuned was the scintillator Birks constant. A well defined Birks value is necessary due the proton elastic scattering by neutrons, otherwise its spectrum would have an incorrect shape. Figure~\ref{fig:AmBe_calib} shows the neutron-like and gamma-like events for data and MC when using an Americium-Beryllium (AmBe) source, where the MC value for the Birks constant is 0.100~mm/MeV, while a referenced value is 0.107~mm/MeV~\cite{NE213}. %
The AmBe source is a well used source for neutron studies. When an Am nucleus emits an alpha particle, it can interact with the Be producing free neutrons and gammas. The method to separate the two components of the AmBe spectrum, was to divide the tail integral of the electric pulse by the total integral. Since the slow decay component of the scitillator has a higher contribution for neutrons than for gammas, this ratio variable ($Q_{tail}/Q_{tot}$) has a higher value as well.

\begin{figure}[htpb!]
	\includegraphics[width=\textwidth]{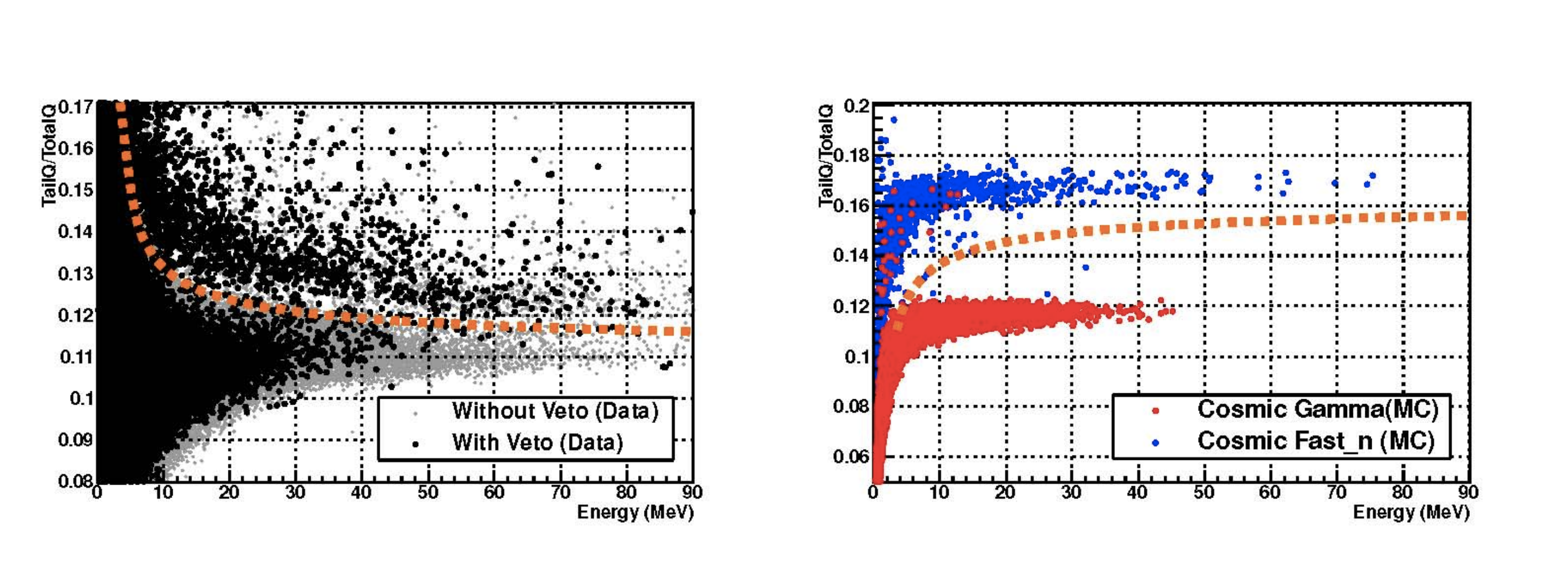}
	\caption{\setlength{\baselineskip}{4mm}
AmBe scatter plot of $Q_{tail}/Q_{tot}$ versus the event reconstructed energy. The left plot shows the data and the right one the MC. The doted horizontal lines represent the cut values to separate the neutron and gamma components.}
	\label{fig:AmBe_scatter}
\end{figure}

As shown in figure~\ref{fig:AmBe_scatter}, the neutron component is distinguishable from the gamma one, when plotting the $Q_{tail}/Q_{tot}$ versus the event reconstructed energy. Although the shape of the data and MC scatter plots does not agree well, due to difficulties on calibrating the MC for timing and pulse creation, both particles can be selected when applying a correct energy and PID cut.

\begin{figure}[htpb!]
	\includegraphics[width=0.95\textwidth]{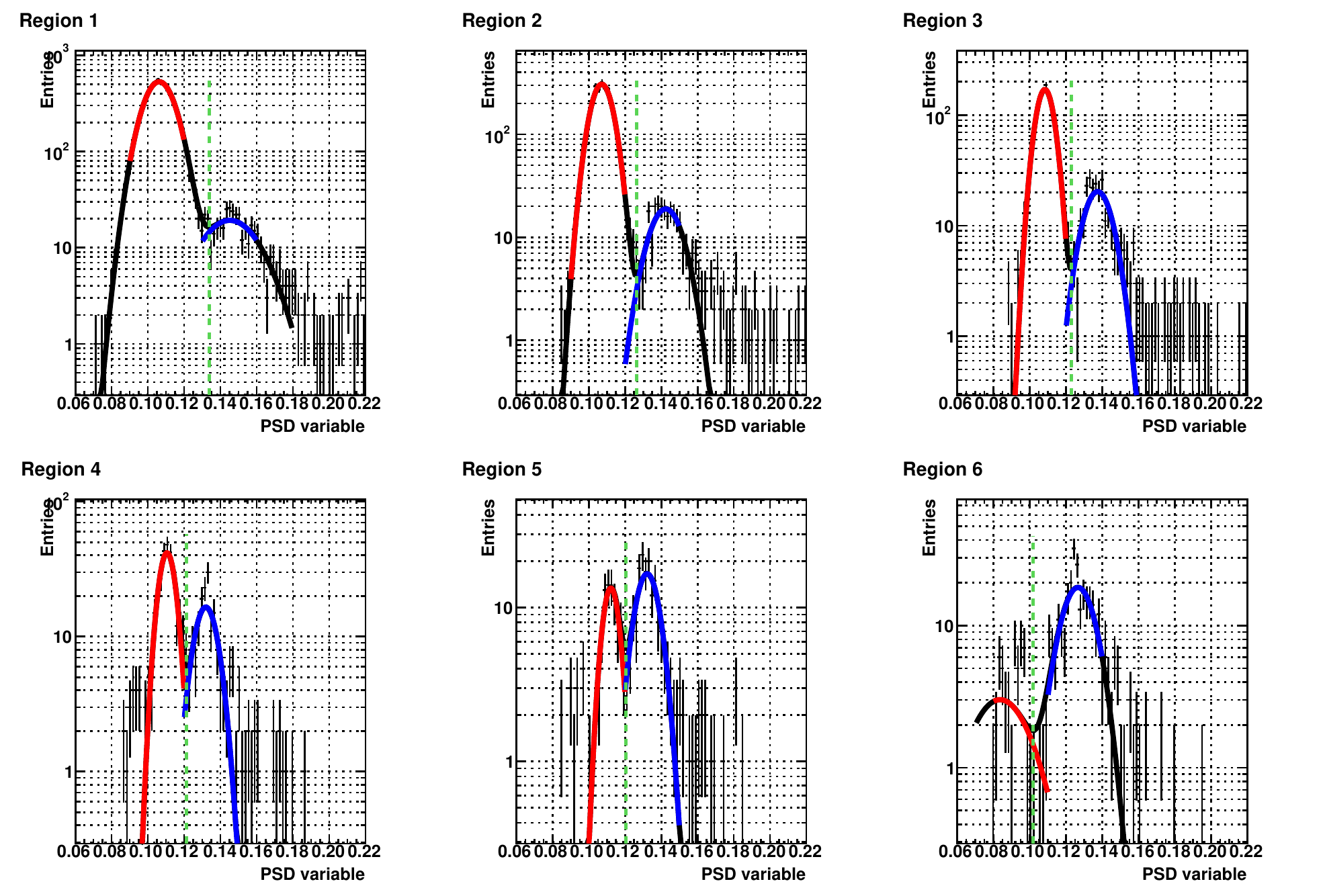}
	\caption{\setlength{\baselineskip}{4mm}
$Q_{tail}/Q_{tot}$ histograms for each region defined in the text. The data was fitted by a two peak Gaussian. The red and blue lines represent the gamma and neutron component respectively. The vertical line is the cut value to select the neutron events from the data for each region.}
	\label{fig:AmBe_regions}
\end{figure}

The next step was to calculate the efficiency of the data sample's PSD cut. The data sample in the energy range (from 7 to 61~MeV) was divided in six regions: 1) 7 to 11, 2) 11 to 16, 3) 16 to 23, 4) 23 to 30, 5) 30 to 38, and 6) 38 to 61~MeV. For each region an one dimension histogram was constructed for the PSD variable and fitted by a two peak Gaussian function. Each peak of the Gaussian function represents the neutron and gamma components. Then, a cut value was chosen to distinguish both components and its efficiency was calculated based on the fitted Gaussian. This method is depicted by figure~\ref{fig:AmBe_regions}, where the histogram for each region is displayed together with the fitted Gaussian and the neutron selection cut value represented by a vertical line. The neutron and gamma component of the Gaussian is shown in blue and red respectively.

\begin{figure}[htpb!]
	\includegraphics[width=\textwidth]{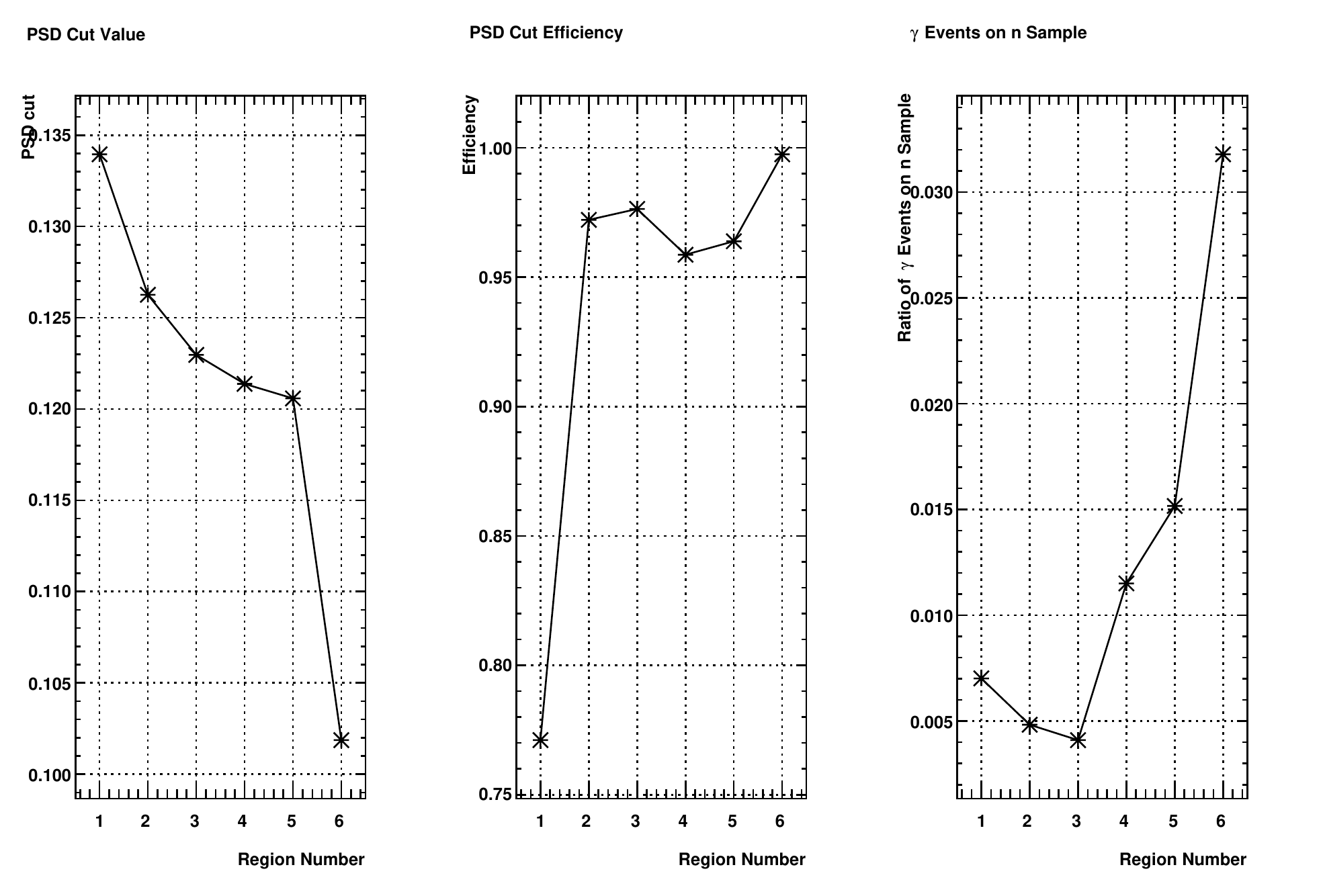}
	\caption{\setlength{\baselineskip}{4mm}
$Q_{tail}/Q_{tot}$ cut value (left), cut efficiency for neutrons selection (center) and gamma contamination at the neutron sample (right), for each regions defined in the text. The cut efficiency for neutron events selection is defined as the ratio between the events above the cut value and the total integral of the fitted Gaussian function. Similarly, the gamma contamination of the neutron sample is the ration between the gamma events above the neutron cut and the selected neutrons.}
	\label{fig:PSD_cut_eff_con}
\end{figure}

In figure~\ref{fig:PSD_cut_eff_con} the $Q_{tail}/Q_{tot}$ cut value, cut efficiency for neutrons and gamma contamination at the neutron sample, for each region are shown.

\begin{figure}[htpb!]
	\includegraphics[width=\textwidth]{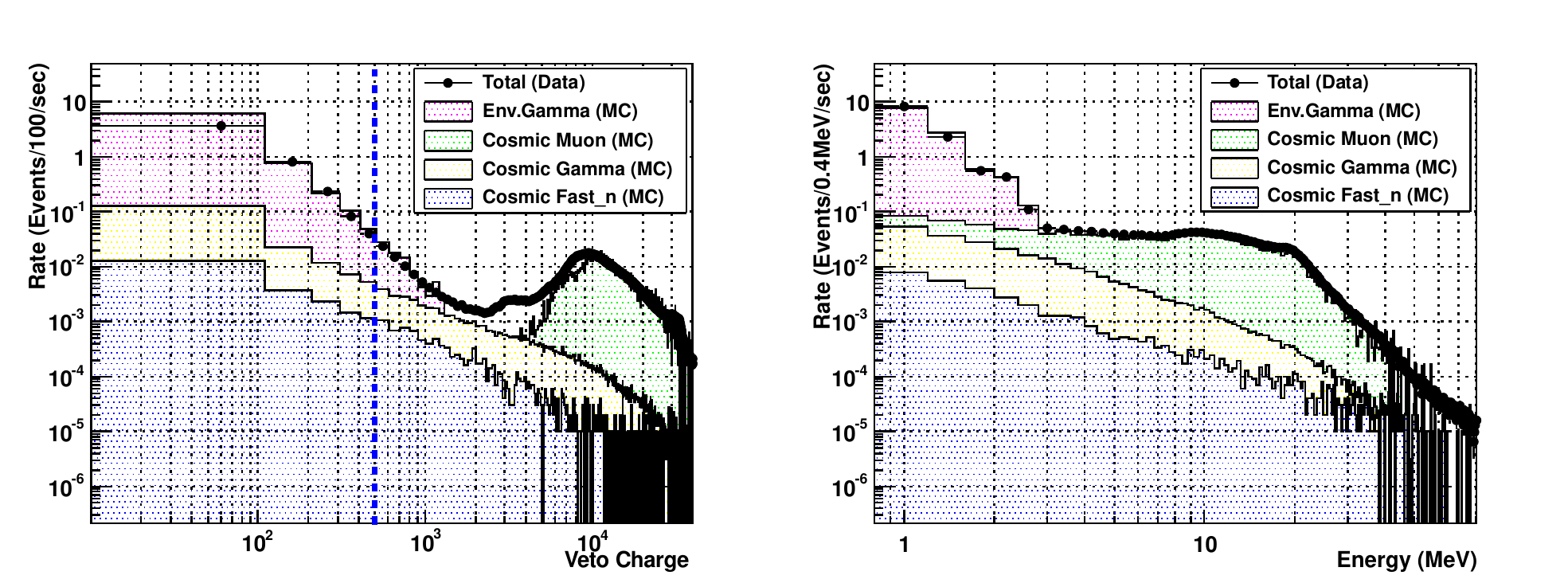}
	\caption{\setlength{\baselineskip}{4mm}
Data and MC comparison for the NE213 set-up. The left plot shows the veto charge distribution and the right plots shows the NE213 reconstructed energy distribution.}
	\label{fig:BKG_Meigo_Veto}
\end{figure}

\begin{figure}[htpb!]
	\centering \includegraphics[width=.75\textwidth]{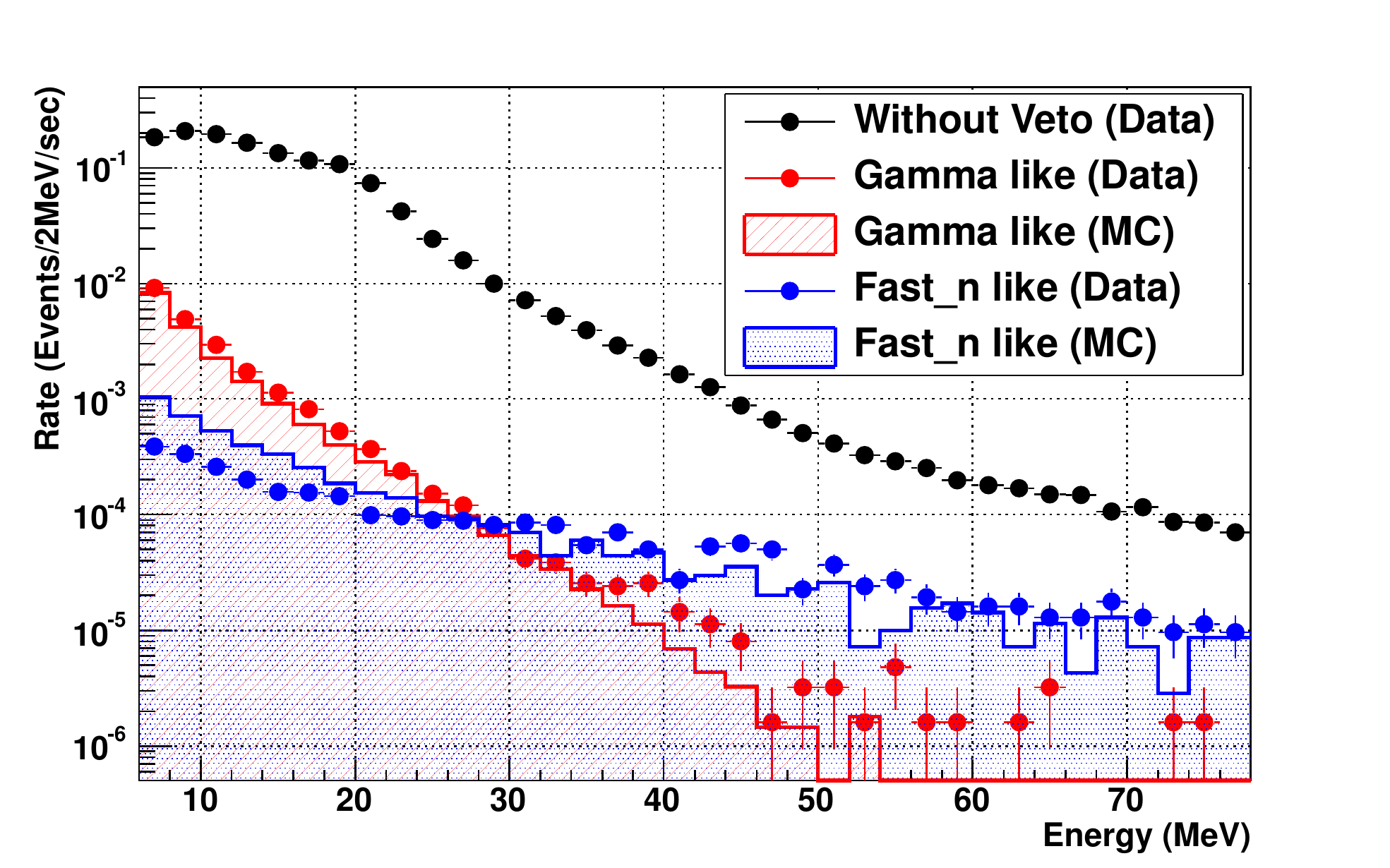}
	\caption{\setlength{\baselineskip}{4mm}
Reconstructed energy distribution for events with low or none energy deposition in the veto. The data and MC components of neutrons and gammas, selected as explained in the text, are also compared. Data events without veto applied are also shown for comparison.}
	\label{fig:BKG_Meigo_Spec}
\end{figure}

Finally, as performed with the NaI, figure~\ref{fig:BKG_Meigo_Veto} shows the veto charge distribution on the left and NE213 reconstructed energy on the right, where the data and MC are presented. %
On figure~\ref{fig:BKG_Meigo_Spec} the vetoed selected neutron and gamma components, as explained above, can also be seen. On this figure it is possible to see that the components agree well between the data and the MC model.

With this set-up, the fast-n flux detected above 20~MeV is of $1.28 \pm 0.05 \times 10^{-3}$~Hz (statistic uncertainty), while the MC gives a rate of $1.12 \times 10^{-3}$~Hz. These numbers are close to the one used in the first proposal. For gammas, in the same energy range, the measured rate is of $1.18 \pm 0.04 \times 10^{-3}$~Hz (statistic uncertainty), while the MC gives $0.95 \times 10^{-3}$~Hz. Therefore, the data and MC above 20~MeV for both gammas and fast-n agrees within 20\% of the uncertainty. For the MC presented here and for the one in the first proposal, the same cosmic fast-n generator was used. The generator's flux and spectrum were tuned with Tohoku University's Reactor Monitor detector, composed of 200~litters of liquid scitillator.

%% file: 13detector_tohoku24.tex
\section{\setlength{\baselineskip}{4mm}
Gamma Ray Measurements with a Small Plastic Scintillator}

\subsection{Setup}
\label{sec:tohoku24_setup}
\begin{figure}
\begin{center}
\includegraphics[width=8cm]{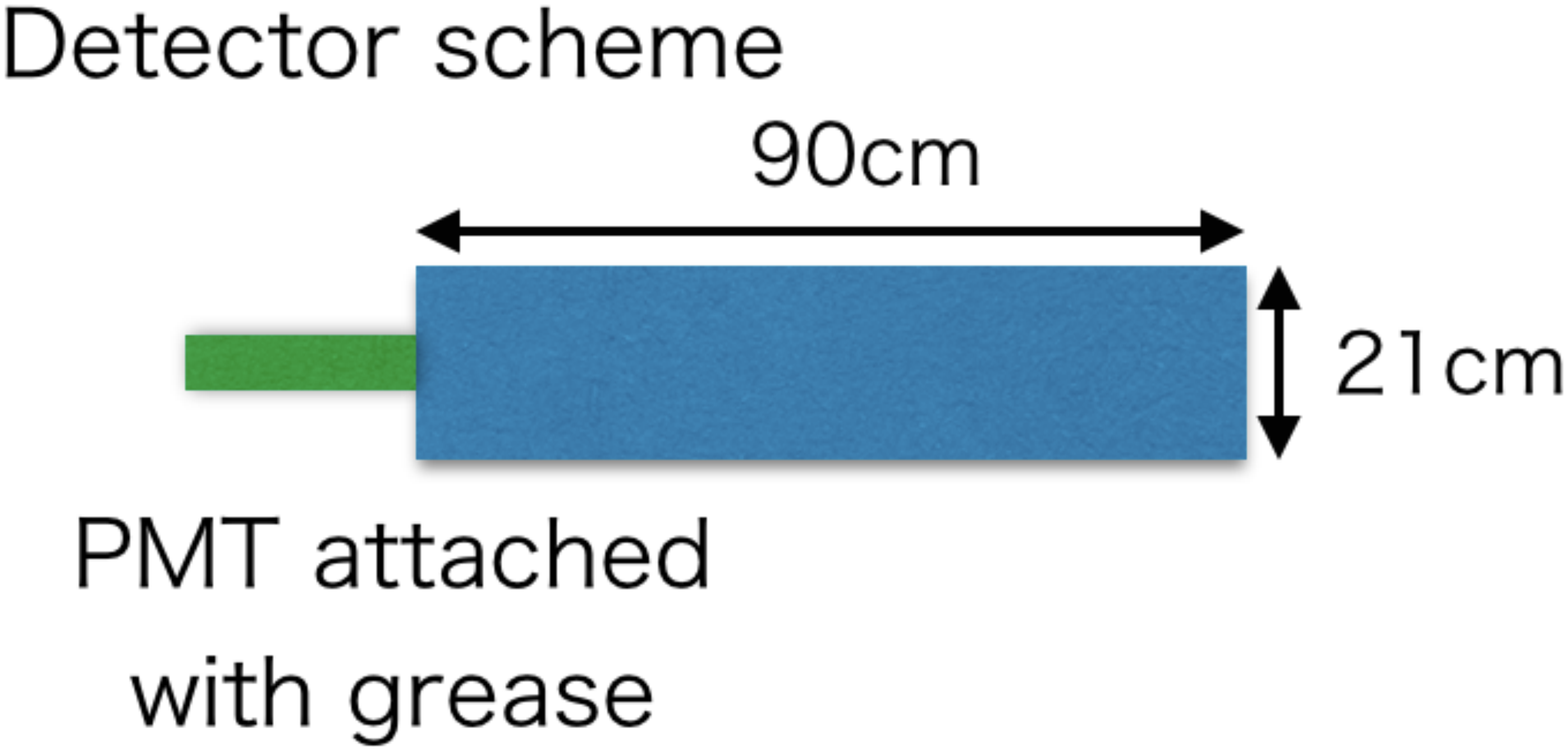}
\caption{Schematic of the small plastic scintillator}
\label{fig:tohoku24_scheme}
\end{center}
\end{figure}

\indent
\indent
To validate the MC model for gamma background, 
we measured the event rate at Point 2 with a small plastic scintillator.
The scintillator dimenssions are $90\times 21\times 4.5$ cm$^3$ and 
a PMT is attached to side of the scintillator with grease as shown in Fig.~\ref{fig:tohoku24_scheme}.
The PMT waveforms were digitized with a 6 $\mu$s time window and recorded through a FINESSE 65 MHz FADC.
The trigger system was same as the 500 kg detector.

\subsection{Lead Shield Configurations}
\label{sec:tohoku24_configuration}
\begin{figure}
\begin{center}
\includegraphics[width=14cm]{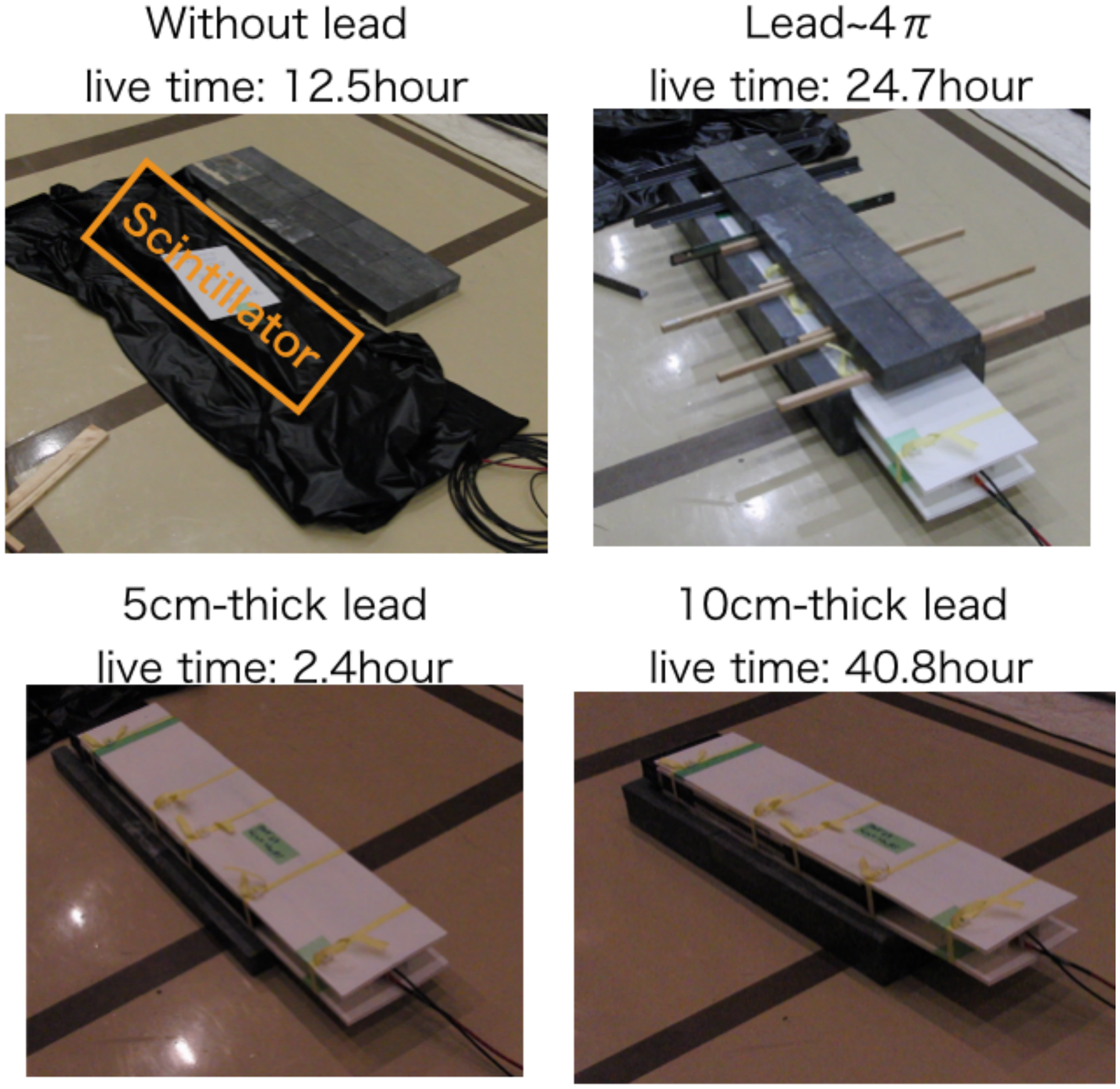}
\caption{Shield configurations of the small scintillator.}
\label{fig:tohoku24_shieldconfig}
\end{center}
\end{figure}

\indent
\indent
We measured the number of events induced by gammas with 4 shield configurations as shown in Fig.~\ref{fig:tohoku24_shieldconfig};
\begin{enumerate}
\item Without lead shield (top left of Fig.~\ref{fig:tohoku24_shieldconfig}, labeled ``Without lead")
\item 5-cm thick lead shield surrounding nearly all over the scintillator (top right of Fig.~\ref{fig:tohoku24_shieldconfig}, labeled ``Lead~4$\pi$")
\item 5-cm thick lead shield under the scintillator (bottom left of Fig.~\ref{fig:tohoku24_shieldconfig}, labeled ``5-cm thick lead")
\item 10-cm thick lead shield under the scintillator (bottom right of Fig.~\ref{fig:tohoku24_shieldconfig}, labeled ``10-cm thick lead")
\end{enumerate}

\subsection{Energy Calibration}
\label{sec:tohoku24_calibration}
\begin{figure}
\begin{center}
\includegraphics[width=8cm]{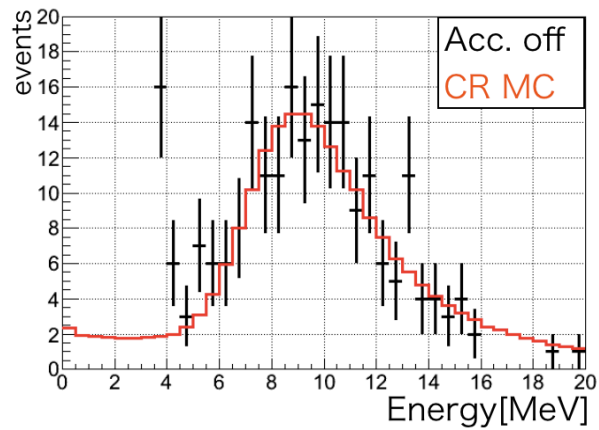}
\caption{\setlength{\baselineskip}{4mm}
Energy calibration using cosmic ray muon. The black line shows the measurement and the red line shows MC.}
\label{fig:tohoku24_calibration}
\end{center}
\end{figure}

\indent
\indent
For the energy calibration, we used the peak of cosmic muons.
Using MC, we estimated that the peak energy is about 9 MeV.
We that the assumed energy resolution around 9 MeV for MC is 18 \% mainly caused by attenuation of light.
A calibrated energy spectrum taken while the accelerator was off and a MC energy spectrum made by cosmic ray muons with energy resolution are shown in Fig.~\ref{fig:tohoku24_calibration}.

%% file: 14bkg_delayed_tohoku24.tex
\subsection{Measurements}
\label{sec:tohoku24_measurement}
\subsubsection{Energy spectra of 4 configurations}
\label{sec:tohoku24_spectra}
\begin{figure}
\begin{center}
\includegraphics[width=12cm]{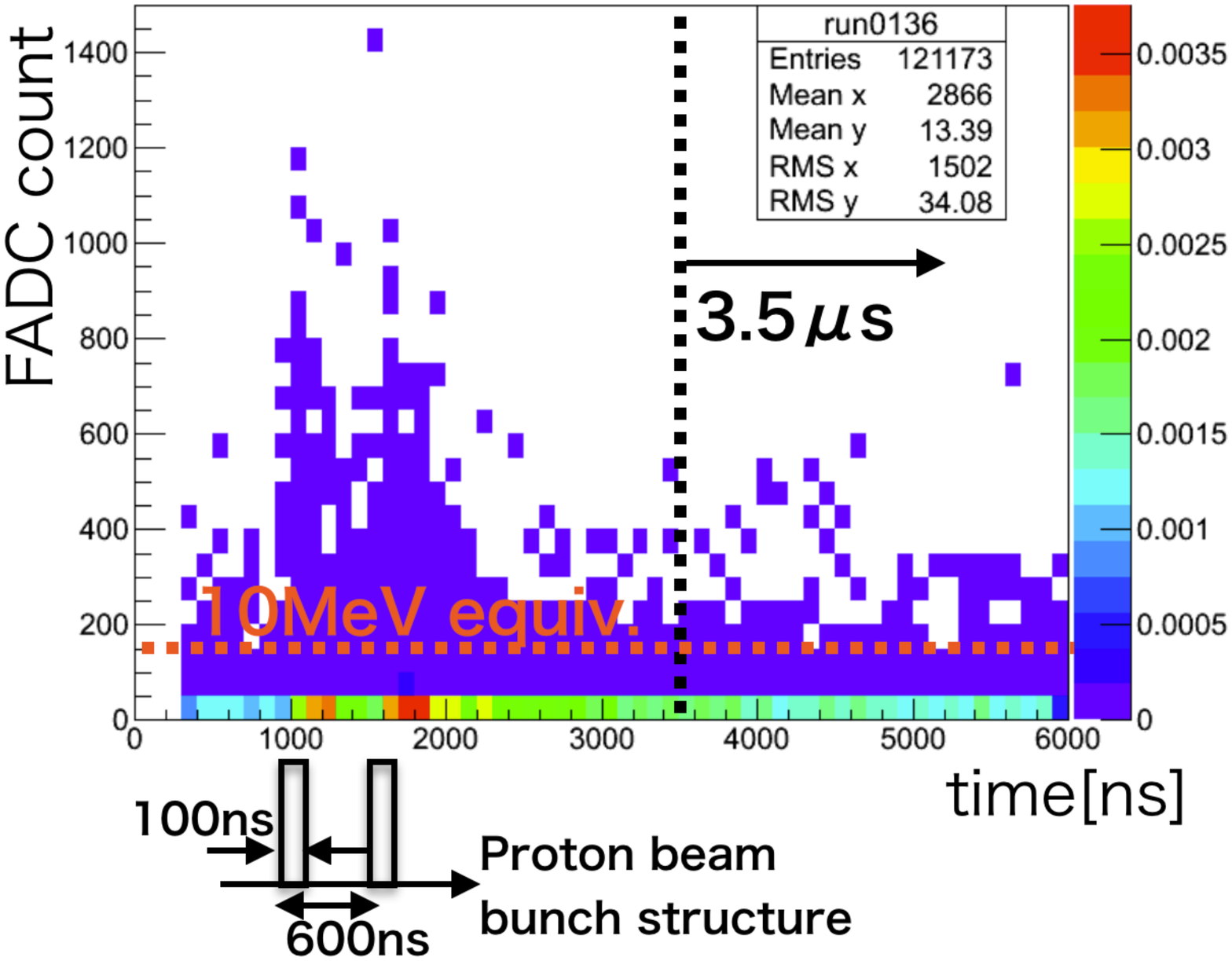}
\caption{\setlength{\baselineskip}{4mm}Example of FADC count vs time plot ("Without lead" configuration). Around 1-2$\mu$s, beam bunch structure is clearly shown. To avoid effects of beam activity, we use data 3.5 $\mu$s after the gate start timing. FADC count of 150 is equivalent to $\sim$10 MeV.}
\label{fig:tohoku24_FADCvstime}
\end{center}
\end{figure}

\begin{figure}
\begin{center}
\includegraphics[width=12cm]{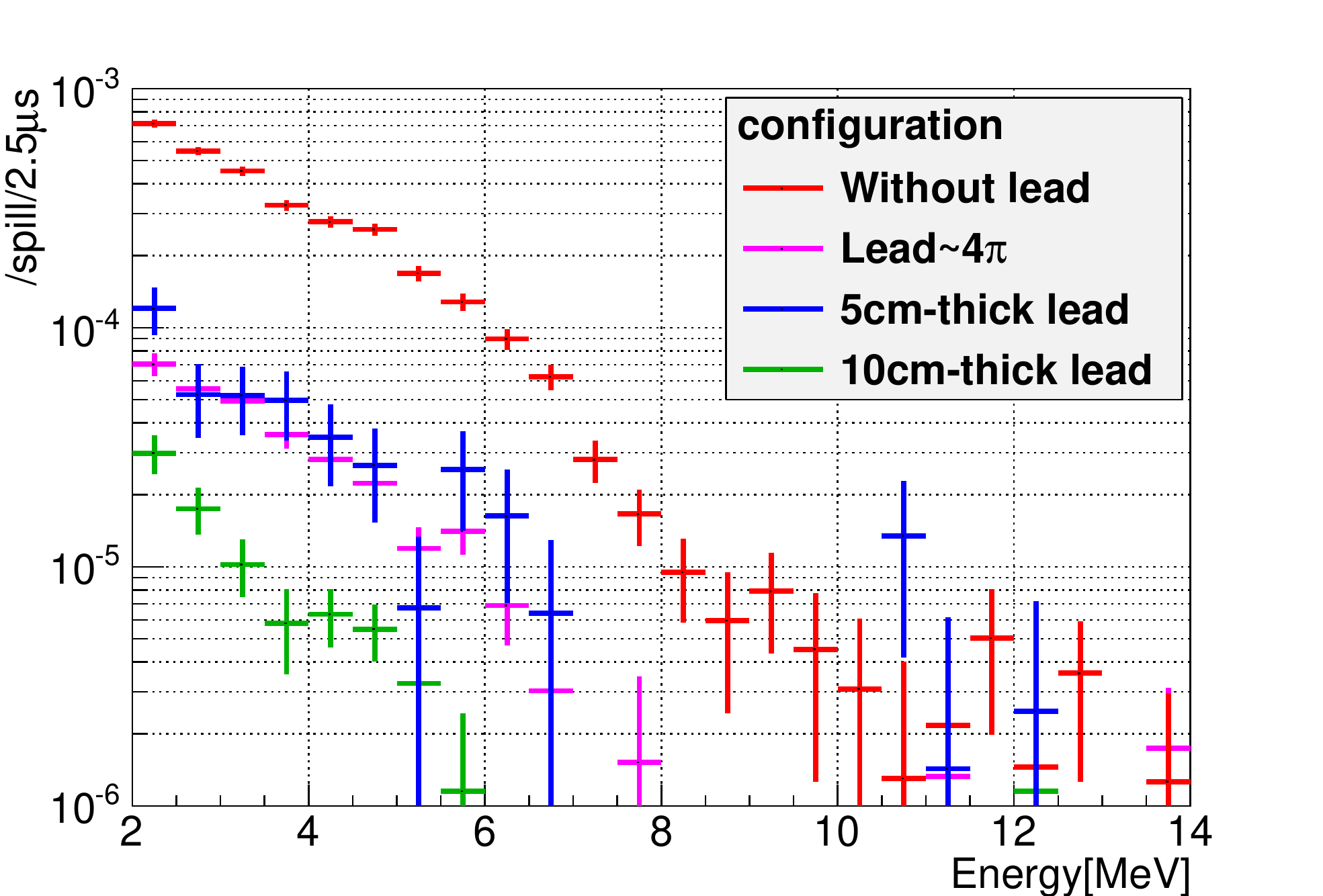}
\caption{Energy spectra of 4 configurations after subtraction of accelerator off data.}
\label{fig:tohoku24_spectrasubtracted}
\end{center}
\end{figure}

\indent
\indent
As shown in Fig.~\ref{fig:tohoku24_FADCvstime}, data includes beam bunch timing.
To avoid effects of beam activity, we use data 3.5 $\mu$s after the gate start time.
Converting FADC count into energy, the energy spectra of 4 configurations were made.
To remove the effect of environmental gammas and cosmic rays, 
we subtracted the accelerator off data from these spectra.
The energy spectra after the subtraction are shown in Fig.~\ref{fig:tohoku24_spectrasubtracted}.


\subsubsection{Validation of the MC model: rejection power and energy spectra}
\label{sec:tohoku24_validation}
\indent
\indent
For each configuration, the energy spectrum was reproduced by the MC.
According to the MC model, gammas are generated uniformly in a $5\times 5{\rm m}^2$ floor surface, where the small scintillator was centred.
Deposit energy was smeared with a 18\% energy resolution.

\begin{figure}
\begin{center}
\includegraphics[width=12cm]{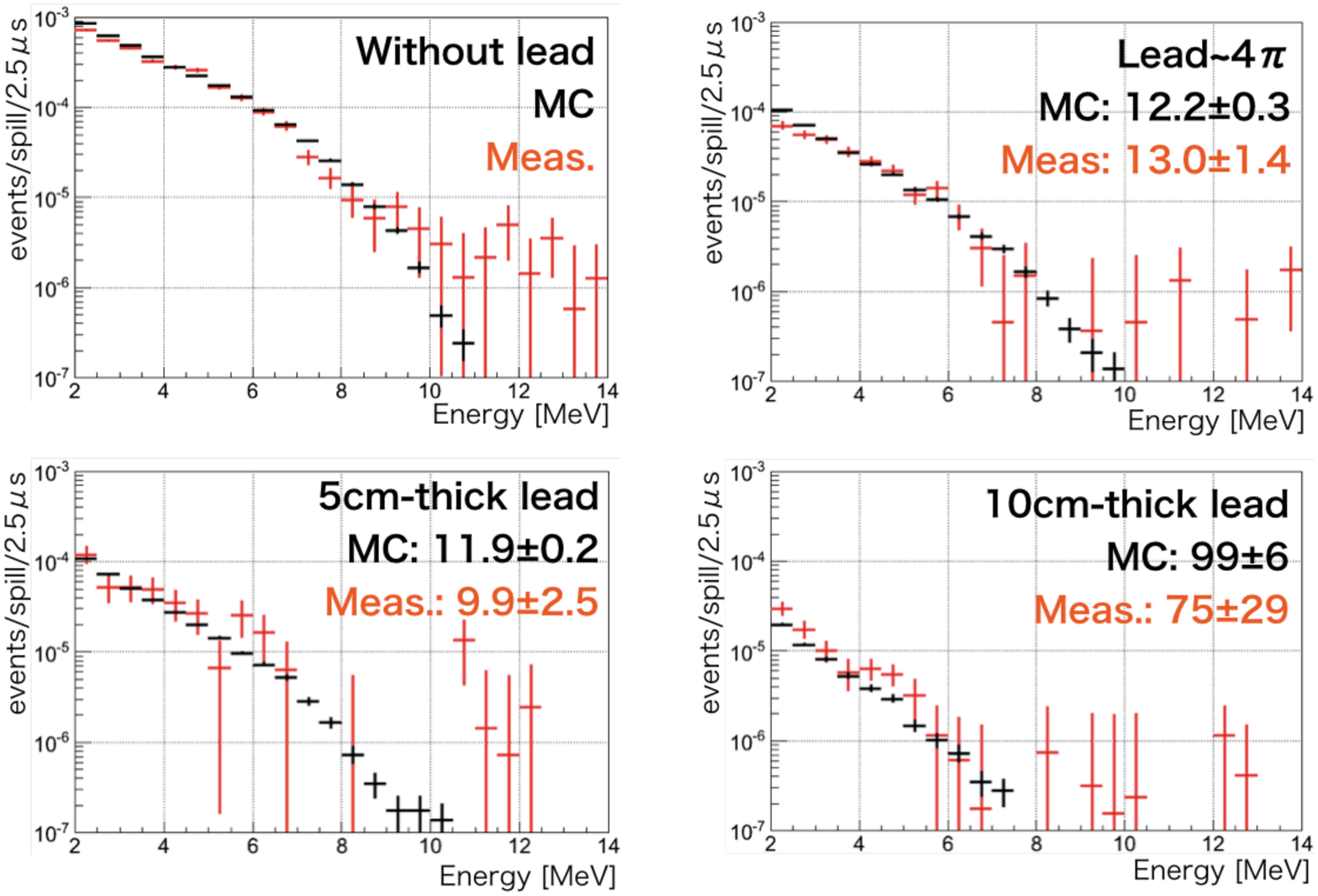}
\caption{\setlength{\baselineskip}{4mm}
Comparison between measurements and MC. The legend numbers are the rejection powers.}
\label{fig:tohoku24_comparison}
\end{center}
\end{figure}

To validate the MC model for gamma background, we compared rejection powers of measurements and MC.
The definition of rejection power, $r$, is;
\begin{equation}
r=\frac{N_{\rm Without \, lead}}{N_{\rm config}},
\end{equation}
where $N_{\rm Without \, lead}$ is the event rate of 4 $< E [{\rm MeV}] <$ 10 at ``Without lead" configuration and $N_{\rm config}$ is that at other configurations (``Lead$\sim 4\pi$", ``5cm-thick lead", ``10cm-thick lead").
In addition, we checked the energy spectra.
The rejection powers and energy spectra comparing between measurements and MC are shown in Fig.~\ref{fig:tohoku24_comparison}.
For normalization, the energy spectra of MC were multiplied by a factor, 
$N^{\rm meas.}_{\rm Without \, lead}/N^{\rm MC}_{\rm Without \, lead}$, 
where $N^{\rm meas.}_{\rm Without \, lead}$ is event rate of measurement and $N^{\rm MC}_{\rm Without \, lead}$ is event number of MC in energy region of 4 $< E [{\rm MeV}] <$ 10 at ``Without lead" configuration.

The MC model explains the measurements well because the rejection powers for 3 configurations are consistent between measurements and MC, and the energy spectra shape of MC also reproduce the measurements below 9 MeV.
Therefore the MC model was validated and the lead shield effect was confirmed.
